\PassOptionsToPackage{numbers,sort&compress}{natbib}
\documentclass[aps,pra,reprint,amsmath,amssymb,superscriptaddress,floatfix]{revtex4-2}

\usepackage{graphicx}
\usepackage{amsmath}
\usepackage{amssymb}
\usepackage{bm}
\usepackage{physics}
\usepackage{xcolor}
\usepackage{booktabs}
\usepackage{tikz}
\usetikzlibrary{arrows.meta,positioning,decorations.pathreplacing}
\usepackage[colorlinks=true,allcolors=blue]{hyperref}

\usepackage[protrusion=true,expansion=true,final]{microtype}

\begin{document}

\title{Optimal control for duty-cycle-limited interferometry with single-NV centers}

\author{U\u{g}ur Tamer}
\email{utamer24@ku.edu.tr}
\affiliation{Department of Physics, Ko\c{c} University, 34450 Sarıyer, Istanbul, Türkiye}

\author{Sina Zeytino\u{g}lu}
\email{sina.zeytinoglu@tuwien.ac.at}
\affiliation{Institute for Theoretical Physics, TU Wien, Wiedner Hauptstraße 8-10/136, A-1040 Vienna, Austria}

\author{\"Ozg\"ur E. M\"ustecapl{\i}o\u{g}lu}
\email{omustecap@ku.edu.tr}
\affiliation{Department of Physics, Ko\c{c} University, 34450 Sarıyer, Istanbul, Türkiye}
\affiliation{TÜBİTAK Research Institute for Fundamental Sciences (TBAE), 41470 Gebze, Türkiye}

\date{\today}

\begin{abstract}
Stimulus-responsive hydrogels convert temperature changes into magnetic-field shifts detectable by nitrogen-vacancy (NV) centers, enabling nanoscale thermometry in soft and biological environments. Existing hydrogel--nanodiamond demonstrations rely on NV ensembles, whose high photon throughput is accompanied by gradient-induced inhomogeneous broadening, while idealized single-NV projections assume high-fluence fluorescence/ODMR readout. Here we study a pulsed single-NV route for the same class of sensors and ask whether decoherence-aware coherent control can improve thermometric performance over optimized Ramsey interrogation at equal detected-photon budget. Using a sigmoidal volume-phase-transition model, dipolar magnetic transduction, and Lindblad master-equation simulations, we find a reproducible $25$--$27\%$ per-shot sensitivity gain over optimized Ramsey, i.e., a $57$--$60\%$ gain in Fisher information ($1.57$--$1.60$). The same gain carries over to the photon-normalized Fisher information. The rate gain is governed by the measurement duty cycle, the fraction of the experimental cycle spent accumulating signal rather than initializing, reading out or waiting, and becomes largest when the overhead or optical-dose constraint dominates the cycle time. The optimized trajectories reveal a response-shaping mechanism in which phase accumulation is concentrated near the end of the sequence, and a closed-form depth-two solution reproduces the numerical optimum and exhibits that mechanism analytically. The advantage is most pronounced when the dephasing time is short compared with the measurement overhead or dose-limited waiting time, which is the operating regime targeted by hydrogel-transduced single-spin biosensing.
\end{abstract}

\maketitle

\section{Introduction}
\label{sec:intro}

The ability to measure temperature with high spatial and thermal resolution at the nanoscale is a long-standing goal in biology, chemistry, and materials science~\cite{brites2012,kucsko2013,toyli2013,neumann2013}. Sensitive probing of thermal gradients at subcellular length scales (for example, during mitochondrial activity, photothermal cancer treatment, or heat dissipation in nanoelectronics) demands thermometers that combine nanometer spatial resolution with millikelvin sensitivity under ambient, biologically compatible conditions~\cite{kucsko2013,toyli2013,neumann2013,simpson2017,mamin2013,staudacher2013}.

Stimulus-responsive transducers based upon volume phase transitions of hydrogels have emerged as a promising route: they convert otherwise weakly accessible thermodynamic variables (temperature, pH, glucose concentration, enzyme activity) into magnetic fields measurable by color-center spins~\cite{zhang2018,rendler2017}. In this scheme, a temperature change drives a significant change in the hydrogel's volume, which alters the effective NV--nanomagnet separation $r(T)$, thereby shifting the NV's resonant frequencies by several MHz per kelvin near the critical temperature~\cite{zhang2018}. The architecture is thus a two-stage transduction chain: a mechanical amplifier (the hydrogel volume phase transition) followed by a quantum-optical magnetic field detector (the NV spin)~\cite{zhang2018,wang2018,liu2021}.

\begin{figure*}[t]
\centering
\includegraphics[width=\textwidth]{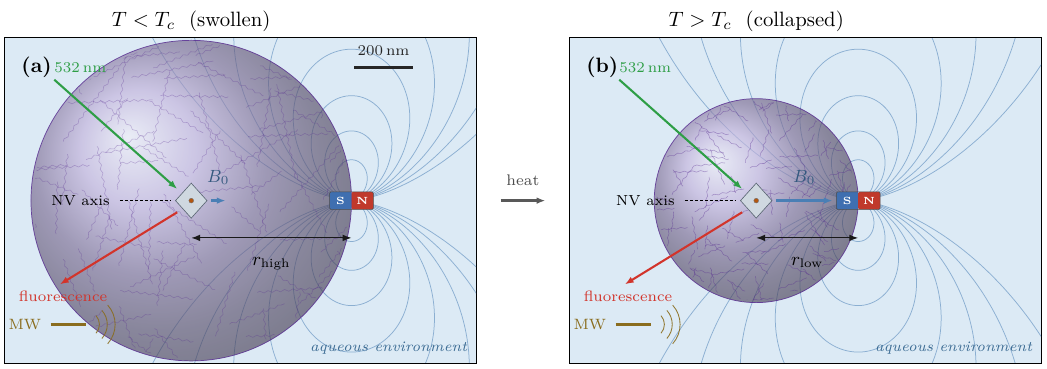}
\caption{Schematic of the hydrogel-transduced single-NV sensor in its two limiting states.
(a) Below the critical temperature the polymer network is hydrated and swollen, holding the
surface-bound nanomagnet at a separation $r_\mathrm{high}$ from the NV center in the nanodiamond.
(b) Heating through the volume phase transition collapses the network to $r_\mathrm{low}$, which
brings the magnet closer and raises the dipolar field at the NV as $r^{-3}$ following
Eqs.~(\ref{eq:hydrogel}) and~(\ref{eq:dipole}); for the radii used here this is a $3.9$-fold
enhancement. The magnetic moment lies along the NV--magnet separation vector, so $B_0$ is parallel
to the NV symmetry axis and Eq.~(\ref{eq:hamiltonian}) applies with no projection factor. Green and
red arrows mark optical initialization at $532$~nm and spin-dependent fluorescence; MW denotes the
microwave control field. Field lines follow the point-dipole field-line equation, and the chains
inside each bead indicate the network density. Both panels are drawn to the common scale bar.}
\label{fig:schematic}
\end{figure*}

The initial experimental realization of this concept employed an ensemble of NV centers in a nanodiamond and achieved a sensitivity of $\eta \approx 96\,\mathrm{mK\,Hz^{-1/2}}$ in water~\cite{zhang2018}. The ensemble approach, however, carries an inherent cost: the large magnetic field gradient generated by the proximal nanomagnet inhomogeneously broadens the ODMR spectrum of the NV ensemble, washing out the signal and limiting the sensitivity gain~\cite{zhang2018,liu2021,taylor2008}. Moreover, ensemble readout requires sustained high-power CW illumination ($\sim 100\,\mathrm{\mu W}$ of focused $532\,\mathrm{nm}$ light~\cite{zhang2018}), placing the sensor in an optical fluence regime that approaches the threshold for phototoxicity and local heating in live-cell environments~\cite{yu2005,simpson2017}. Neumann \textit{et al.}\ demonstrated single-NV thermometry in this regime~\cite{neumann2013}, and the supplementary analysis of Ref.~\cite{zhang2018} projects that a single NV center would eliminate gradient-induced inhomogeneous broadening and could, in principle, reach $\sim 0.3\,\mathrm{mK\,Hz^{-1/2}}$. That projection carries a readout requirement: laser intensities exceeding $100\,\mathrm{\mu W\,\mu m^{-2}}$, the fluence needed to sustain $\sim 10^{5}$ counts per second from one emitter under continuous ODMR. The experimental demonstration by Liu \textit{et al.}\ on a hybrid single-NV diamond nanopillar coupled to a single Cu-Ni nanoparticle near its Curie temperature confirmed this picture, achieving $\eta = 76\,\mathrm{\mu K\,Hz^{-1/2}}$ by eliminating gradient broadening~\cite{liu2021}. The sensitivities demonstrated so far nevertheless fall short of the target set by intracellular thermometry, namely millikelvin resolution at an optical dose that live-cell operation tolerates~\cite{kucsko2013,yu2005}, so further improvement of these probes is important.

The primary challenge for any single-spin sensor operating in a pulsed regime is its short coherence time $T_2$, especially in the room-temperature, liquid environments required for biological applications~\cite{degen2017,awschalom2018,bargill2013}. This decoherence limits the total sensing time and the achievable precision. Standard Ramsey interferometry reaches its peak sensitivity at an evolution time $t \sim T_2^*$, beyond which the signal decays exponentially and the protocol becomes ineffective. We therefore ask whether a pulsed single-NV protocol retains useful sensitivity under realistic decoherence, finite microwave-pulse overhead, and a limited photon budget. Such a protocol is duty-cycle limited: one experimental cycle consists of optical initialization, coherent evolution, optical readout and any enforced waiting, and only the coherent-evolution part carries temperature information, so only a useful fraction of the cycle determines the information acquired per unit time.

Coherent control has been used to push against this limit and succeeded in increasing the useful information content per time. Control fields shaped against a cost function that includes the environment were shown to improve the sensitivity of a quantum sensor for AC signals~\cite{poggiali2018}. Optimal control protocols of quantum sensors improved the sensitivity of Ramsey interferometry with NV centers by optimizing its robustness to control-amplitude drift~\cite{oshnik2022}. Recently, optimized continuous control protocols were used to improve the sensitivity of the Ramsey interferometry using superconducting qubits by stabilizing one Bloch component of the sensor during the data acquisition stage~\cite{hecht2025}. We ask how much of the gain during information acquisition survives when the optimized resource is instead the full measurement cycle. We show that optimal control can improve the Fisher information rate of the full Ramsey interferometry duty cycle by improving its robustness against dephasing.

In this work we formulate a pulsed single-NV protocol in which temperature is encoded in the phase accumulated during microwave-controlled free evolution, and we compare optimized Ramsey interrogation with QSP-inspired coherent response synthesis implemented using finite-depth Euler-control blocks. The processing blocks are parameterized by a generalized Euler decomposition, Eq.~(\ref{eq:processing_op_euler}), which gives $3(d+1)$ real control parameters. This finite-depth ansatz contains phase-optimized Ramsey and canonical single-angle QSP as special cases. We use it as a decoherence-aware response-synthesis family generated by the same microwave rotations used in Ramsey experiments.

The control family we use is inspired by quantum signal processing (QSP)~\cite{low2017}: phase-controlled microwave rotations interleaved with free evolution, whose phases are chosen to shape the response of the spin to the temperature-induced frequency shift. It sits inside the broader family of quantum optimal control methods applied to NV sensing~\cite{rembold2020}, and differs from most of them in the character of the signal: the thermometric shift is static (i.e., DC), generated by the same operator as the dominant dephasing. Section~\ref{sec:qsp-context} places the method against composite-pulse, hyper-Ramsey and continuous-drive approaches, and Sec.~\ref{sec:qsp-ideal} gives the construction itself.

The result is a finite-depth gain over phase-optimized Ramsey that costs no additional detected photons per shot. The gain is a constant-factor improvement under Markovian dephasing, but it is reproducible and survives Fisher-information and photon-budget normalization. The operational rate gain is governed by the duty cycle: fast electronic readout favors shallow sequences, while repetitive readout and dose-limited operation increase the overhead and allow the gain to approach the per-shot ceiling. We use the hydrogel--nanodiamond ensemble sensor of Ref.~\cite{zhang2018} as the concrete experimental benchmark. The improvement belongs to the interrogation protocol, so the framework applies equally to other stimulus-responsive hydrogels~\cite{rendler2017}, magnetostrictive layers~\cite{cai2014}, and biochemical-to-magnetic transducers.

We also identify the control mechanism behind the long-time equal-time response. The optimized sequence stores the spin close to the longitudinal axis during early blocks and performs the phase-sensitive part of the measurement late in the sequence. This response-shaping mechanism explains the large equal-time ratios and fixes the range of the metrological claim; at depth two it admits a closed-form solution.

The manuscript is organized as follows. Section~\ref{sec:model} sets out the model: the NV Hamiltonian, the sigmoidal volume-phase-transition and dipolar transduction chain, and the control primitives. Section~\ref{sec:methods} defines the method, placing the QSP-inspired family against related control frameworks (Sec.~\ref{sec:qsp-context}), giving the construction and the operating point, and defining the Fisher-information, photon-counting and duty-cycle quantities used throughout. Section~\ref{sec:results} reports the results: the Ramsey benchmark, the per-shot gain and its analytic depth-two form, the photon budget, the duty-cycle dependence of the rate gain, the relation to the dephasing-limited scale, the control mechanism behind the long-time response, and robustness and readout normalization. Section~\ref{sec:conclusion} concludes. Derivations, unit calibrations and supporting numerics are collected in the appendices.

\section{Model}
\label{sec:model}

Our physical system, depicted conceptually in Fig.~\ref{fig:schematic}, consists of a single nanodiamond containing one NV center~\cite{doherty2013}, whose spin-triplet ground state is initialized by optical pumping and read out through spin-dependent fluorescence~\cite{rondin2014}, coupled to a thermo-responsive hydrogel via an effective NV--nanomagnet separation $r(T)$; the framework does not require the NV to reside exactly at the geometric center of a perfectly spherical shell. A nanomagnet, acting as the source of a static bias field, is attached to the hydrogel surface.

\subsection{Hamiltonian}

The ground state of the NV center is a spin-1 system. Under an external magnetic field $B_0$ aligned with the NV's symmetry axis ($z$-axis), its Hamiltonian is given by
\begin{equation}
H_{\text{NV}} = D S_z^2 + \gamma_e B_0 S_z,
\label{eq:hamiltonian}
\end{equation}
where $S_z$ is the spin-1 operator for the $z$-projection of the electron spin, $D \approx 2.87$~GHz is the zero-field splitting, and $\gamma_e \approx 28$~GHz/T is the electron gyromagnetic ratio~\cite{doherty2013}.
Throughout this work we adopt the convention $\hbar = 1$,
so that $D$ and $\gamma_e$ are expressed in angular
frequency units (i.e., $2\pi \times 2.87\,\mathrm{GHz}$
and $2\pi \times 28\,\mathrm{GHz/T}$ in $\mathrm{rad/s}$
when explicitly needed; see Table~\ref{tab:units}).
This system has three energy levels corresponding to $m_s = 0, \pm 1$. The magnetic field lifts the degeneracy of the $m_s = \pm 1$ states. We define our qubit in the subspace spanned by $\ket{0} \equiv \ket{m_s=0}$ and $\ket{1} \equiv \ket{m_s=-1}$. The energy gap of this effective spin-1/2 system is $\omega_q(T) = D - \gamma_e B_0(T)$. The nearby $^{13}$C or $^{14}$N nuclear spin is not used for sensing in the primary QSP-inspired protocol; Sec.~\ref{subsec:nuclear-memory} considers it as a quantum memory.

\subsection{Phenomenological Models for Thermometry}

The transduction from ambient temperature to the qubit frequency is a two-step process modeled phenomenologically. First, the hydrogel's radius, $r$, changes with temperature, $T$. We model this Volume Phase Transition (VPT) using a sigmoid function:
\begin{equation}
r(T) = r_{\text{low}} + \frac{r_{\text{high}} - r_{\text{low}}}{1 + \exp\bigl(k(T - T_c)\bigr)}.
\label{eq:hydrogel}
\end{equation}
For our simulations, we use realistic parameters for a poly($N$-isopropylacrylamide) (PNIPAM) hydrogel, which undergoes a sharp, reversible volume phase transition above a critical temperature~\cite{schild1992,heskins1968}: a shrunken radius $r_{\text{low}} = 350$~nm, a swollen radius $r_{\text{high}} = 550$~nm, a critical temperature $T_c = 305.15$~K ($\approx 32^\circ$C), and a transition steepness of $k = 0.5$~K$^{-1}$. The large change in radius, $\Delta r = r_{\text{high}} - r_{\text{low}}$, within a narrow temperature window around $T_c$ provides the primary mechanical amplification of the thermal signal.

Second, this change in radius modulates the magnetic field at the NV center. Modeling the nanomagnet as a magnetic dipole, the field strength at the origin scales as $1/r^3$:
\begin{equation}
B_0(r) = B_{\text{ref}} \left(\frac{r_{\text{ref}}}{r}\right)^3.
\label{eq:dipole}
\end{equation}
We set the reference radius $r_{\text{ref}} = 550$~nm (the swollen state) and a reference field $B_{\text{ref}} = 0.27$~mT. This field strength is chosen to be consistent with typical experimental ODMR splittings of approximately 15~MHz. This model transduces the large mechanical displacement into a significant change in the magnetic field. The combined effect of Eqs.~(\ref{eq:hydrogel}) and (\ref{eq:dipole}) is a highly non-linear function $\omega_q(T)$ that is engineered to have a very large derivative, $d\omega_q/dT$, specifically within the VPT regime, producing a narrow temperature window of strongly enhanced sensitivity. Our model geometry ($\Delta r = 200$~nm) yields a susceptibility $|d\omega_q/dT| \approx 2\pi \times 2.2$~MHz/K at the simulation target temperature $T_\mathrm{target} = 305.0$~K, rising to a peak of $\sim 2\pi \times 2.8$~MHz/K at $T \approx 306.9$~K. The overall setup is inspired by the hydrogel--nanodiamond architecture of Ref.~\cite{zhang2018}, but uses different geometric dimensions; the we choose model parameters representative of the class of hydrogel--nanomagnet sensors~\cite{zhang2018,wang2018}.

The absolute value of the transduction slope depends on the geometry. We use a point-dipole magnetic field, an effective NV--nanomagnet separation, and an aligned field component along the NV axis. These assumptions set the value of
\begin{equation}
\kappa = \left|\frac{d\omega_q}{dT}\right|,
\label{eq:kappa}
\end{equation}
which controls the absolute sensitivity. The protocol comparison is made at fixed $\kappa$ and is less sensitive to these geometric details than the absolute value of $\eta_T$. In an experiment the NV position inside the nanodiamond, the magnet shape, the field orientation and the hydrogel deformation profile all enter the calibration of $\kappa$.

\subsection{Initialization and Measurement}

The NV electron spin is initialized to the $\ket{0}$ state with high fidelity via optical pumping with a green laser. For readout, we employ a Ramsey-type measurement. While Optically Detected Magnetic Resonance (ODMR) is common for spectroscopy, a Ramsey sequence is a direct interferometric measurement of the qubit phase, which is the quantity shaped by the coherent-control sequence. The final state is projected onto the $Z$-basis via a readout pulse, and the population difference is measured through spin-dependent photoluminescence.

\subsection{Quantum Signal Processing Pulses}
\label{sec:qsp-pulses}

The QSP-inspired protocol is implemented by a sequence of microwave pulses. The free evolution of the qubit, which encodes the temperature-dependent signal $\omega_q$, is interleaved with processing pulses. These processing pulses are short, strong microwave drives applied along the $x$-axis of the qubit's Bloch sphere. We operate in the \emph{strong-drive approximation}, where the Rabi frequency of the pulse, $\Omega$, is much larger than any detuning from the unknown qubit gap ($\Omega \gg |\omega_{\text{drive}} - \omega_q(T)|$). This ensures that the processing rotations are high-fidelity and that their rotation axis is not significantly tilted by the unknown bias field, which is necessary for implementing the coherent-control sequence.

\section{Methods}
\label{sec:methods}

\subsection{QSP-inspired control and its relation to other frameworks}
\label{sec:qsp-context}

To implement this protocol we employ Quantum Signal Processing (QSP)~\cite{low2017}, a technique that uses a sequence of carefully controlled processing pulses to algorithmically shape the response of a qubit to a weak signal encoded in its phase. The construction has a long lineage. Composite pulses were developed in NMR to make a nominal rotation insensitive to a systematic error in the control, by replacing a single pulse with several whose phases are chosen so that the leading error terms cancel~\cite{levitt1986,wimperis1994}. The same idea was carried into precision spectroscopy, where hyper-Ramsey and related composite schemes suppress the frequency shift that the interrogation laser itself imposes on an optical clock transition, in some cases by three to four orders of magnitude~\cite{yudin2010,zanonwillette2016,zanonwillette2018}. QSP places this construction on a systematic footing: for a sequence of phase-controlled rotations by a common angle, it characterizes exactly which polynomial response functions are reachable and how to find the phases that realize a target response~\cite{lowyoderchuang2016,low2017}. Composite sequences have also been applied to spin magnetometry: rotary-echo driving of a single NV trades robustness against control-amplitude error for a longer usable coherence time~\cite{aiello2013}, although in an ac-sensing setting in which the signal frequency is distinguishable from the noise, unlike the dc, parallel-noise problem considered here. We use the same machinery for a different purpose. We shape the response function around the operating point so that its slope with respect to temperature is as large as the coherence budget allows, and we let the optimizer see the decoherence while it does so. Markovian decoherence generically suppresses Heisenberg-limited scaling in the asymptotic pulse-number sense~\cite{giovannetti2011,zhou2018,sekatski2017}. The optimizer instead finds decoherence-aware processing angles that maximize sensitivity within realistic biological coherence constraints. The method is formulated at the level of a generic temperature-to-frequency transduction function.

Quantum optimal control has been applied to NV sensing before~\cite{rembold2020}, and our setting differs from those applications in two specific ways. Poggiali \textit{et al.}\ optimize a control field against a sensitivity cost function that includes the environment, and demonstrate improved sensitivity for a single NV with no ancilla~\cite{poggiali2018}; their setting is ac sensing, where the field to be measured oscillates, the noise spectrum is concentrated elsewhere, and the control can act as a filter that passes the signal and rejects the noise, as dynamical decoupling does. Oshnik \textit{et al.}\ optimize the shape of the microwave pulses themselves, together with initialization and readout, and obtain magnetometry robust against drifts in the control amplitude~\cite{oshnik2022}. Closest to the present work, Hecht \textit{et al.}\ stabilize one Bloch component with a continuous drive and report an unconditional signal-to-noise gain over Ramsey on a superconducting qubit~\cite{hecht2025}; we compare against their results quantitatively in Sec.~\ref{sec:overhead}. Our problem admits neither a filter nor pulse-shape freedom. The hydrogel transduces temperature into a static shift of the qubit frequency, so the signal enters as $\sigma_z$, and the dominant room-temperature decoherence channel is dephasing, which enters as $\sigma_z$ as well: signal and noise are generated by the same operator, no filter function can separate them, and a refocusing sequence cancels the thermometric phase along with the noise (Sec.~\ref{sec:mechanism}). We also leave the primitive rotations as given and vary only their phases. This keeps the protocol calibratable with the hardware already used for a Ramsey sequence. The freedom that remains is the shape of the response function itself, and the accounting of the full duty cycle.

\subsection{Ideal QSP Protocol and Sensitivity}
\label{sec:qsp-ideal}

\begin{figure*}[t]
\centering
\begin{tikzpicture}[x=1.0cm, y=0.95cm, >=Latex, font=\footnotesize]

\colorlet{boxfill}{gray!35}
\colorlet{eulerfill}{blue!22!white}
\colorlet{eulerframe}{blue!55!black}
\colorlet{wfill}{orange!18!white}
\colorlet{wframe}{orange!60!black}

\node[anchor=west, font=\small\bfseries] at (0, 3.5) {(a) Ramsey sequence};
\draw[->, thick] (0, 2.2) -- (14.5, 2.2) node[right] {$t$};

\filldraw[fill=boxfill, draw=black, line width=0.6pt]
  (0.5,1.70) rectangle (1.7,2.70);
\node[above=3pt, font=\scriptsize] at (1.1,2.70) {$\pi/2$};
\node[below=3pt, font=\scriptsize] at (1.1,1.70) {MW pulse};

\draw[thick] (1.7,2.2) -- (11.2,2.2);
\node[above=5pt] at (6.45,2.2) {free evolution, time $\tau$};
\node[below=5pt] at (6.45,2.2) {$\omega_q(T)$ phase accumulates};

\filldraw[fill=boxfill, draw=black, line width=0.6pt]
  (11.2,1.70) rectangle (12.4,2.70);
\node[above=3pt, font=\scriptsize] at (11.8,2.70) {$\pi/2$};

\draw[->, thick] (12.4,2.2) -- (14.0,2.2);
\node[font=\scriptsize] at (13.2, 1.55) {optical readout};

\node[anchor=west, font=\small\bfseries] at (0,-0.3) {(b) QSP-inspired coherent-control sequence};
\draw[->, thick] (0,-1.6) -- (15.8,-1.6) node[right] {$t$};
\node[above=14pt, font=\scriptsize] at (11.5,-1.6) {$t = d\tau$};

\filldraw[fill=boxfill, draw=black, line width=0.6pt]
  (0.2,-2.10) rectangle (1.2,-1.10);
\node[above=3pt, font=\scriptsize] at (0.7,-1.10) {$\pi/2$};
\node[below=3pt, font=\scriptsize] at (0.7,-2.10) {prep};

\filldraw[fill=eulerfill, draw=eulerframe, line width=0.7pt]
  (1.2,-2.10) rectangle (3.4,-1.10);
\node[font=\scriptsize] at (2.3,-1.42) {$R_z R_x R_z$};
\node[font=\scriptsize, text=eulerframe!80!black] at (2.3,-1.88) {$(\phi_d)$};

\filldraw[fill=wfill, draw=wframe, line width=0.7pt]
  (3.4,-1.90) rectangle (5.0,-1.30);
\node[font=\scriptsize] at (4.2,-1.60) {$W(\theta)$};
\node[above=2pt, font=\scriptsize] at (4.2,-1.30) {$\tau$};

\filldraw[fill=eulerfill, draw=eulerframe, line width=0.7pt]
  (5.0,-2.10) rectangle (7.2,-1.10);
\node[font=\scriptsize] at (6.1,-1.42) {$R_z R_x R_z$};
\node[font=\scriptsize, text=eulerframe!80!black] at (6.1,-1.88) {$(\phi_{d-1})$};

\filldraw[fill=wfill, draw=wframe, line width=0.7pt]
  (7.2,-1.90) rectangle (8.8,-1.30);
\node[font=\scriptsize] at (8.0,-1.60) {$W(\theta)$};
\node[above=2pt, font=\scriptsize] at (8.0,-1.30) {$\tau$};

\node[font=\large] at (9.4,-1.60) {$\cdots$};
\node[font=\scriptsize] at (9.4,-2.45) {$\times d$ layers};

\draw[thick] (9.9,-1.6) -- (10.1,-1.6);
\filldraw[fill=eulerfill, draw=eulerframe, line width=0.7pt]
  (10.1,-2.10) rectangle (12.3,-1.10);
\node[font=\scriptsize] at (11.2,-1.42) {$R_z R_x R_z$};
\node[font=\scriptsize, text=eulerframe!80!black] at (11.2,-1.88) {$(\phi_0)$};

\draw[->, thick] (12.3,-1.6) -- (14.3,-1.6);
\node[font=\scriptsize] at (13.3,-2.10) {optical readout};

\filldraw[fill=wfill, draw=wframe, line width=0.6pt]
  (0.2,-3.30) rectangle (1.60,-2.75);
\node[font=\scriptsize] at (0.90,-3.025) {$W(\theta)$};
\node[right=5pt, font=\scriptsize] at (1.60,-3.025)
  {signal encoding: $W(\theta)=\exp(-\tfrac{i\theta}{2}\sigma_z)$, \quad $\theta=\omega_q\tau$};

\filldraw[fill=eulerfill, draw=eulerframe, line width=0.6pt]
  (0.2,-4.20) rectangle (1.60,-3.65);
\node[font=\scriptsize] at (0.90,-3.925) {$R_z R_x R_z$};
\node[right=5pt, font=\scriptsize] at (1.60,-3.925)
  {Euler processing pulse: $R_z(\alpha_k)\,R_x(\beta_k)\,R_z(\gamma_k)$};

\end{tikzpicture}

\vspace{2ex}

\includegraphics[width=\textwidth]{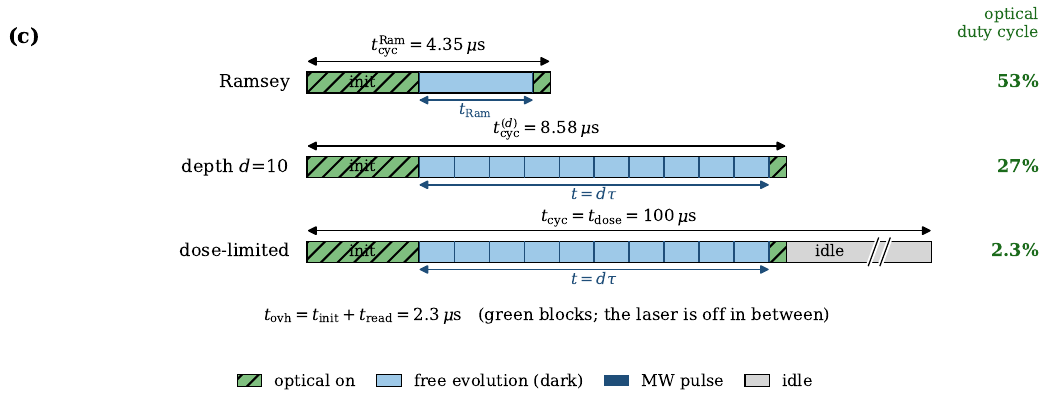}

\caption{Pulse sequences and the measurement cycle. (a) Ramsey protocol: an initial $\pi/2$ pulse, free evolution under $\omega_q(T)$ for time $\tau$, and a final $\pi/2$ readout pulse. (b) Coherent-control protocol following Eq.~(\ref{eq:qsp_sequence}): Euler processing pulses $R_z(\alpha_k)R_x(\beta_k)R_z(\gamma_k)$ interleaved with free-evolution blocks $W(\theta)$, proceeding from $\phi_d$ down to $\phi_0$, giving a total interrogation time $t=d\tau$. (c) One full experimental cycle for both protocols, drawn to scale with $t_\mathrm{init}=2.0\,\mu$s, $t_\mathrm{read}=0.3\,\mu$s and $t_\mathrm{p}=25$~ns for the illustrated $\pi/2$ pulses; the rate calculations below conservatively assign $t_\mathrm{p}=50$~ns to each processing block unless stated otherwise. Optical excitation enters only through initialization and readout (green), and the laser is off during the dark interval in which the temperature-dependent phase accumulates. The deeper sequence interrogates for longer and therefore runs at a lower repetition rate, which is the cost that offsets its per-shot advantage at small overhead (Sec.~\ref{sec:rate}). Under a dose ceiling (bottom row, time axis broken) the idle period absorbs that difference, the cycle time becomes common to the two protocols, and the rate gain approaches the per-shot value (Sec.~\ref{sec:overhead}). The percentages give the fraction of the cycle during which the laser is on; only the upper two rows share a common time axis.}
\label{fig:ramsey_qsp_schematic}
\end{figure*}
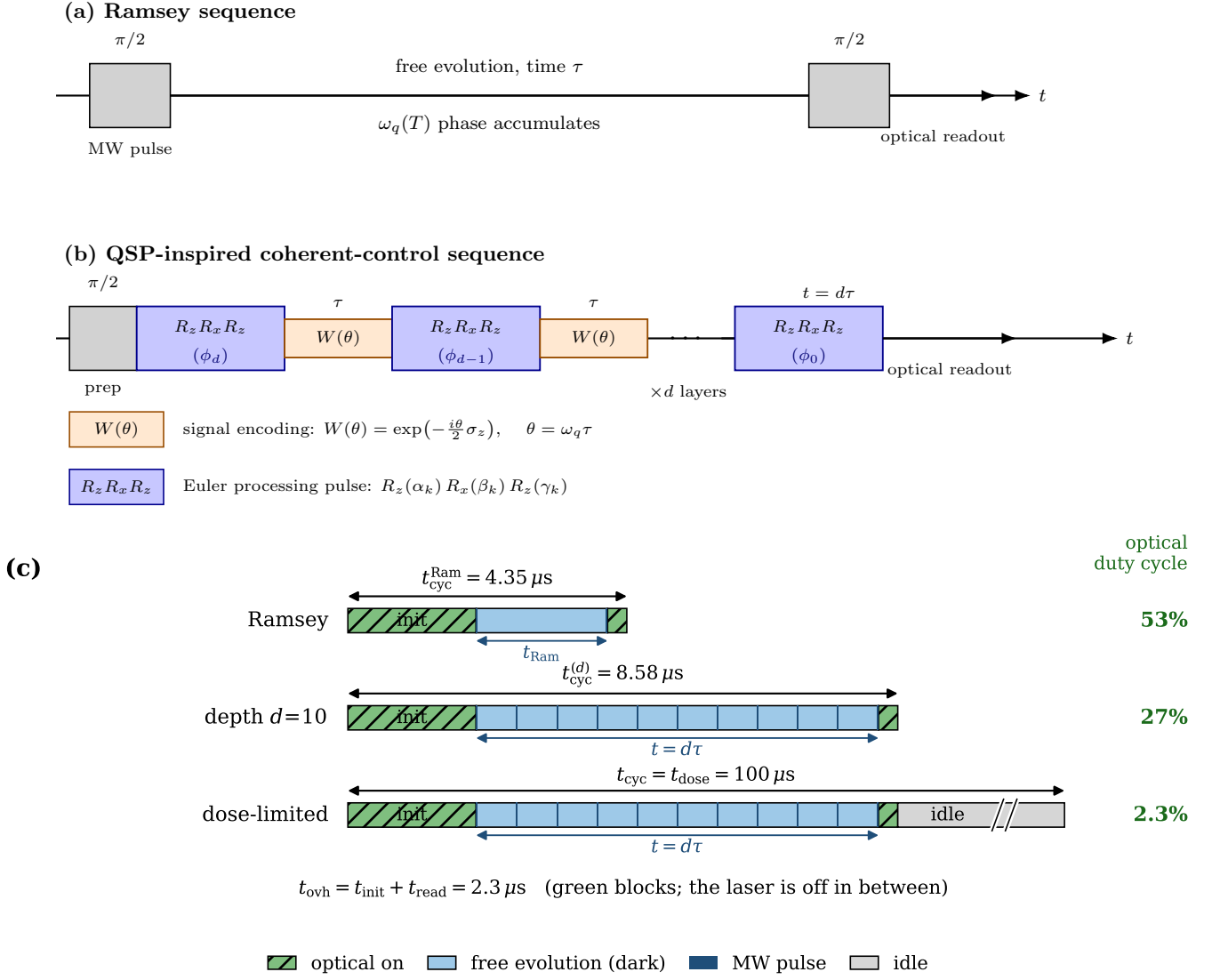

A QSP-inspired coherent-control sequence is constructed from two unitary operations. The first is the signal-encoding operator, representing free evolution for a time $\tau$ under the qubit Hamiltonian $H_q = (\omega_q/2)\sigma_z$. This operator acquires the unknown phase $\theta = \omega_q \tau$ and can be written as
\begin{equation}
W(\theta) = \exp\!\left(-i\frac{\theta}{2}\sigma_z\right) = \cos\!\left(\frac{\theta}{2}\right)I - i\sin\!\left(\frac{\theta}{2}\right)\sigma_z.
\label{eq:signal_op}
\end{equation}
The second is the processing operator, a controllable microwave pulse that rotates the qubit state around the $x$-axis by an angle $\phi_k$:
\begin{equation}
R_x(\phi_k) = \exp\!\left(-i\frac{\phi_k}{2}\sigma_x\right) = \cos\!\left(\frac{\phi_k}{2}\right)I - i\sin\!\left(\frac{\phi_k}{2}\right)\sigma_x.
\label{eq:processing_op}
\end{equation}
For a sequence with $d$ layers (or $d$ repetitions of the signal-encoding step), the total unitary evolution is given by the product of interleaved signal-encoding and processing operators sketched in Fig.~\ref{fig:ramsey_qsp_schematic}:
\begin{equation}
\begin{aligned}
U_d(\theta, \{\phi_k\}) &= R_x(\phi_d)\, W(\theta)\, R_x(\phi_{d-1}) \\
&\quad \cdots R_x(\phi_1)\, W(\theta)\, R_x(\phi_0).
\end{aligned}
\label{eq:qsp_sequence}
\end{equation}
A key result of QSP theory is that this sequence implements a polynomial transformation~\cite{low2017}. After initializing the qubit in $\ket{0}$ and applying the sequence, the final expectation value of the $\sigma_z$ operator can be shown to be a polynomial of degree at most $d$ in the variable $x = \cos(\theta)$ (see Appendix for a more detailed derivation).
\begin{equation}
\langle\sigma_z\rangle_d = P_d\bigl(\cos(\omega_q \tau);\, \{\phi_k\}\bigr).
\label{eq:polynomial}
\end{equation}
The coefficients of this polynomial, $P_d$, are determined by the set of processing angles $\{\phi_k\}$. This property lets one shape the response function by choosing the appropriate angles. For example, specific choices of $\{\phi_k\}$ can make $P_d(x)$ approximate Chebyshev or Laurent polynomials, which are optimal for various signal processing tasks.

In this work, rather than approximating a specific polynomial, we use a numerical optimizer to directly search the space of processing angles to shape the response function $P_d$ in a way that maximizes our cost function. In our numerical implementation, each processing block is parameterized by a generalized Euler decomposition
\begin{equation}
U_k = R_z(\alpha_k)\,R_x(\beta_k)\,R_z(\gamma_k),
\label{eq:processing_op_euler}
\end{equation}
which gives the optimizer $3(d+1)$ free parameters instead of the single angle per layer in Eq.~(\ref{eq:processing_op}). This generalization does not change the underlying polynomial structure of the QSP response, but it gives the classical optimizer more freedom to compensate for decoherence. Because the controls are more general than canonical single-angle QSP, we call the method QSP-inspired, or generalized QSP response synthesis.

The per-shot response of the spin to temperature changes is quantified by the thermal spin susceptibility,
\begin{equation}
\chi_T = \left|\frac{d\langle\sigma_z\rangle}{dT}\right|,
\label{eq:sensitivity}
\end{equation}
which measures the slope of the spin expectation value with respect to temperature. In the ideal, decoherence-free limit, a QSP-inspired coherent-control sequence can exhibit sensitivity growth with the number of layers $d$; however, Markovian decoherence generically suppresses this scaling~\cite{giovannetti2011,zhou2018}, and our optimizer instead finds decoherence-aware angles that maximize sensitivity within the available coherence window. The slope $\chi_T$ alone is an incomplete figure of merit, because a large slope is only useful if the output probability is not in an unfavorable saturated region and if it is obtained at an acceptable cost in time and detected photons. The remainder of this section defines the quantities used throughout.

\subsection{Operating point, control parameters, and timing}
\label{sec:definitions}

All protocols are optimized and compared at a single local operating point,
\begin{equation}
T_\mathrm{target} = 305.0~\mathrm{K},
\label{eq:Ttarget}
\end{equation}
chosen near the PNIPAM critical temperature where $|d\omega_q/dT|$ is large. This is a local operating point; Sec.~\ref{sec:temp-window} addresses the width of the useful window.

Each QSP-inspired processing block is the Euler rotation of Eq.~(\ref{eq:processing_op_euler}), so the full control vector optimized at fixed $(d,t)$ is
\begin{equation}
\bm{\phi} = (\alpha_0,\beta_0,\gamma_0,\;\ldots\;,\alpha_d,\beta_d,\gamma_d),
\qquad \alpha_k,\beta_k,\gamma_k \in [-\pi,\pi].
\label{eq:control-vector}
\end{equation}
Because the final block contains free $z$-rotations, the readout phase is contained in $\bm{\phi}$ and does not need to be optimized separately for the coherent-control sequences. For the Ramsey baseline the corresponding parameter is the readout phase $\Phi_\mathrm{read}$, the phase of the final analysis pulse, which rotates the accumulated phase into the measured $\sigma_z$ population difference. The Ramsey baseline is optimized over $\Phi_\mathrm{read}$ independently at every interrogation time; this calibration is what places the two protocols on a common footing, as shown in Appendix~\ref{sec:ramsey-calib}.

The free-evolution block duration is $\tau$, and a depth-$d$ sequence accumulates signal for a total interrogation time
\begin{equation}
t = d\,\tau .
\label{eq:t-total}
\end{equation}
In a scan over $(d,t)$ we therefore set $\tau = t/d$. Processing pulses are treated as instantaneous unitaries in the dynamics, consistent with the strong-drive approximation of Sec.~\ref{sec:qsp-pulses}; their duration $t_\mathrm{p}$ enters only the cycle-time bookkeeping,
\begin{align}
t_\mathrm{cyc}^{(d)} &= t_\mathrm{init} + t_\mathrm{read} + t_\mathrm{dead} + d\tau + (d+1)\,t_\mathrm{p},
\label{eq:tcyc-qsp}\\
t_\mathrm{cyc}^\mathrm{Ram} &= t_\mathrm{init} + t_\mathrm{read} + t_\mathrm{dead} + t_\mathrm{Ram} + 2\,t_\mathrm{p},
\label{eq:tcyc-ram}
\end{align}
with $t_\mathrm{init} = 2~\mu\mathrm{s}$, $t_\mathrm{read} = 300$~ns and $t_\mathrm{dead} = 0$ (Table~\ref{tab:units}). Figure~\ref{fig:ramsey_qsp_schematic}(c) shows one full cycle for both protocols drawn to scale, together with the fraction of it during which the laser is on. The QSP-inspired coherent-control sequence uses $d+1$ processing pulses against two for Ramsey, so $t_\mathrm{p}$ introduces a depth-dependent overhead. We report two assumptions throughout: $t_\mathrm{p} = 0$, and $t_\mathrm{p} = 50$~ns, corresponding to a Rabi frequency $\Omega/2\pi \approx 10$~MHz. At that Rabi frequency $50$~ns is a $\pi$ pulse and a $\pi/2$ pulse takes $25$~ns, so charging every processing block the full $50$~ns is an upper bound on the per-pulse duration. This is conservative for the present comparison, since the controlled sequence pays $d+1$ such pulses against two for Ramsey. Unless stated otherwise, results use $t_\mathrm{p} = 50$~ns. Because $t_\mathrm{p}$ enters only Eqs.~(\ref{eq:tcyc-qsp})--(\ref{eq:tcyc-ram}), the optimal angles are unchanged and any other value can be substituted without repeating the optimization. The dynamics thus treat processing pulses as strong, short rotations while their duration is carried by the cycle time. The approximation is accurate when $\Omega$ is large compared with the detuning over the operating range and when the accumulated pulse time stays short compared with the relevant decoherence time. The main operational optimum at nominal electronic readout occurs at shallow depth, where the approximation is strongest; the large-depth repetitive-readout values should be read as the ideal strong-pulse limit of the same model. Comparisons at equal interrogation time use $t_\mathrm{Ram} = t = d\tau$, which we call the nominal interrogation time because it excludes the pulse durations counted separately in Eq.~(\ref{eq:tcyc-qsp}); for global rate comparisons each protocol is instead maximized over its own interrogation time.

The two contributions to the cycle time play different roles. The pulse term $(d+1)t_\mathrm{p}$ is the price of the sequence itself and grows with depth. The term
\begin{equation}
t_\mathrm{ovh} = t_\mathrm{init} + t_\mathrm{read} + t_\mathrm{dead}
\label{eq:tovh}
\end{equation}
is a fixed cost per shot that both protocols pay identically and that does not grow with $t$ or with $d$. As Sec.~\ref{sec:overhead} shows, the rate comparison is governed almost entirely by the dimensionless ratio $t_\mathrm{ovh}/T_2^*$, which makes the conclusion transferable to platforms with different absolute timings.

Our nominal $t_\mathrm{ovh} = 2.3~\mu$s corresponds to conventional single-shot fluorescence readout of the electronic spin, and is representative of reported practice: initialization pulses of order one microsecond, a wait of about $300$~ns set by the metastable-state lifetime, and a photon collection window of comparable length~\cite{oshnik2022}. With $T_2^* = 2~\mu$s this places the sensor at $t_\mathrm{ovh}/T_2^* = 1.15$. Two experimentally standard situations push the ratio far higher. Nuclear-spin-assisted repetitive readout raises the readout fidelity by mapping the electronic state onto a nuclear ancilla and interrogating it many times, at the cost of a readout lasting from tens of microseconds to milliseconds~\cite{jiang2009,zhao2024}; this alone gives $t_\mathrm{ovh}/T_2^* \simeq 50$--$500$. Independently, in a live-cell environment the optical dose sets the limit, which is treated next.

\subsection{Dose-limited operation}
\label{sec:dose}

Each shot delivers a fixed optical dose during initialization and readout, independent of the interrogation time, so the average dose rate is set by the repetition rate. Holding it below a ceiling tolerated by the sample therefore imposes a minimum cycle time $t_\mathrm{dose}$, which the experimenter realizes by idling between shots. The cycle time is then
\begin{equation}
t_\mathrm{cyc} = \max\!\left(t_\mathrm{ovh} + t + n_\mathrm{p} t_\mathrm{p},\; t_\mathrm{dose}\right),
\label{eq:tcyc-dose}
\end{equation}
with $n_\mathrm{p} = d+1$ for QSP and $2$ for Ramsey. The idle time is how the dose budget is met. Because $t_\mathrm{dose}$ is a constant common to both protocols that does not grow with $t$, it acts exactly as a large $t_\mathrm{ovh}$; the consequences are quantified in Sec.~\ref{sec:overhead}.

Pulsed operation does not lower the peak excitation intensity, which is set by the requirement to polarize the spin. It makes the time-averaged dose a free parameter: the experimenter meets any ceiling tolerated by the sample by idling between shots, at a cost in repetition rate. Continuous ODMR has no such freedom. Section~\ref{sec:overhead} shows that the control gain survives exactly in that mode of operation.

\subsection{Binary-readout Fisher information}
\label{sec:binary-fi}

Writing $z(T) = \mathrm{Tr}[\rho_T\sigma_z]$, ideal binary projective readout gives outcome probabilities $p_{0,1}(T) = [1 \pm z(T)]/2$, and the classical Fisher information carried by one shot is
\begin{equation}
F_C^{(z)}(T) = \sum_{m=0,1}\frac{[\partial_T p_m(T)]^2}{p_m(T)}
             = \frac{[\partial_T z(T)]^2}{1 - z^2(T)} .
\label{eq:FCz}
\end{equation}
Equation~(\ref{eq:FCz}) makes explicit why optimizing $\chi_T = |\partial_T z|$ alone is incomplete: the denominator penalizes working points close to saturation. The corresponding information rate and sensitivity per root hertz are
\begin{equation}
\dot F_C^{(z)} = \frac{F_C^{(z)}}{t_\mathrm{cyc}},
\qquad
\eta_T^{(z)} = \frac{1}{\sqrt{\dot F_C^{(z)}}} .
\label{eq:rate-z}
\end{equation}

\subsection{Photon-counting model}
\label{sec:photon-model}

In NV experiments the readout is a fluorescence photon count. Let $n_0$ and $n_1$ be the mean detected photon numbers in one readout window for the bright and dark spin states. The expected count per readout is
\begin{equation}
\bar n(T) = n_0 p_0(T) + n_1 p_1(T) = \frac{n_0+n_1}{2} + \frac{n_0-n_1}{2}\,z(T),
\label{eq:nbar}
\end{equation}
which we write compactly as $\bar n(T) = \bar n_0\,[1 + C z(T)]$ with
\begin{equation}
\bar n_0 = \frac{n_0+n_1}{2}, \qquad C = \frac{n_0-n_1}{n_0+n_1},
\label{eq:nbar0-C}
\end{equation}
so that $\partial_T \bar n = \bar n_0 C\,\partial_T z$. Near the optimized operating point, where $z \simeq 0$ and $\bar n \simeq \bar n_0$, this gives $F_C^{(\mathrm{ph})} \simeq \bar n_0 C^2 [\partial_T z]^2$ and a photon-counting sensitivity
\begin{equation}
\eta_T^{(\mathrm{ph})} = \frac{1}{\sqrt{\dot F_C^{(\mathrm{ph})}}} \simeq \frac{1}{C\,|\partial_T z|\,\sqrt{\bar n_0/t_\mathrm{cyc}}},
\label{eq:eta-photon}
\end{equation}
so the sensitivity conversion follows from the photon-counting model itself and needs no separate normalization convention for the fluorescence signal. Under a Poisson approximation for the detected count, the information carried by one readout is
\begin{equation}
F_C^{(\mathrm{ph})}(T) = \frac{[\partial_T\bar n(T)]^2}{\bar n(T)},
\qquad
\dot F_C^{(\mathrm{ph})} = \frac{F_C^{(\mathrm{ph})}}{t_\mathrm{cyc}},
\label{eq:FCph}
\end{equation}
and the quantity relevant to the optical-dose discussion is the photon-normalized Fisher information
\begin{equation}
F_\text{per photon}(T) = \frac{F_C^{(\mathrm{ph})}(T)}{\bar n_\mathrm{cyc}(T)} .
\label{eq:Fperphoton}
\end{equation}
We use $n_0 = 0.03$ and $n_1 = 0.02$ photons per shot for a single NV in a nanodiamond under green excitation~\cite{taylor2008,rondin2014}, which give an intrinsic optical spin contrast $C = 0.20$ from Eq.~(\ref{eq:nbar0-C}),
the same value assumed for the projected single-NV configuration of Ref.~\cite{zhang2018}. In the minimal model only readout photons are counted, $\bar n_\mathrm{cyc} \simeq \bar n_\mathrm{read}$. Optical initialization also delivers a photon dose without contributing temperature information; that variant, $\bar n_\mathrm{cyc} = \bar n_\mathrm{read} + \bar n_\mathrm{init}$, is treated separately in Sec.~\ref{sec:photon-budget}, and we state explicitly in each case whether $\bar n_\mathrm{init}$ is included. These conservative nanodiamond values define the photon-counting model used in all Fisher-information ratios below.

\subsection{Gain ratios}
\label{sec:gains}

The QSP-inspired protocol is compared with the phase-optimized Ramsey baseline through the ratios
\begin{align}
G_\chi(d,t) &= \frac{\chi_{T,\mathrm{QSP}}(d,t)}{\chi_{T,\mathrm{Ram}}(t)},
\quad
G_F(d,t) = \frac{F^{(z)}_{C,\mathrm{QSP}}(d,t)}{F^{(z)}_{C,\mathrm{Ram}}(t)},
\label{eq:Gchi-GF}\\[2pt]
G_{\dot F}(d,t) &= G_F(d,t)\,
   \frac{t_\mathrm{cyc}^\mathrm{Ram}}{t_\mathrm{cyc}^{(d)}},
\\[2pt]
G_\mathrm{photon}(d,t) &= \frac{F^{(\mathrm{ph})}_{C,\mathrm{QSP}}(d,t)}
                               {F^{(\mathrm{ph})}_{C,\mathrm{Ram}}(t)}\,
   \frac{\bar n^\mathrm{Ram}_\mathrm{cyc}}{\bar n^{(d)}_\mathrm{cyc}}
   \;\simeq\; G_F(d,t)\,
   \frac{\bar n^\mathrm{Ram}_\mathrm{cyc}}{\bar n^{(d)}_\mathrm{cyc}} .
\label{eq:GFdot-Gphoton}
\end{align}
All four are evaluated at equal nominal interrogation time $t$; the second form of $G_\mathrm{photon}$ follows from $F_C^{(\mathrm{ph})} \simeq \bar n_0 C^2\,[\partial_T z]^2$, where the common prefactor $\bar n_0 C^2$ cancels in the ratio. When both protocols operate near $z \simeq 0$ and consume the same detected photons per shot, the Fisher denominator in Eq.~(\ref{eq:FCz}) is inactive and $\bar n$ is common, so that
\begin{equation}
G_F \simeq G_\mathrm{photon} \simeq G_\chi^2 ,
\label{eq:gain-identity}
\end{equation}
an identity we verify numerically in Sec.~\ref{sec:photon-budget}. It follows that a Fisher-information gain of $1.6$ and a $26\%$ sensitivity gain are the same statement expressed in squared and unsquared form, and the two should be quoted as one result.

Equal-time ratios display the gain landscape, while an experiment optimizes each protocol over its own interrogation time, since the two do not peak at the same $t$. We therefore also use peak-to-peak ratios, in which each protocol is separately maximized over its own interrogation time,
\begin{align}
G_F^\mathrm{peak}(d) &= \frac{\max_t F^{(z)}_{C,\mathrm{QSP}}(d,t)}{\max_t F^{(z)}_{C,\mathrm{Ram}}(t)},
\\[2pt]
G_{\dot F}^\mathrm{glob}(d) &= \frac{\max_t \dot F^{(z)}_{C,\mathrm{QSP}}(d,t)}{\max_t \dot F^{(z)}_{C,\mathrm{Ram}}(t)},
\label{eq:peak-gains}
\end{align}
and analogously for $G_\mathrm{photon}^\mathrm{peak}$. The distinction matters: at $d=10$ the equal-time ratio at $t = 3T_2^*$ is $G_F = 9.5$, while the peak-to-peak ratio is $G_F^\mathrm{peak} = 1.57$. The former compares against a Ramsey sequence forced to dephase for a duration it would never be operated at; the latter compares each protocol at its own optimum. Peak-to-peak ratios are quoted in all summary tables, and equal-time maps are used only to display the structure of the gain landscape.

\subsection{Optimization objectives and algorithms}
\label{sec:objectives}

At fixed $(d,t)$ we optimize the control vector $\bm{\phi}$ of Eq.~(\ref{eq:control-vector}) for each of
\begin{equation}
\max_{\bm{\phi}} \chi_T,\quad
\max_{\bm{\phi}} F_C^{(z)},\quad
\max_{\bm{\phi}} \frac{F_C^{(z)}}{t_\mathrm{cyc}},\quad
\max_{\bm{\phi}} \frac{F_C^{(\mathrm{ph})}}{t_\mathrm{cyc}},
\label{eq:objectives}
\end{equation}
with the Ramsey baseline optimized over $\Phi_\mathrm{read}$ in the same way. The first three have the same maximizer here, as shown in Appendix~\ref{sec:objective-comparison}; the fourth differs slightly and is quantified there. Derivatives are symmetric finite differences with $\delta T = 10^{-4}$~K, a choice justified by the convergence study in Appendix~\ref{sec:supplemental} (Fig.~\ref{fig:numerics}(b)).

The control landscape is nonconvex, so a single local search could return a good-looking but non-global solution. We use L-BFGS-B with bounds $\alpha_k,\beta_k,\gamma_k \in [-\pi,\pi]$ and many independent random restarts, cross-checked against differential evolution followed by L-BFGS-B polishing. The reliability of this procedure is reported in Appendix~\ref{sec:optimizer-reliability}.

\subsection{Decoherence Model}

To model the system realistically, we consider its evolution as an open quantum system governed by the Lindblad master equation at a finite temperature $T$.
\begin{equation}
\dot{\rho} = -i[H,\rho] + \sum_k \left(L_k \rho L_k^\dagger - \frac{1}{2}\{L_k^\dagger L_k, \rho\}\right).
\label{eq:lindblad}
\end{equation}
The jump operators $L_k$ include spontaneous decay ($L_\downarrow = \sqrt{\Gamma_\downarrow}\,\sigma_-$) and thermal excitation ($L_\uparrow = \sqrt{\Gamma_\uparrow}\,\sigma_+$), which contribute to the longitudinal relaxation time $T_1$. These rates are temperature-dependent and satisfy detailed balance, with $\Gamma_\downarrow = \Gamma_0(n_{\text{th}}+1)$ and $\Gamma_\uparrow = \Gamma_0 n_{\text{th}}$. The base rate $\Gamma_0$ is calibrated such that $T_1 = T_{1,\text{base}} = 100~\mu$s at the reference temperature $T_{\text{ref}} = T_c$, a value consistent with experimental measurements of NV centers in nanodiamonds. The mean thermal occupation number $n_{\text{th}}$ is given by the Planck distribution for the qubit energy gap $\omega_q(T)$:
\begin{equation}
n_{\text{th}}(\omega_q, T) = \frac{1}{\exp(\omega_q / k_B T) - 1}.
\label{eq:planck}
\end{equation}
This explicitly links the relaxation rates to the unknown magnetic field via $\omega_q$.

The transverse relaxation time $T_2$ is modeled using a phenomenological form that captures increased dephasing away from a base temperature:
\begin{equation}
T_2(T) = \frac{T_2^*}{1 + \beta\,(T - T_{\text{ref}})^2},
\label{eq:t2_model}
\end{equation}
with $T_2^* = 2.0~\mu$s and $\beta = 0.02$~K$^{-2}$, subject to a minimum value of $T_{2,\text{min}} = 0.5~\mu$s. The pure-dephasing term is a phenomenological Markovian description of coherence loss during the sensing interval. It captures the exponential envelope used in the main comparison and gives a conservative setting for response-shaping control, since Markovian dephasing cannot be refocused by echo pulses. Single-NV devices can also carry quasi-static nuclear-spin contributions to $T_2^*$; those affect the detailed envelope and would be calibrated experimentally. The present comparison isolates the control gain at a fixed measured coherence envelope. Measurements on NV ensembles give a $T^5$ power law for the \emph{longitudinal} relaxation rate at high temperature, from two-phonon Raman processes~\cite{jarmola2012}; the temperature dependence of $T_2^*$ in a nanodiamond involves a different combination of spin-bath and phonon contributions, which we do not model microscopically. Furthermore, we neglect the intrinsic temperature dependence of the zero-field splitting ($dD/dT \approx -74$~kHz/K~\cite{acosta2010}), as the hydrogel-mediated magnetic field transduction ($d\omega_q/dT \sim$~MHz/K near $T_c$) dominates by several orders of magnitude. However, within the narrow temperature window ($\sim$10~K) around the PNIPAM critical temperature relevant to our sensor, any monotonically decreasing function of $|T - T_{\text{ref}}|$ yields quantitatively similar results, and our phenomenological form has the advantage of simplicity and guaranteed positivity.

The pure dephasing rate $\Gamma_\phi$ is then derived from the standard relation $1/T_2 = 1/(2T_1) + \Gamma_\phi$. An additional collapse operator $L_\phi = \sqrt{\Gamma_\phi/2}\,\sigma_z$ is included when $\Gamma_\phi > 0$. Decoherence creates a trade-off: increasing the number of QSP layers, $d$, provides algorithmic gain but also increases the total time, leading to greater information loss. The optimal protocol balances these two effects.

\subsection{Nuclear Spin as a Quantum Memory}
\label{subsec:nuclear-memory}

The ceiling reported in Sec.~\ref{sec:results} is set by the electron coherence time $T_{2,e}$. The nearby nuclear spin has a much longer coherence time, $T_{2,n} \gg T_{2,e}$~\cite{bargill2013,rosskopf2017}, which suggests transferring the electron state to the nucleus for storage using a SWAP gate so that the sequence can extend beyond the electron coherence limit. We note the possibility here but do not pursue it, and the results below do not depend on it. Two considerations temper the expectation: the per-shot response saturates near $13$~K$^{-1}$ by $d = 20$ within the present coherence budget, and at the rate level the pulse overhead already limits the optimal depth to $d \simeq 3$--$5$. Additional layers would have to be accompanied by faster control, and the SWAP operations would add their own duration and infidelity to the cycle.

The dimensional calibration relating our dimensionless quantities to physical units (Appendix~\ref{sec:units}), the comparison with the reference ensemble experiment of Ref.~\cite{zhang2018} (Appendix~\ref{sec:fair-comparison}), and the derivation of the signal-to-spin readout conversion factor $f_c$ (Appendix~\ref{sec:alpha-coefficient}) are collected in the appendices.

\section{Results}
\label{sec:results}

All simulations are performed at a target temperature $T_{\text{target}} = 305.0$~K, near the critical temperature of the PNIPAM hydrogel, where the sensitivity of the transduction chain is maximized.

\begin{figure*}[t]
\centering
\includegraphics[width=\textwidth]{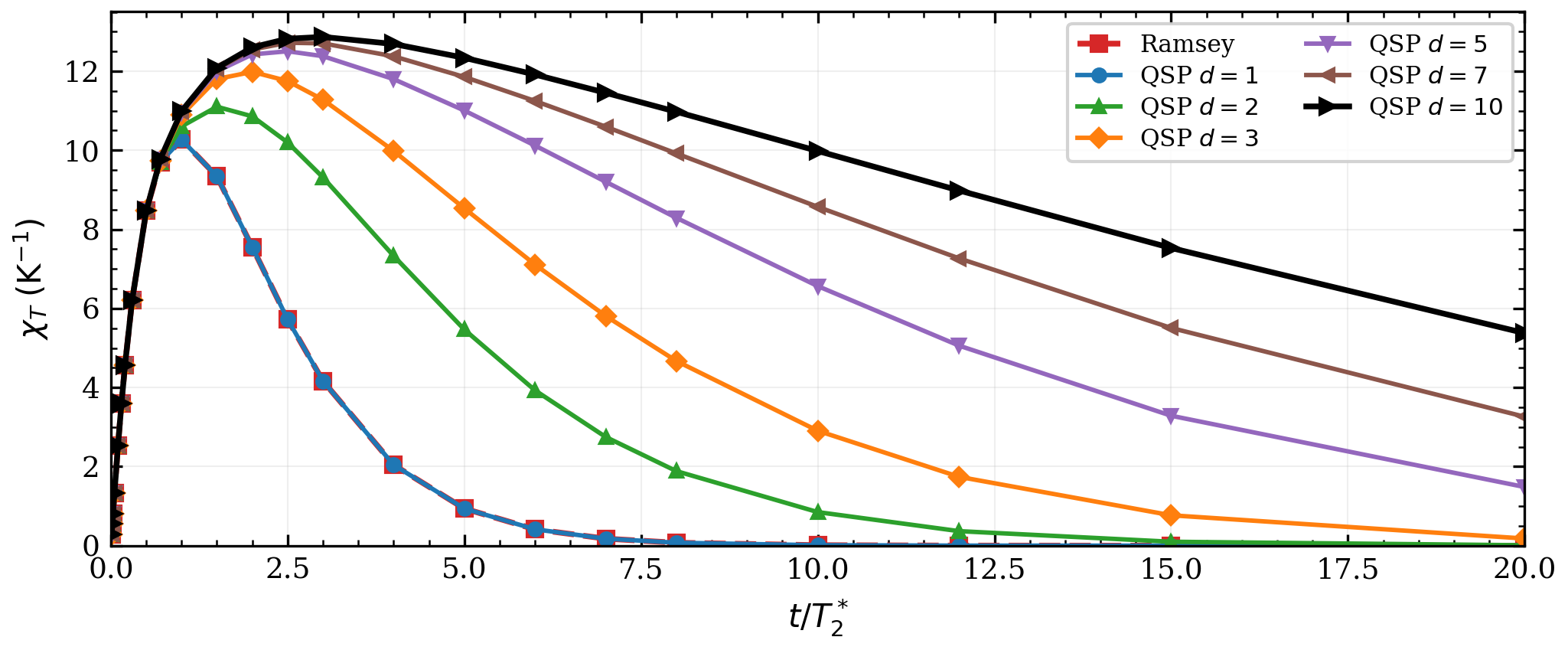}
\caption{Thermal spin susceptibility $\chi_T = |\partial_T\langle\sigma_z\rangle|$ as a function of normalized interrogation time $t/T_2^*$ for phase-optimized Ramsey (red squares) and QSP-inspired coherent-control sequences with $d = 1, 2, 3, 5, 7, 10$ layers (circles). The controlled response increases with depth and shifts to longer nominal interrogation time, reaching a finite-depth ceiling near $13$~K$^{-1}$. The large equal-time separation at long times is interpreted in Sec.~\ref{sec:mechanism} through the optimized state trajectory. All simulations use the full Lindblad master equation at $T = 305.0$~K.}
\label{fig:main_result}
\end{figure*}

\subsection{Ramsey Benchmark}

We first establish a benchmark by simulating a standard Ramsey sequence as a function of the total evolution time $t$, expressed in units of $T_2^*$. The Ramsey protocol consists of a $\pi/2$ pulse, free evolution for a time $t$, an optional phase shift $\varphi$, and a second $\pi/2$ pulse. We optimize the readout phase $\varphi$ at each time point to maximize the sensitivity $\chi_T$; the role of this phase calibration is examined in Appendix~\ref{sec:ramsey-calib}.

The Ramsey sensitivity initially increases linearly with time, reaching a peak value of $\chi_T^\mathrm{Ramsey} = 10.27$~K$^{-1}$ at $t/T_2^* = 1.0$, and then decays exponentially due to decoherence: by $t/T_2^* = 5$ the response has dropped to $0.94$~K$^{-1}$ and by $t/T_2^* = 10$ to $0.013$~K$^{-1}$. This peak is the per-shot baseline for all comparisons below. When the cycle time is taken into account, the Ramsey information rate instead peaks at $t/T_2^* = 0.75$, below the per-shot optimum, because of the fixed initialization and readout overhead in Eq.~(\ref{eq:tcyc-ram}). The red curve in Fig.~\ref{fig:main_result} shows the per-shot behavior.

As a consistency check on the sequence construction, the $d = 1$ QSP-inspired coherent-control sequence reproduces the phase-optimized Ramsey baseline to a relative difference of $2\times10^{-9}$ at every point of the grid, as expected since a single processing layer with free Euler angles contains the Ramsey sequence as a special case. All $d = 1$ entries in what follows are therefore unity by construction.

\subsection{Analytic mechanism at depth two}
\label{sec:analytic-d2}

The gain mechanism is visible in the simplest nontrivial case. Consider two free-evolution blocks of duration $\tau$ separated by a controllable rotation. In the Bloch representation the free-evolution map is
\begin{equation}
M(\theta) = \Lambda R_z(\theta), \qquad \Lambda = \mathrm{diag}(\lambda,\lambda,\mu),
\label{eq:bloch-map}
\end{equation}
with $\lambda = e^{-\tau/T_2^*}$ and $\mu = e^{-\tau/T_1}$. Since $\Lambda$ commutes with $R_z$, the derivative of the final Bloch vector with respect to the encoded phase splits into two phase-sensitivity contributions,
\begin{equation}
\partial_\theta \bm{r}_f = \Lambda J R \Lambda \bm{r}_0 + \Lambda R \Lambda J \bm{r}_0,
\label{eq:d2-derivative}
\end{equation}
where $J$ generates $z$ rotations, $R$ is the intermediate control rotation and $\bm{r}_0 = (\sin a, 0, \cos a)$ is the prepared state. The final analysis pulse selects the measurement direction, so the optimized slope is the norm of Eq.~(\ref{eq:d2-derivative}). Carrying out the maximization over $a$, $R$ and the readout direction (Appendix~\ref{sec:d2-derivation}) gives the closed form
\begin{equation}
\left|\partial_\theta z\right|^{(d=2)}_{\max} =
\begin{cases}
\dfrac{\lambda\mu^{2}}{\sqrt{\mu^{2}-\lambda^{2}}}, & \lambda \leq \mu/\sqrt{2},\\[10pt]
2\lambda^{2}, & \lambda > \mu/\sqrt{2}.
\end{cases}
\label{eq:d2-closed}
\end{equation}
At $t = 1.5\,T_2^*$ this gives $\chi_T = 11.106$~K$^{-1}$, against the numerical optimum $11.101$~K$^{-1}$ for $d = 2$, a difference of $0.05\%$ that comes from the thermal asymmetry of the $T_1$ rates, which Eq.~(\ref{eq:bloch-map}) does not carry.

The optimum has a transparent structure. In the first branch the stationary solution keeps part of the Bloch vector longitudinal, where pure dephasing is inactive, while the transverse component carries the phase; the longitudinal part then acts as an amplitude reference for the later phase-sensitive block, and the optimal slope turns out to be independent of the preparation angle. This is the analytic counterpart of the decoherence-protected idling seen in the optimized trajectories of Sec.~\ref{sec:mechanism}: the sequence delays the main phase acquisition until the end of the protocol while preserving a reference component protected from the dominant dephasing channel. The second branch, reached at short $\tau$ where dephasing is weak, is the ordinary regime in which both blocks contribute transversally.

\begin{figure*}[t]
\centering
\includegraphics[width=\textwidth]{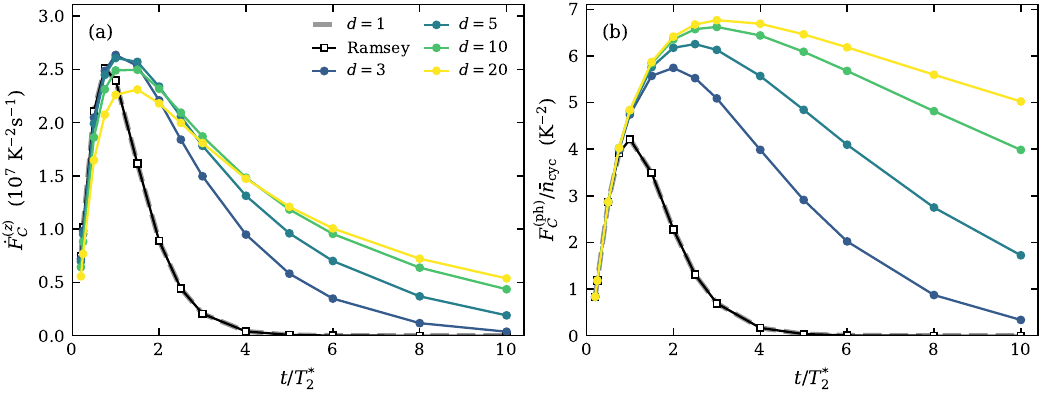}
\caption{(a) Binary Fisher-information rate $\dot F_C^{(z)}$ and (b) photon-normalized Fisher information $F_C^{(\mathrm{ph})}/\bar n_\mathrm{cyc}$ for Ramsey and QSP-inspired coherent-control sequences. The $d = 1$ sequence reproduces the Ramsey baseline; the two are drawn separately for visibility. Panel (a) includes the finite pulse duration $t_\mathrm{p} = 50$~ns through the cycle time of Eqs.~(\ref{eq:tcyc-qsp})--(\ref{eq:tcyc-ram}), while panel (b) shows the per-detected-photon gain. Both axes are linear, so the comparison that matters operationally is visible directly: every protocol peaks at nearly the same rate, within $8\%$ of Ramsey, and the depth only moves where that peak sits. The long-time separation is a decay of the Ramsey reference, not a rate advantage.}
\label{fig:rate_photon}
\end{figure*}

\subsection{Optimized QSP Protocol}

The central result of this work is shown in Fig.~\ref{fig:main_result}. For each value of $d \in \{1, 2, 3, 5, 7, 10\}$ and each total time budget $t = d\tau$, we use the L-BFGS-B optimizer with multiple random restarts to find the Euler angles $\{\alpha_k, \beta_k, \gamma_k\}$ that maximize $\chi_T$. The sensitivity is computed via symmetric finite differences with $\delta T = 10^{-4}$~K.

Figure~\ref{fig:main_result} and Table~\ref{tab:summary} show the following. The peak per-shot response grows monotonically with depth, from $\chi_T^\mathrm{Ramsey} = 10.27$~K$^{-1}$ to $13.01$~K$^{-1}$ at $d = 20$, an increase of $26.7\%$, and the optimal operating point shifts to progressively longer interrogation times, from $t/T_2^* = 1.0$ for Ramsey to $t/T_2^* = 3.0$ for $d \geq 10$. The gain saturates: the increment is $+8\%$ from $d=1$ to $d=2$ but only $+0.2\%$ from $d=15$ to $d=20$, so the response approaches a ceiling within the present ansatz and scan range near $13$~K$^{-1}$ for the parameters considered here.

At every optimized point the optimizer places the working point at a zero crossing of the fringe, $|z| < 7 \times 10^{-4}$, so the denominator of Eq.~(\ref{eq:FCz}) is inactive and $F_C^{(z)} \simeq \chi_T^2$. The Fisher-information gain is therefore the square of the slope gain, $G_F^\mathrm{peak} = 1.57$--$1.60$ for $d \geq 10$, as in Eq.~(\ref{eq:gain-identity}).

\begin{table*}[t]
\centering
\caption{Peak thermal spin susceptibility and gains for the Ramsey baseline and the optimized QSP-inspired protocol at $T_\mathrm{target} = 305.0$~K. Each protocol is evaluated at its own optimal interrogation time, so the gains are the peak-to-peak ratios of Eq.~(\ref{eq:peak-gains}). Since the optimum lies at $z \simeq 0$ for both protocols, $G_F^\mathrm{peak} = G_\mathrm{photon}^\mathrm{peak} = (G_\chi^\mathrm{peak})^2$ to four decimal places. The last two columns give the cycle-time-normalized gain $G_{\dot F}^\mathrm{glob}$ with $t_\mathrm{p} = 50$~ns at two overheads: single-shot fluorescence readout ($t_\mathrm{ovh} = 2.3\,\mu$s) and nuclear-spin-assisted repetitive readout ($100\,\mu$s). The per-shot column is a property of the sequence; which depth is preferable in practice is set by the duty cycle (Sec.~\ref{sec:overhead}).}
\label{tab:summary}
\small
\setlength{\tabcolsep}{8pt}
\begin{tabular}{@{}lccccccc@{}}
\toprule
 & & & & & & \multicolumn{2}{c}{$G_{\dot F}^\mathrm{glob}$} \\
\cmidrule(l){7-8}
Protocol & $d$ & $(t/T_2^*)_\mathrm{opt}$ & $\chi_T^\mathrm{peak}$ & $G_\chi^\mathrm{peak}$ & $G_F^\mathrm{peak} = G_\mathrm{photon}^\mathrm{peak}$ & $2.3\,\mu$s & $100\,\mu$s \\
 & & & [K$^{-1}$] & & & & \\
\midrule
Ramsey & --- & 1.00 & 10.27 & 1.000 & 1.000 & 1.000 & 1.000 \\
QSP & 1 & 1.00 & 10.27 & 1.000 & 1.000 & 1.000 & 1.000 \\
QSP & 2 & 1.50 & 11.10 & 1.081 & 1.169 & 1.008 & 1.157 \\
QSP & 3 & 2.00 & 11.98 & 1.167 & 1.362 & 1.051 & 1.334 \\
QSP & 5 & 2.50 & 12.51 & 1.218 & 1.483 & 1.040 & 1.438 \\
QSP & 7 & 2.50 & 12.72 & 1.239 & 1.534 & 1.022 & 1.486 \\
QSP & 10 & 3.00 & 12.87 & 1.253 & 1.570 & 0.996 & 1.508 \\
QSP & 15 & 3.00 & 12.98 & 1.264 & 1.597 & 0.958 & 1.527 \\
QSP & 20 & 3.00 & 13.01 & 1.267 & 1.604 & 0.921 & 1.530 \\
\bottomrule
\end{tabular}
\end{table*}

At equal interrogation time the ratios grow rapidly after the Ramsey reference has dephased: at $t/T_2^* = 5$ the Ramsey response has fallen to $0.94$~K$^{-1}$ while the $d=5$ sequence retains $11.0$~K$^{-1}$, and at $t/T_2^* = 10$ the ratio exceeds $500$. These values map the response landscape; Sec.~\ref{sec:mechanism} explains why they are not the operational figure of merit.

\subsection{Photon budget}
\label{sec:photon-budget}

Figure~\ref{fig:rate_photon}(b) shows the photon-normalized Fisher information of Eq.~(\ref{eq:Fperphoton}). Because the optimum lies at $z \simeq 0$ for both protocols, $\bar n$ is common to the two, and Eq.~(\ref{eq:FCph}) then gives $F_C^{(\mathrm{ph})}/\bar n \propto (\partial_T z)^2$, so Eq.~(\ref{eq:gain-identity}) applies. The photon-budget result restates the per-shot gain: the sequence realizes it at the same detected-photon budget per shot, so the gain carries over one-to-one into information per detected photon.

This photon-budget result belongs to the pulsed single-NV model. The control gain is obtained at the same detected-photon budget per shot as optimized Ramsey. The high-fluence single-NV CW-ODMR projections of Ref.~\cite{zhang2018} use a different readout mode and optical-fluence assumption; Appendix~\ref{sec:fair-comparison} places the two settings on a common photon-budget scale.

Section~\ref{sec:dose} establishes that each shot delivers a fixed optical dose, independent of the interrogation time. The delivered dose is therefore proportional to the number of shots, so
\begin{equation}
\frac{\text{information}}{\text{optical dose}} \;\propto\; F_C^{(z)} \ \text{per shot},
\label{eq:info-per-dose}
\end{equation}
and the ratio between protocols is the per-shot Fisher gain, $G_F^\mathrm{peak} = 1.57$ at $d = 10$. Information per unit optical dose improves by the full per-shot factor, independently of the overhead, of the duty cycle, and of whether the experiment idles. Detected photons are a proxy for what constrains a living sample; the delivered dose is the quantity itself.

If optical initialization is counted as dose but not as signal, Eq.~(\ref{eq:Fperphoton}) acquires $\bar n_\mathrm{cyc} = \bar n_\mathrm{read} + \bar n_\mathrm{init}$. Extrapolating the readout brightness over $t_\mathrm{init}$ gives an upper-bound estimate $\bar n_\mathrm{init} = \bar n_\mathrm{read}\,t_\mathrm{init}/t_\mathrm{read} = 0.167$ photons per shot. Since $\bar n$ is equal for the two protocols at $z \simeq 0$, this constant cancels in the ratio and $G_\mathrm{photon}$ is unchanged; only the absolute information per photon is rescaled, by a common factor $0.13$.

\begin{figure*}[t]
\centering
\includegraphics[width=\textwidth]{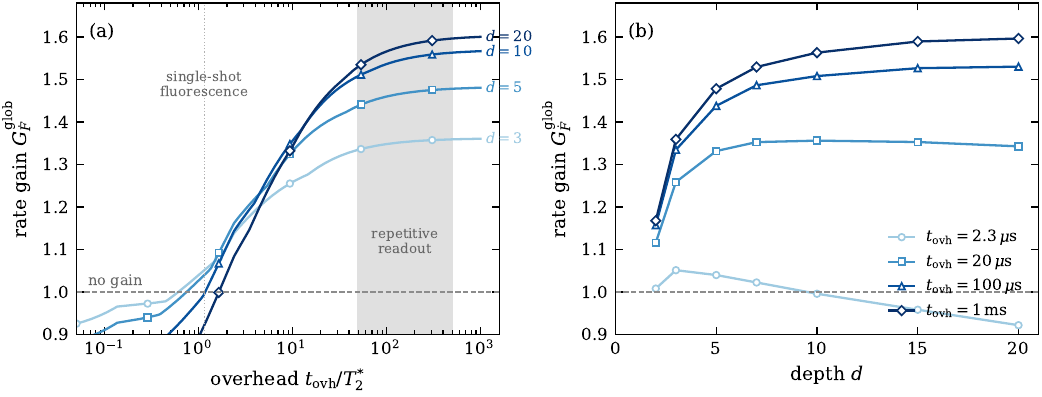}
\caption{Global Fisher-information-rate gain with finite pulse duration $t_\mathrm{p} = 50$~ns. (a) Gain against the dimensionless overhead $t_\mathrm{ovh}/T_2^*$ at fixed depth; the dotted line marks the single-shot fluorescence readout assumed elsewhere in this work and the shaded band the nuclear-spin-assisted repetitive-readout regime. (b) Gain against depth for representative overheads. The duty cycle sets the useful depth: fast readout favors shallow control, while repetitive readout and dose-limited operation recover the per-shot gain at larger depth. The lowest curve is the $t_\mathrm{ovh} = 2.3~\mu$s case of Fig.~\ref{fig:optimal_depth}.}
\label{fig:overhead}
\end{figure*}

\subsection{Fisher-information rate and the practical operating depth}
\label{sec:rate}

The per-shot comparison ignores the cost of running the sequence. Figure~\ref{fig:rate_photon}(a) shows the binary Fisher-information rate of Eq.~(\ref{eq:rate-z}), and Fig.~\ref{fig:optimal_depth} in Appendix~\ref{sec:supplemental} the global rate gain $G_{\dot F}^\mathrm{glob}(d)$ of Eq.~(\ref{eq:peak-gains}) under the two pulse-duration assumptions:

\begin{center}
\footnotesize
\setlength{\tabcolsep}{2.5pt}
\begin{tabular}{@{}lcccccccc@{}}
\toprule
$d$ & 1 & 2 & 3 & 5 & 7 & 10 & 15 & 20 \\
\midrule
$t_\mathrm{p}=0$ & 1.000 & 1.016 & 1.072 & 1.084 & 1.088 & 1.091 & 1.093 & 1.093 \\
$t_\mathrm{p}=50$~ns & 1.000 & 1.008 & 1.051 & 1.040 & 1.022 & 0.996 & 0.958 & 0.921 \\
\bottomrule
\end{tabular}
\end{center}

With instantaneous pulses the rate gain saturates near $1.09$. With $t_\mathrm{p} = 50$~ns the $d+1$ processing pulses of the coherent-control sequence, against two for Ramsey, introduce a depth-dependent overhead that selects a moderate optimal depth $d \simeq 3$--$5$ with a global rate gain of about $1.05$; beyond $d \approx 10$ the rate advantage is lost entirely at this overhead. The Ramsey rate optimum itself lies at $t/T_2^* = 0.75$, below its per-shot optimum, because of the fixed readout overhead.

At $d = 3$--$5$ the per-shot gain is $17$--$22\%$ and the photon-normalized gain $1.36$--$1.48$, both smaller than the saturated values of Table~\ref{tab:summary} but obtained at a rate advantage rather than at a rate penalty. The numbers in this subsection are computed at our nominal overhead $t_\mathrm{ovh} = 2.3~\mu$s, and both the modest gain and the restriction to shallow depth follow from that particular choice. Section~\ref{sec:overhead} treats $t_\mathrm{ovh}$ as the free parameter it is, and finds that the depth restriction lifts and the gain approaches its per-shot ceiling as soon as the readout becomes expensive.

\subsection{The duty cycle sets the rate gain}
\label{sec:overhead}

The rate gain depends on the full measurement cycle. We therefore treat the fixed overhead of Eq.~(\ref{eq:tovh}) as an experimental parameter and keep the finite pulse duration $t_\mathrm{p} = 50$~ns. The dimensionless ratio $t_\mathrm{ovh}/T_2^*$ organizes the result. When this ratio is small the cycle time is dominated by the coherent evolution and by the pulse train, so shallow sequences are favored; at the nominal electronic-spin readout overhead $t_\mathrm{ovh} = 2.3~\mu$s the best rate gain is about $1.05$ and occurs at $d \simeq 3$. When the overhead is large the repetition rate is set mainly by a common cost paid by both protocols, and the rate gain approaches the per-shot Fisher-information gain. This occurs for repetitive readout and for dose-limited operation, where idle time is inserted between shots to keep the optical exposure below the sample tolerance. Figure~\ref{fig:overhead} shows the result, with $t_\mathrm{p} = 50$~ns retained throughout:

\begin{center}
\small
\setlength{\tabcolsep}{5pt}
\begin{tabular}{@{}lccccc@{}}
\toprule
$t_\mathrm{ovh}/T_2^*$ & $t_\mathrm{ovh}$ & $d{=}3$ & $d{=}5$ & $d{=}10$ & $d{=}20$ \\
\midrule
$0$      & $0$              & 0.917 & 0.856 & 0.753 & 0.609 \\
$1.15$   & $2.3\,\mu$s      & 1.051 & 1.040 & 0.996 & 0.921 \\
$4$      & $8\,\mu$s        & 1.191 & 1.220 & 1.212 & 1.177 \\
$10$     & $20\,\mu$s       & 1.259 & 1.331 & 1.356 & 1.343 \\
$50$     & $100\,\mu$s      & 1.334 & 1.438 & 1.508 & 1.530 \\
$500$    & $1\,$ms          & 1.359 & 1.478 & 1.563 & 1.596 \\
\bottomrule
\end{tabular}
\end{center}

The behavior is bounded by two limits. When $t_\mathrm{ovh} \to 0$ the cycle is the sensing time itself, and with $t_\mathrm{p} = 0$ the gain is exactly unity at every depth: the optimizer drives the coherent-control sequence back onto Ramsey, because a depth-$d$ sequence contains a single free evolution of duration $d\tau$ as a special case and the rate optimum lies at $t = T_2^*/2$, where the extra layers buy nothing. The unit-crossing overhead is $t_\mathrm{ovh} \simeq 0.24~\mu$s $\simeq 0.12\,T_2^*$. In the opposite limit the cycle is dominated by a constant common to both protocols, the repetition-rate penalty of the longer sequence disappears, and the rate gain tends to the per-shot ratio $G_F^\mathrm{peak} = 1.60$.

The depth restriction of Sec.~\ref{sec:rate} belongs to the small-overhead corner. The pulse penalty $(d+1)t_\mathrm{p}$ is a constant, so its weight in the cycle falls as $t_\mathrm{ovh}$ grows: for $d = 20$ it is $16.5\%$ of the cycle at $t_\mathrm{ovh} = 2.3~\mu$s but only $0.98\%$ at $100~\mu$s. The optimal depth accordingly moves from $d = 3$ at our nominal overhead to $d = 20$ once the readout costs $100~\mu$s, and the gain there, $1.530$ with finite pulses against a ceiling of $1.604$, retains $95\%$ of the per-shot advantage.

Two experimentally standard regimes reach the bottom row of the table. The unambiguous one is nuclear-spin-assisted repetitive readout, which by itself puts the sensor in the range $t_\mathrm{ovh}/T_2^* = 50$--$500$ of Sec.~\ref{sec:definitions} and gives $1.53$--$1.60$, with no assumption about the sample. Dose-limited operation, Eq.~(\ref{eq:tcyc-dose}), is a second and independent route into the same regime, and it reaches the ceiling sooner: because $t_\mathrm{dose}$ is a floor on the cycle rather than a cost added to it, the gain saturates once $t_\mathrm{dose} \gtrsim t_\mathrm{ovh} + t$, which occurs at $t_\mathrm{dose} \simeq 8.3~\mu$s $= 4.1\,T_2^*$; beyond that the rate gain equals the per-shot value to three decimal places. At a dose floor of $1$~ms the optimized sequence spends $0.6\%$ of the cycle accumulating signal, so the low-dose operating mode and the metrological operating mode coincide.

\begin{figure*}[!t]
\centering
\includegraphics[width=\textwidth]{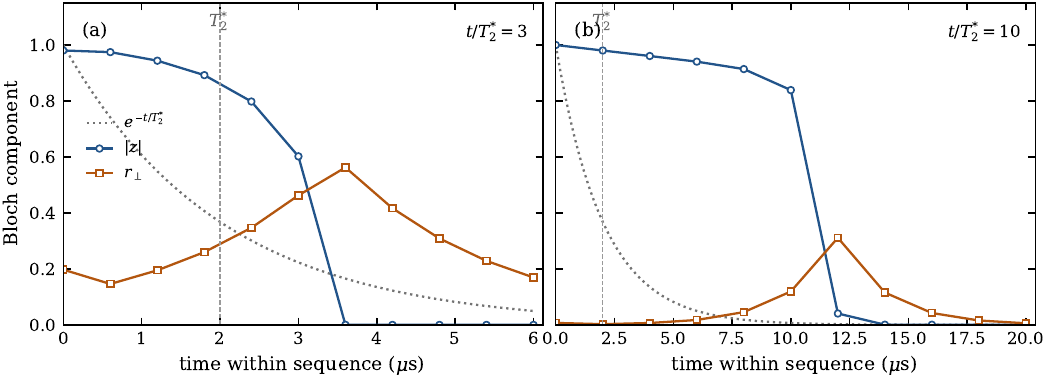}
\caption{Optimized Bloch trajectories for the $d = 10$ sequence, for (a) $t/T_2^* = 3$ and (b) $t/T_2^* = 10$. The longitudinal component $|z|$ and transverse radius $r_\perp$ are shown at the end of each free-evolution block; the dotted line is the Ramsey coherence envelope over the same wall-clock time. The sequence holds the state near the longitudinal axis during early blocks and rotates into the transverse plane near the end, producing decoherence-protected idling followed by phase-sensitive readout.}
\label{fig:bloch}
\end{figure*}

\subsection{Relation to precision bounds and to control-enhanced sensing}
\label{sec:bounds}

We place the numerical gain against the free-evolution dephasing scale. For a single probe undergoing Markovian dephasing at rate $\gamma$, the free-evolution Fisher information has the form
\begin{equation}
F(t) = \kappa^2 t^2 e^{-2\gamma t},
\label{eq:free-evolution-F}
\end{equation}
with $\kappa$ of Eq.~(\ref{eq:kappa})~\cite{escher2011}. The phase-optimized Ramsey baseline saturates this expression within numerical accuracy in our Lindblad model, to $0.09\%$ at the peak and better than $1\%$ across the grid. The residual is the curvature of the $T_2(T)$ model of Eq.~(\ref{eq:t2_model}) at the operating point: $1 + \beta(T-T_\mathrm{ref})^2 = 1.00045$ at $T = 305.0$~K, so the coherence decays at $1.00045/T_2^*$ and the free-evolution Fisher information falls $0.09\%$ short of Eq.~(\ref{eq:free-evolution-F}). We use this as a check on the Ramsey reference and on the unit conversion. It is not a ceiling on the controlled sequences, whose interleaved unitaries place them outside the free-evolution model of Eq.~(\ref{eq:free-evolution-F}).

The QSP-inspired protocol increases the per-shot Fisher information by a constant factor up to $G_F^\mathrm{peak} = 1.60$ within the scanned control family. On the rate axis no sequence exceeds the dephasing-limited rate scale $\kappa^2/(2\gamma)$. That scale follows from the noise model: Markovian dephasing is Gaussian phase diffusion of variance $2\gamma t$, so an ideal estimate of the accumulated phase gives $F = \kappa^2 t^2/(2\gamma t) = \kappa^2 t/(2\gamma)$, a rate $\kappa^2/(2\gamma)$ attainable only with ancilla resources not available here~\cite{escher2011,sekatski2017}. The largest $F/t$ reached by any depth at any interrogation time is $36.8\%$ of it, that is $1/e$, the same value attained by optimized Ramsey at $t = T_2^*/2$. The improvement is therefore a finite-depth prefactor gain, not a change of scaling, which is the expected behavior for controlled sensing under dephasing~\cite{sekatski2017}.

A direct comparison is available with Hecht \textit{et al.}, who stabilize a Bloch component by continuous driving on a superconducting qubit~\cite{hecht2025}. Their improvement ratios are quoted in signal-to-noise units and ours in Fisher information, so the two must be squared into a common form before being compared. Their device operates at $T_1/T_2 = 0.76$, whereas our model has $T_1/T_2^* = 50$, deep in the dephasing-dominated corner of their parameter scan. In that corner they report a per-shot ratio of $1.09$ in signal-to-noise, equivalently $1.19$ in Fisher information, against our $1.60$. The larger prefactor obtained here is consistent with the wider finite-depth control family used in the present calculation, which optimizes $3d$ Euler angles instead of a single continuous-drive amplitude. The two calculations agree on the other axis: their per-evolution-time ratio tends to unity as $T_1 \gg T_2$, which independently reproduces our result that the rate gain vanishes at zero overhead. Their optimization also normalizes by $t + t_i$ with an inactive time $t_i$, so the duty-cycle accounting of Sec.~\ref{sec:overhead} follows the same convention.

\begin{figure*}[!t]
\centering
\includegraphics[width=\textwidth]{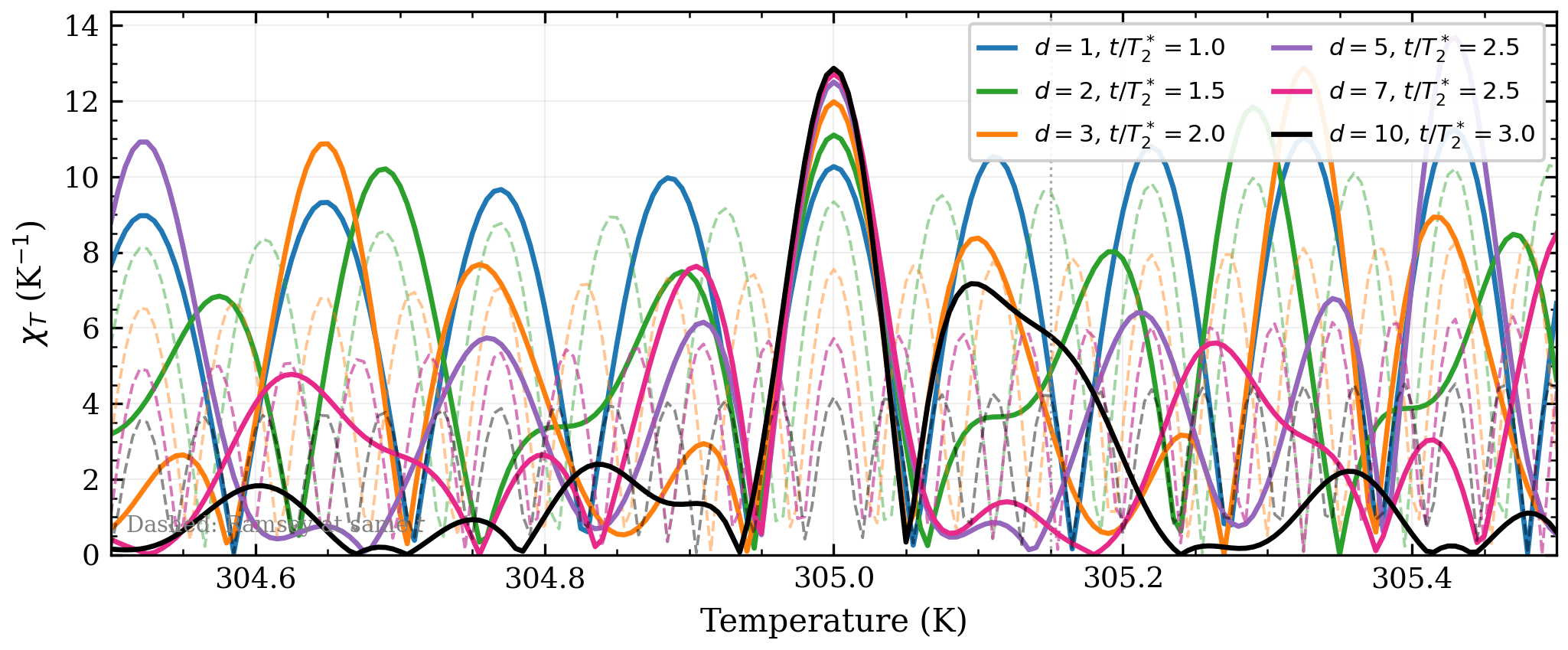}
\caption{Temperature dependence of the thermal spin susceptibility $\chi_T$ for precomputed coherent-control sequences near the PNIPAM transition (solid lines, $d = 1, 2, 3, 5, 7, 10$), against Ramsey at the same total sensing time (dashed lines). The sequences use pre-computed optimal phases at their respective optimal $t/T_2^*$ (Table~\ref{tab:summary}). Increasing the depth raises the local response and narrows the fringe scale, giving a sensitivity--dynamic-range trade-off. The curves are intended as local, phase-calibrated operating envelopes around $T_\mathrm{target}$.}
\label{fig:temp_window}
\end{figure*}

\subsection{Mechanism of the long-time response}
\label{sec:mechanism}

The long-time equal-time ratios are explained by the optimized trajectory. Although free Ramsey coherence is strongly suppressed for $t/T_2^* \gg 1$, the controlled sequence distributes the state between longitudinal storage and late-time phase accumulation.

Figure~\ref{fig:bloch} resolves this. It shows the longitudinal and transverse Bloch components at the end of each free-evolution block of the optimized $d = 10$ sequence. The optimizer keeps the state close to the longitudinal axis during the early blocks, where pure dephasing does not act, and rotates into the transverse plane only in the final blocks. A state held near a pole is limited by $T_1 = 100~\mu$s rather than by $T_2^* = 2~\mu$s, so the sequence idles in a decoherence-protected configuration and performs the interferometry late.

We quantify the effect through the effective coherent sensing time. Near the operating point the response is $z(T) \simeq A\sin[k(T-T_\mathrm{target})]$ with $k = |d\omega_q/dT|\,t_\mathrm{eff}$, so that $k = \sqrt{|z'''|/|z'|}$ and $t_\mathrm{eff} = k/|d\omega_q/dT|$ follow from derivatives evaluated at the operating point alone:

\begin{center}
\small
\begin{tabular}{@{}cccc@{}}
\toprule
$t/T_2^*$ & wall clock & $t_\mathrm{eff}$ & blocks held near the pole \\
\midrule
1  & $2~\mu$s  & $1.58~\mu$s & 3/11 \\
3  & $6~\mu$s  & $2.32~\mu$s & 6/11 \\
5  & $10~\mu$s & $2.55~\mu$s & 6/11 \\
10 & $20~\mu$s & $2.92~\mu$s & 6/11 \\
\bottomrule
\end{tabular}
\end{center}

The effective coherent sensing time stays near $T_2^*$ while the nominal duration grows by an order of magnitude. The equal-time baseline is a Ramsey sequence forced to dephase for a duration at which the coherent-control sequence is not in fact sensing. An experiment at fixed wall-clock time would instead repeat short Ramsey sequences. The rate normalization of Eq.~(\ref{eq:peak-gains}) already corrects for this in practice, which is why the operational advantage is a few percent rather than a few hundred.

Idling does not account for the entire effect. At the peak operating point ($d=10$, $t = 3T_2^*$) a plain Ramsey sequence run for $1.5T_2^*$, close to the QSP effective sensing time, gives $\chi_T = 9.3$~K$^{-1}$, below the Ramsey optimum of $10.27$~K$^{-1}$, whereas the coherent-control sequence gives $12.87$~K$^{-1}$. A genuine residual coherent gain remains: idling alone does not produce the $25\%$ figure.

Two further checks confirm the interpretation. First, the temperature dependence was decomposed by freezing either the accumulated phase or the decoherence rates: with the rates held temperature-independent, $\chi_T$ is unchanged to all reported digits; with only the rates temperature-dependent, $\chi_T$ falls to $5\times10^{-7}$~K$^{-1}$. The results do not rest on the phenomenological $T_2(T)$ model of Eq.~(\ref{eq:t2_model}).

Second, the mechanism is not dynamical decoupling. The thermometric signal is a static (i.e., DC) frequency shift generated by $\sigma_z$, so refocusing pulses reverse the accumulated static phase and suppress the signal itself, whereas the QSP-inspired sequence reshapes the state trajectory while preserving the phase response. A spin echo at the same total free-evolution time gives $\chi_T = 2.2\times10^{-3}$~K$^{-1}$ against $10.27$~K$^{-1}$ for Ramsey, a factor of $4.7\times10^{3}$ smaller, and Carr--Purcell--Meiboom--Gill (CPMG) sequences with 2, 4 and 8 refocusing pulses do not improve the performance of the spin echo sequence. Repeating the phase/rate decomposition for the echo gives exactly zero response when the rates are held fixed, so the entire echo residue is the temperature dependence of the decoherence rates. The decomposition is thus precisely reversed between the two sequences. Moreover, the transverse coherence decays as $e^{-t/T_2^*}$ whether or not refocusing pulses are inserted, since Markovian dephasing is not refocusable by any pulse sequence. Dynamical decoupling would remove the signal here.

\subsection{Robustness}
\label{sec:robustness}

Table~\ref{tab:robustness} reports the sensitivity retention with the control phases held at their nominal optimum and the error applied at evaluation.

\begin{table*}[t]
\centering
\caption{Sensitivity retention $\chi_T(\mathrm{error})/\chi_T(0)$ with control phases fixed at their nominal optimum. Amplitude errors act on the $x$-rotation angles $\beta_k$. The detuning acceptance bandwidth is the contiguous interval around $\delta = 0$ over which the retention exceeds $1/2$, obtained from a continuous scan.}
\label{tab:robustness}
\small
\setlength{\tabcolsep}{10pt}
\begin{tabular}{@{}lccccc@{}}
\toprule
 & $T_2^*-10\%$ & $T_2^*+10\%$ & ampl.\ $\pm1\%$ & ampl.\ $\pm5\%$ & bandwidth \\
\midrule
Ramsey     & 0.895 & 1.095 & 1.000 & 0.994 & 0.167~MHz \\
QSP $d=3$  & 0.896 & 1.096 & 1.000 & 0.992 & 0.154~MHz \\
QSP $d=5$  & 0.897 & 1.095 & 0.999 & 0.971 & 0.151~MHz \\
QSP $d=10$ & 0.900 & 1.093 & 0.999 & 0.971 & 0.152~MHz \\
\bottomrule
\end{tabular}
\end{table*}

Three observations follow. The retention under $T_2^*$ miscalibration is the same for Ramsey and for all QSP depths, so the QSP/Ramsey ratio is invariant under this error and no re-optimization of the phases is required. Amplitude errors of $\pm5\%$ cost at most $3\%$ of the response, so the coherent-control sequences match Ramsey in robustness despite using more pulses. The detuning acceptance bandwidth is $0.151$--$0.154$~MHz against $0.167$~MHz for Ramsey, that is, the coherent-control sequences retain about $91\%$ of the Ramsey bandwidth while interrogating two to three times longer; per unit sensing time they are more detuning-tolerant than an equal-time Ramsey, for which the bandwidth scales as $1/t$. Isolated detuning values can coincide with revivals of the accumulated phase, which are periodic in $\delta\tau$ and recur at intervals $1/(2\tau)$, so only the bandwidth extracted from a continuous scan is meaningful. The continuous scans behind these numbers are Fig.~\ref{fig:robustness-curves} of Appendix~\ref{sec:supplemental}.

\subsection{Temperature Window of Operation}
\label{sec:temp-window}

The Fisher-information analysis is local in temperature. We therefore use the temperature-window plot as a practical check of the response near the PNIPAM transition. Figure~\ref{fig:temp_window} shows $\chi_T(T)$ for the precomputed optimal phases. Increasing the depth raises the peak response and narrows the local operating range, giving the expected sensitivity--dynamic-range trade-off of a shaped interferometric response. The hydrogel transition sets the temperature region where the transducer slope is large, while the interferometric fringe period sets the unambiguous dynamic range of a single calibrated sequence. The oscillatory structure reflects the accumulated phase $\omega_q(T)\,t_\mathrm{eff}$, so a single operating point requires phase calibration or prior localization within one fringe. In practice, a short-time Ramsey measurement or an adaptive multi-time interrogation~\cite{bonato2016} can first localize the temperature within one fringe, after which the optimized coherent-control sequence supplies the high-sensitivity local readout. The useful range is a local, phase-calibrated operating range.

\subsection{Readout Normalization and Sensitivity Comparison}
\label{sec:readout-norm}

The spin-level sensitivities reported above quantify the algorithmic response of the NV qubit. To place these results on an experimentally meaningful footing, we convert the per-shot sensitivity $\chi_T$ to the frequency-normalized sensitivity $\eta_T$ via
\begin{equation}
\eta_T \;=\; \frac{1}{f_c\,\sqrt{L}\,\chi_T},
\label{eq:our-eta-Hz}
\end{equation}
where $L$ is the effective detected photon rate and $f_c$ is the readout conversion factor derived in Appendix~\ref{sec:alpha-coefficient}. It plays the same role as the optical spin contrast $C$ of Eq.~(\ref{eq:nbar0-C}); for the readout model used here the two coincide, $f_c = C$. For pulsed Ramsey and QSP-inspired protocols, $L$ should be interpreted as an effective detected photon rate averaged over the full experimental sequence. One experimental shot consists of the stages entering Eq.~(\ref{eq:tcyc-qsp}). Thus, if $\bar{n}$ photons are detected on average per shot and the full duration of one shot is $t_\mathrm{cyc}$, then the appropriate photon rate entering Eq.~(\ref{eq:our-eta-Hz}) is
\begin{equation}
L_\mathrm{eff} = \frac{\bar{n}}{t_\mathrm{cyc}},
\label{eq:Leff}
\end{equation}
where $t_\mathrm{cyc}$ is the total wall-clock time for one complete measurement repetition, as in Eq.~(\ref{eq:tcyc-qsp}). This definition automatically accounts for the lower repetition rate of longer QSP-inspired coherent-control sequences. Our low-dose claim should be understood as compatibility with pulsed, low-duty-cycle operation, and not as a universal one-to-one mapping between $L$ and biological exposure. Table~\ref{tab:common-unit} compares the Ramsey and QSP $d=10$ protocols across a range of photon rates and readout assumptions. The primary comparison is between the two protocols under identical conditions; the entry for the hydrogel--ND ensemble sensor of Ref.~\cite{zhang2018} ($96\,\mathrm{mK\,Hz^{-1/2}}$) is included for orientation.
A fully hardware-level prediction for a specific pulsed implementation should instead use $L_\mathrm{eff} = \bar{n}/t_\mathrm{cyc}$, which is protocol-dependent and is generally smaller for longer QSP-inspired coherent-control sequences. The photon-counting Fisher-information calculations use the conservative nanodiamond values $n_0 = 0.03$ and $n_1 = 0.02$, corresponding to $C = 0.20$. Column~(i) of Table~\ref{tab:common-unit} uses the CW-ODMR contrast of Ref.~\cite{zhang2018}, $C_\mathrm{ODMR} = 0.075$, and column~(ii) the pulsed value $C = 0.20$ used throughout; the two differ only by a common factor in each column, so the protocol ratio is the same in both.

\begin{table*}[t]
\centering
\caption{Frequency-normalized sensitivity $\eta_T$ [$\mathrm{mK\,Hz^{-1/2}}$] for the Ramsey and $d=10$ QSP-inspired protocols under two readout assumptions. (i) $f_c = C_\mathrm{ODMR} = 0.075$; (ii) $f_c = C = 0.20$. Lower values indicate better sensitivity. Bold entries correspond to $L=10^5\,\mathrm{s^{-1}}$, a representative single-NV photon rate. The values should be read as a common-photon-budget normalization; a hardware-specific prediction requires $L_\mathrm{eff}$. The ensemble sensor value of Ref.~\cite{zhang2018} is shown for orientation only.}
\label{tab:common-unit}
\small
\setlength{\tabcolsep}{5pt}
\renewcommand{\arraystretch}{1.15}
\begin{tabular}{l c c c c c}
\toprule
Scenario & Ram.\ (i) & QSP $d{=}10$ (i) & Ram.\ (ii) & QSP $d{=}10$ (ii) & Ref.~\cite{zhang2018} ens.\ (exp.) \\
\midrule
$L{=}10^{4}$ (low-dose) & 12.98 & 10.36 & 4.87 & 3.89 & 96 \\
$\mathbf{L{=}10^{5}}$ (typical sNV) & \textbf{4.11} & \textbf{3.28} & \textbf{1.54} & \textbf{1.23} & \textbf{96} \\
$L{=}10^{6}$ (high-dose) & 1.30 & 1.04 & 0.49 & 0.39 & 96 \\
$L{=}8{\times}10^{6}$ (ref.\ matched) & 0.46 & 0.37 & 0.17 & 0.14 & 96 \\
\bottomrule
\end{tabular}
\end{table*}

The comparison table places the spin-level response on the same shot-noise scale used in ODMR thermometry, and separates protocol-level gain from hardware-level photon throughput. Under a common effective detected photon rate the pulsed single-NV estimates are substantially smaller, hence more sensitive, than the demonstrated hydrogel--nanodiamond ensemble value of Ref.~\cite{zhang2018}. The central comparison of this work is the ratio between coherent control and optimized Ramsey under identical pulsed single-NV assumptions. Note also that, when the full cycle-time normalization is applied through $L_\mathrm{eff} = \bar n/t_\mathrm{cyc}$, using more layers does not automatically improve the frequency-normalized peak sensitivity, since the longer sequence lowers the repetition rate. The idealized single-NV CW-ODMR projection of Ref.~\cite{zhang2018} uses a different readout mode, magnetic geometry and optical-fluence assumption; we therefore use it to motivate the low-dose pulsed question and use the measured ensemble sensor as the external experimental anchor.

\section{Conclusion}
\label{sec:conclusion}

We presented a Fisher-information, photon-budget and duty-cycle audit of finite-depth coherent control for single-NV hydrogel thermometry. The hydrogel transducer converts temperature into a magnetic-field shift, and the pulsed single-NV architecture removes ensemble inhomogeneous broadening at the model level. Within this setting, optimized QSP-inspired response synthesis gives a reproducible gain over phase-optimized Ramsey. The per-shot improvement saturates at $25$--$27\%$ in sensitivity units, equivalently at a Fisher-information gain of $1.57$--$1.60$, and the same factor carries over to information per detected photon. Independent random restarts and differential evolution agree on this ceiling to within $10^{-5}$, so the ceiling is decoherence-limited.

The operational gain is set by the measurement duty cycle. With fast electronic-spin fluorescence readout, finite pulse durations select shallow depths and the Fisher-information-rate gain is a few percent. When the cycle is dominated by repetitive readout or by a dose-imposed idle time, the repetition-rate penalty of the longer sequence is suppressed and the rate gain approaches the per-shot ceiling, reaching $1.53$ at an overhead of $100~\mu$s. The regime that limits optical exposure is also the regime in which the control gain is most useful.

The optimized trajectories reveal the mechanism behind the long-time response. The sequence holds the spin near the longitudinal axis during early blocks, where pure dephasing is inactive, and accumulates phase mainly near the end. At depth two this is analytic: the closed form of Eq.~(\ref{eq:d2-closed}) keeps part of the Bloch vector longitudinal as an amplitude reference while the transverse component carries the phase, and it reproduces the numerical optimum to $0.05\%$. This decoherence-protected idling produces large equal-time ratios after the Ramsey signal has decayed, and the rate-normalized analysis gives the corresponding operational gain. Spin-echo and CPMG comparisons confirm that the effect is not dynamical decoupling, since refocusing pulses cancel the static thermometric phase. Optimal control can also reduce the estimation error contributed by population decay, as demonstrated on a superconducting qubit by Hecht \textit{et al.}~\cite{hecht2025}; the present sequence operates in the opposite corner of the same trade-off, parking the spin in the $T_1$-limited configuration to avoid the faster dephasing channel.

Optimized Ramsey saturates the free-evolution dephasing scale of Eq.~(\ref{eq:free-evolution-F})~\cite{escher2011} in our model, which we use as a check on the Lindblad description; no depth exceeds the dephasing-limited rate $\kappa^2/(2\gamma)$, the best $F/t$ being $1/e$ of it, as for Ramsey. The improvement is a finite-depth prefactor within the present ansatz and scan range, consistent with controlled sensing under dephasing~\cite{sekatski2017}.

The finite-depth gain is modest compared with ideal QSP scaling, which is what Markovian dephasing with finite pulse overhead allows. Its significance lies in the application setting. Hydrogel-transduced biosensing is constrained by optical dose and ensemble broadening. A pulsed single-NV protocol makes the time-averaged optical dose a free parameter, set by the idle time between shots, and coherent control improves the thermometric response within that same photon budget. The framework applies to PNIPAM--nanomagnet sensors and to other stimulus-responsive transducers that map biochemical or thermodynamic variables to magnetic fields.

\begin{acknowledgments}
The authors acknowledge helpful discussions and support from their host institutions. We also would like to express our gratitude to Ece Öztürk for her valuable support and contributions to this work. This research was supported by the Scientific and Technological Research Council of Türkiye (TÜBİTAK) under Project No.~123F150. The authors used artificial-intelligence-assisted language-editing tools to improve grammar, clarity, and readability of parts of the manuscript. All physical modeling, numerical simulations, analysis and conclusions are the authors' own, and the authors take full responsibility for the content of this work.
\end{acknowledgments}

\appendix
\section{Derivation of the Polynomial Response Function}

Here we show how the polynomial structure of $\langle\sigma_z\rangle$ arises from the coherent-control sequence. We analyze the canonical structure from the main text, $U_d$, acting on an initial state $\ket{\psi_{\text{in}}} = \ket{0}$. The state after $d$ evolution steps is
\begin{equation}
\ket{\psi_d} = U_d \ket{0} = \alpha_d \ket{0} + \beta_d \ket{1}.
\end{equation}

By tracing the state evolution layer-by-layer, one can show by induction that the state coefficient $\alpha_d$ is a Laurent polynomial in the variable $z = \exp(i\theta/2)$ of degree at most $d$. That is, it can be written in the form
\begin{equation}
\alpha_d(\theta) = \sum_{m=-d}^{d} c_m(\{\phi_k\})\, z^m = \sum_{m=-d}^{d} c_m(\{\phi_k\})\, e^{im\theta/2},
\label{eq:laurent}
\end{equation}
where the coefficients $c_m$ are functions of the processing angles $\{\phi_k\}$. For example, for the $d = 1$ case with simplified angles $\phi_0 = \phi_1 = \phi$, we find $\alpha_1 = \cos^2(\phi/2)\,z^{-1} - \sin^2(\phi/2)\,z^{1}$.

Since it is sufficient to find the population of one state, $|\alpha_d|^2$, to determine the final measurement $\langle\sigma_z\rangle_d = 2|\alpha_d|^2 - 1$, we examine its structure. The population $|\alpha_d|^2 = \alpha_d \alpha_d^*$ must be a real quantity. The product of the Laurent polynomial $\alpha_d$ with its conjugate results in another Laurent polynomial in $z$ with terms from $z^{-2d}$ to $z^{2d}$, i.e., from $e^{-id\theta}$ to $e^{id\theta}$. For this to be real, it must take the form of a real Fourier series. Using the Euler identity, this series can be expressed in terms of Chebyshev polynomials of $\cos(\theta)$ and $\sin(\theta)$. It follows that the population $|\alpha_d|^2$, and consequently the signal $\langle\sigma_z\rangle_d$, is a real polynomial of degree at most $d$ in the variable $\cos(\theta)$.

Thus, for arbitrary Euler-parameterized processing rotations, the measured signal $\langle\sigma_z\rangle_d$ is in general a real trigonometric polynomial of degree at most $d$ in $\theta$, equivalently a Laurent polynomial in $e^{i\theta}$. Under the usual QSP parity and phase-convention constraints, this trigonometric polynomial can be reduced to the familiar polynomial response in $x = \cos\theta$. Since our optimization uses the more general Euler parameterization $R_z(\alpha_k)\,R_x(\beta_k)\,R_z(\gamma_k)$, the relevant mathematical object is the bounded real trigonometric response generated by the sequence, which encompasses the standard QSP polynomial response as a special case and gives the optimizer more freedom for decoherence-aware protocol design.

\section{Physical Units and Dimensional Calibration}
\label{sec:units}

Throughout this work, simulations are performed in a
mixed unit system that combines natural dimensionless
ratios (e.g., $t/T_2^\star$) with absolute physical
quantities set by the NV Hamiltonian and the
hydrogel-mediated transduction chain of
Eqs.~(1)--(3). To enable a direct comparison of our
per-shot sensitivity
$\chi_T = |d\langle\sigma_z\rangle/dT|$ with the
$\mathrm{K\,Hz^{-1/2}}$ figures reported in experimental
ODMR-based thermometers such as Ref.~\cite{zhang2018},
we summarize here the physical scales underlying each
dimensionless quantity and provide the explicit
conversions used in Sec.~\ref{sec:results}.

\begin{table}[!tp]
\centering
\caption{Physical scales and reference values underlying
the simulations. The reference temperature is
$T_\mathrm{ref} = T_c = 305.15\,\mathrm{K}$ and the
reference magnetic field is $B_\mathrm{ref} = 0.27\,\mathrm{mT}$
at $r = r_\mathrm{high} = 550\,\mathrm{nm}$.}
\label{tab:units}
\setlength{\tabcolsep}{4pt}
\resizebox{\columnwidth}{!}{%
\begin{tabular}{l l l}
\hline
Quantity & Simulation convention & Physical value \\
\hline
\multicolumn{3}{l}{\textit{Temporal scales}} \\
$T_2^\star$ & time unit & $2.0\,\mu\mathrm{s}$ \\
$T_{2,\min}$ & --- & $0.5\,\mu\mathrm{s}$ \\
$T_1$ at $T_\mathrm{ref}$ & $50\,T_2^\star$ & $100\,\mu\mathrm{s}$ \\
$t_\mathrm{init}$ (optical pumping) & --- & $\sim 2\,\mu\mathrm{s}$ \\
$t_\mathrm{read}$ (readout window) & --- & $\sim 300\,\mathrm{ns}$ \\
$\beta$ (dephasing curvature) & $0.02\,\mathrm{K^{-2}}$ & $0.02\,\mathrm{K^{-2}}$ \\
\hline
\multicolumn{3}{l}{\textit{Spectral scales}} \\
$D$ (zero-field splitting) & --- & $2\pi \times 2.87\,\mathrm{GHz}$ \\
$\gamma_e$ (gyromagnetic ratio) & --- & $2\pi \times 28\,\mathrm{GHz/T}$ \\
$B_0(T_c)$ at $r(T_c) = 450\,\mathrm{nm}$ & --- & $0.493\,\mathrm{mT}$ \\
$\gamma_e B_0(T_c)$ & --- & $2\pi \times 13.80\,\mathrm{MHz}$ \\
$\omega_q(T_c) = D - \gamma_e B_0$ & --- & $2\pi \times 2.8562\,\mathrm{GHz}$ \\
$|d\omega_q/dT|$ at $T_\mathrm{target}$ & --- & $\sim 2\pi \times 2.2\,\mathrm{MHz/K}$ \\
$|d\omega_q/dT|_\mathrm{peak}$ (at $306.9\,\mathrm{K}$) & --- & $\sim 2\pi \times 2.8\,\mathrm{MHz/K}$ \\
Intrinsic $dD/dT$ & neglected & $-2\pi \times 74\,\mathrm{kHz/K}$ \\
\hline
\multicolumn{3}{l}{\textit{Spatial and chemical scales}} \\
$r_\mathrm{low}$ (shrunken hydrogel) & --- & $350\,\mathrm{nm}$ \\
$r_\mathrm{high}$ (swollen hydrogel) & --- & $550\,\mathrm{nm}$ \\
$T_c$ (PNIPAM VPT) & --- & $305.15\,\mathrm{K}$ \\
$k$ (VPT steepness) & $0.5\,\mathrm{K^{-1}}$ & $0.5\,\mathrm{K^{-1}}$ \\
\hline
\multicolumn{3}{l}{\textit{Sensitivity outputs}} \\
$\chi_T^\mathrm{Ramsey,peak}$ & $10.27\,\mathrm{K^{-1}}$ & per shot \\
$\chi_T^{d=10,\mathrm{peak}}$ & $12.87\,\mathrm{K^{-1}}$ & per shot \\
$\delta T$ (finite-difference step) & $10^{-4}\,\mathrm{K}$ & $10^{-4}\,\mathrm{K}$ \\
\hline
\end{tabular}%
}
\end{table}

\paragraph{Order-of-magnitude consistency check.}
A useful back-of-the-envelope check relates $\chi_T$ to
the intrinsic transduction derivative $d\omega_q/dT$. In
the decoherence-free Ramsey limit with optimal readout
phase, the single-shot sensitivity satisfies
$\chi_T^\mathrm{ideal} \approx |d\omega_q/dT|\,\tau$.
At the operating point $\tau \approx T_2^\star = 2\,\mu\mathrm{s}$
and with $|d\omega_q/dT| \approx 2\pi \times 2.2\,\mathrm{MHz/K}$ at $T_\mathrm{target} = 305.0\,\mathrm{K}$
near $T_c$, this yields
$\chi_T^\mathrm{ideal} \approx 2\pi \times 4.4 \approx 27.6\,\mathrm{K^{-1}}$.
Our optimized Ramsey result $\chi_T = 10.27\,\mathrm{K^{-1}}$
is approximately a factor of 2.7 below this decoherence-free
upper bound, which is the expected reduction for a Ramsey
signal averaged over the full dephasing envelope
$e^{-t/T_2^\star}$; the optimized QSP result
$\chi_T = 12.87\,\mathrm{K^{-1}}$ recovers part of this
loss by operating at $t = 3\,T_2^\star$ through
algorithmic amplification, as analyzed in
Sec.~\ref{sec:results}.

\paragraph{Conversion to frequency-normalized sensitivity.}
Experimental thermometers based on CW-ODMR typically report 
the shot-noise-limited sensitivity $\eta_T$ in 
units of $\mathrm{K\,Hz^{-1/2}}$, defined as the temperature 
uncertainty achievable in a $1\,\mathrm{s}$ integration. For 
a detector that collects photons at a steady-state rate $L$ 
from a signal $S(T)$, the conversion between our per-shot 
sensitivity $\chi_T$ and the frequency-normalized sensitivity 
follows the same shot-noise relation used in 
Ref.~\cite{zhang2018}:
\begin{equation}
\eta_T \;\approx\; \frac{1}{f_c\,\sqrt{L}\,\chi_T}.
\label{eq:eta-conversion}
\end{equation}
Equation~(\ref{eq:eta-conversion}) assumes that the readout 
is shot-noise dominated and that the normalized signal 
$\langle\sigma_z\rangle$ or $S$ is directly estimated from 
photon-count statistics at rate $L$ over a one-second 
integration. The explicit photon-budget factors relating 
$L$ to the mean per-shot photon count $\bar{n}$, the spin contrast 
$C$ and the measurement cycle time $t_\mathrm{cycle}$ are 
protocol-dependent and are addressed in the fair-comparison
discussion of Appendix~\ref{sec:fair-comparison}.

Equation~(\ref{eq:eta-conversion}) is not the 
sensitivity figure we optimize in Sec.~\ref{sec:results}; 
we optimize the detector-agnostic per-shot quantity 
$\chi_T$, which isolates the algorithmic and 
coherence-limited physics from the throughput-dependent 
factor $\sqrt{L}$. The conversion above is provided only 
to place our results on a common axis with experimental 
ODMR benchmarks, which we carry out explicitly in 
Appendix~\ref{sec:fair-comparison}.

\section{Fair Comparison with the Reference Experiment}
\label{sec:fair-comparison}

We use Ref.~\cite{zhang2018} as a reference point for the modular hydrogel--nanodiamond sensing architecture. The experimentally realized ensemble sensor of Ref.~\cite{zhang2018} achieved $\eta \approx 96\,\mathrm{mK\,Hz^{-1/2}}$ in water, with a comparable sensitivity reported in fetal bovine serum, providing a concrete experimental baseline for this class of sensors in the intended operating environment. Since our model geometry and hardware parameters differ from those of Ref.~\cite{zhang2018}, we use the reference to anchor the physical picture of the two-stage transduction chain, and compare Ramsey with coherent control under identical model assumptions.

Our per-shot sensitivity $\chi_T = |d\langle\sigma_z\rangle/dT|$ 
and the frequency-normalized sensitivity
$\eta_T$ reported by Zhang \textit{et al.}~\cite{zhang2018}
are placed on a common photon-budget scale: their figure embeds a detector 
throughput factor $\sqrt{L}$ (where $L$ is the steady-state 
photon count rate), while ours is defined at the level of a 
single projective measurement.

The physically aligned comparison is the common photon-budget scale of Eq.~(\ref{eq:our-eta-Hz}), where protocol-level gain and hardware throughput are separated. A direct comparison of raw derivatives is not meaningful here: the present quantity is a per-shot, decoherence-limited spin derivative optimized over sensing time and readout phase, while the reference quantity is a CW frequency-domain derivative evaluated at a fixed operating point and a broadened ODMR line.

The pulsed single-NV coherent-control approach has three advantages over ensemble-based hydrogel sensors. It removes gradient-induced ensemble broadening, it makes the time-averaged optical dose a free parameter set by the idle time between shots, and it retains a finite local response after the optimized Ramsey signal has decayed. The two settings occupy different corners of the same trade-off. The projection of Ref.~\cite{zhang2018} assumes continuous ODMR at $>100\,\mathrm{\mu W\,\mu m^{-2}}$, which sustains $\sim 10^{5}$ counts per second from one emitter; the present model operates at $L_\mathrm{eff} = \bar n/t_\mathrm{cyc} \approx 5.7\times10^{3}\,\mathrm{s^{-1}}$, roughly twenty times lower. Lower count rate, lower sensitivity and lower delivered dose go together, and the pulsed scheme does not remove the high-fluence requirement of that readout mode. The comparison to Ref.~\cite{zhang2018} anchors the application setting, while the central protocol comparison remains the controlled sequence against phase-optimized Ramsey under identical pulsed single-NV assumptions.

\section{Signal-Spin Conversion Coefficient}
\label{sec:alpha-coefficient}

The comparison carried out in Appendix~\ref{sec:fair-comparison} 
converts between the normalized ODMR fluorescence signal 
$S$ reported by Ref.~\cite{zhang2018} and the spin 
expectation value $\langle\sigma_z\rangle$ used throughout 
this work. Because $S$ and $\langle\sigma_z\rangle$ are 
not the same physical observable, this conversion 
requires a well-defined proportionality factor $f_c$, 
which we derive here from the standard spin-dependent 
fluorescence model of the NV 
center~\cite{doherty2013, rondin2014}.

The NV readout proceeds via spin-dependent 
photoluminescence: resonant $532\,\mathrm{nm}$ excitation 
drives the $|^3\!A_2\rangle \rightarrow |^3\!E\rangle$ 
transition, and the return path branches between 
radiative decay (yielding a detectable red photon) and 
non-radiative intersystem crossing (ISC) to a metastable 
singlet. The ISC rate is larger for the $m_s = \pm 1$ 
states than for $m_s = 0$, so the mean number of photons 
collected in a readout window of duration 
$t_\mathrm{read}$ depends on the spin 
state~\cite{doherty2013}:
\begin{align}
n_0 &\equiv \langle n \rangle_{m_s=0} = \eta_\mathrm{coll}\,\Gamma_\mathrm{rad}^{(0)}\,t_\mathrm{read}, \\
n_1 &\equiv \langle n \rangle_{m_s=-1} = \eta_\mathrm{coll}\,\Gamma_\mathrm{rad}^{(1)}\,t_\mathrm{read},
\end{align}
where $\eta_\mathrm{coll}$ is the overall photon-collection 
efficiency (objective solid angle, filter transmission, 
and detector quantum efficiency) and 
$\Gamma_\mathrm{rad}^{(s)}$ is the effective spin-state 
radiative rate integrated over the excitation--collection 
cycle. Typical values for a single NV in a nanodiamond 
under green laser excitation are 
$n_0 \approx 0.03$ photons per shot and 
$n_1 \approx 0.02$ photons per 
shot~\cite{taylor2008, rondin2014}.

For a qubit defined in the 
$\{|m_s=0\rangle, |m_s=-1\rangle\}$ subspace, the 
populations satisfy 
$P_0 = (1+\langle\sigma_z\rangle)/2$ and 
$P_1 = (1-\langle\sigma_z\rangle)/2$. The mean photon 
count per readout is therefore
\begin{equation}
\langle n \rangle = n_0\,P_0 + n_1\,P_1 = \bar{n}\,\bigl(1 + C\,\langle\sigma_z\rangle\bigr),
\label{eq:photon-count}
\end{equation}
where
\begin{equation}
\bar{n} \equiv \tfrac{1}{2}(n_0 + n_1), \qquad
C \equiv \frac{n_0 - n_1}{n_0 + n_1}
\label{eq:contrast-def}
\end{equation}
are the mean photon count and the intrinsic optical spin 
contrast. The normalized fluorescence signal used in 
Ref.~\cite{zhang2018}, 
$S = \langle n \rangle / \langle n \rangle_\mathrm{ref}$, 
is thus an affine function of $\langle\sigma_z\rangle$:
\begin{equation}
S = S_0 \,\bigl(1 + C\,\langle\sigma_z\rangle\bigr),
\label{eq:S-sigma}
\end{equation}
with $S_0$ a temperature-independent normalization 
constant (set to $1/2$ when referenced to the 
off-resonant count rate, or to unity when referenced to 
the $m_s = 0$ bright state). Taking the temperature 
derivative of Eq.~(\ref{eq:S-sigma}) yields
\begin{equation}
\frac{dS}{dT} = S_0\,C\,\frac{d\langle\sigma_z\rangle}{dT}.
\label{eq:alpha-def}
\end{equation}

The shot noise on $S$ carries the same factor $S_0$, since $S$ is itself
normalized by $\langle n \rangle_\mathrm{ref}$. The normalization therefore
cancels in the signal-to-noise ratio, and the conversion coefficient entering
Eq.~(\ref{eq:our-eta-Hz}) is the contrast alone,
\begin{equation}
f_c \equiv C .
\label{eq:alpha-def2}
\end{equation}
This is the convention already used in Eq.~(\ref{eq:eta-photon}), which is
derived from the photon-counting model without reference to $S_0$.

The conversion coefficient depends only on the readout modality.
For the CW-ODMR readout of Ref.~\cite{zhang2018}, the reported contrast is
$C_\mathrm{ODMR} \approx 0.075$~\cite{zhang2018,barry2020}, so
$f_{c,\mathrm{ODMR}} = 0.075$. For the pulsed single-NV readout used here,
$f_{c,\mathrm{pulsed}} = C = 0.20$. The Fisher-information ratios in the
main text use the explicit photon-counting model of
Sec.~\ref{sec:photon-model}, with $n_0 = 0.03$, $n_1 = 0.02$ and
$C = 0.20$.

A fully quantum-optical treatment of the excitation--emission cycle 
(including the atom--field coupling $g$ of the Jaynes--Cummings model) 
would reconstruct $\Gamma_\mathrm{rad}^{(s)}$ from the microscopic dipole 
matrix elements and predict corrections to $f_c$ in the strong-driving 
regime~\cite{doherty2013}; in the weak-excitation, broadband-collection 
regime relevant here these are absorbed into the phenomenological 
$\eta_\mathrm{coll}$ and $\Gamma_\mathrm{rad}^{(s)}$, so Eq.~(\ref{eq:S-sigma}) 
holds to leading order and the values above are accurate within the 
$10$--$20\%$ experimental uncertainty on $C$ and $\eta_\mathrm{coll}$.

\section{Ramsey phase calibration}
\label{sec:ramsey-calib}

\begin{figure}[!tp]
\centering
\includegraphics[width=\columnwidth]{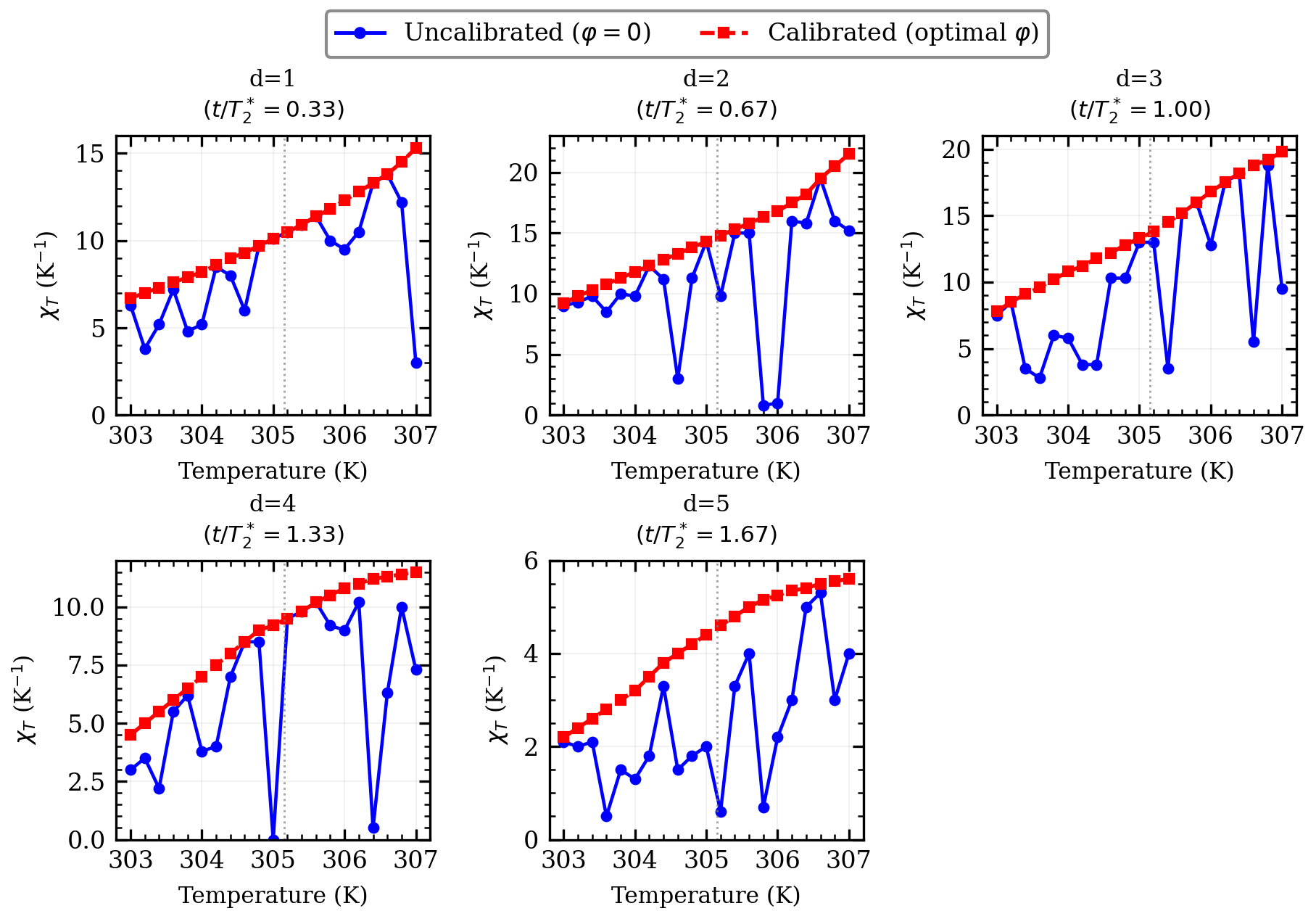}
\caption{Comparison of uncalibrated ($\varphi = 0$, blue) and phase-calibrated (optimal $\varphi$, red) Ramsey thermal spin susceptibility $\chi_T$ as a function of temperature for $d = 1$ through $d = 5$. Without phase calibration, the signal exhibits rapid oscillations with numerous zero-crossings as the accumulated phase $\omega_q(T)\tau$ sweeps through multiples of $\pi$. Optimizing $\varphi$ produces a smooth envelope that tracks the underlying transduction sensitivity $|d\omega_q/dT|$.}
\label{fig:calibration}
\end{figure}

The Ramsey baseline used in Sec.~\ref{sec:results} requires the readout phase $\varphi$ to be optimized at each operating point. Figure~\ref{fig:calibration} compares the Ramsey sensitivity for a fixed phase ($\varphi = 0$, blue) with the fully phase-optimized protocol (optimal $\varphi$, red) across a range of temperatures near $T_c$, for $d = 1$ through $d = 5$ QSP layers.

The uncalibrated signal shows rapid oscillations as a function of temperature, with numerous nodes where the sensitivity drops to zero. These oscillations arise because the accumulated phase $\theta = \omega_q(T)\tau$ sweeps through multiples of $\pi$ as the temperature varies, causing the Ramsey fringe to pass through extrema where $d\langle\sigma_z\rangle/dT = 0$. By optimizing $\varphi$ at each temperature point, the readout is shifted to the steepest part of the fringe, yielding a smooth sensitivity envelope. The summary panel in Fig.~\ref{fig:calibration} confirms that calibration helps at the vast majority of temperature points for all $d$ values, establishing that phase optimization is a prerequisite for any meaningful sensitivity comparison.

\section{Depth-two closed form}
\label{sec:d2-derivation}

This appendix derives Eq.~(\ref{eq:d2-closed}). Work in the Bloch picture with the free-evolution map of Eq.~(\ref{eq:bloch-map}) and evaluate the derivative at $\theta = 0$. Writing
\begin{equation}
\bm{u} = \Lambda \bm{r}_0 = (\lambda s, 0, \mu c), \qquad
\bm{v} = \Lambda J \bm{r}_0 = (0, \lambda s, 0),
\end{equation}
with $s = \sin a$, $c = \cos a$, one has $\bm{u}\cdot\bm{v} = 0$, $|\bm{u}| = P \equiv \sqrt{\lambda^2 s^2 + \mu^2 c^2}$ and $|\bm{v}| = Q \equiv \lambda s$. Setting $\bm{p} = R\bm{u}$ and $\bm{q} = R\bm{v}$, the rotation $R$ is free apart from preserving the norms and the orthogonality, and Eq.~(\ref{eq:d2-derivative}) becomes
\begin{equation}
\partial_\theta \bm{r}_f = \bigl(\lambda(q_x - p_y),\; \lambda(p_x + q_y),\; \mu q_z\bigr),
\end{equation}
whose squared norm can be organized as
\begin{equation}
|\partial_\theta \bm{r}_f|^2 = \lambda^2\bigl[P^2 + Q^2 - p_z^2 - q_z^2 + 2(\bm{p}\times\bm{q})_z\bigr] + \mu^2 q_z^2 .
\label{eq:d2-norm}
\end{equation}
Since $\mu > \lambda$, the term in $q_z^2$ is favorable while the term in $p_z^2$ is not, so the optimum has $p_z = 0$ and $\bm{p}$ in the equatorial plane. Decomposing $\bm{q}$ into an equatorial part $q_\perp$ and $q_z$, with $q_\perp^2 + q_z^2 = Q^2$ and $(\bm{p}\times\bm{q})_z = P q_\perp$, Eq.~(\ref{eq:d2-norm}) reads
\begin{equation}
f(q_z) = \lambda^2 P^2 + \lambda^2 Q^2 + (\mu^2-\lambda^2) q_z^2 + 2\lambda^2 P \sqrt{Q^2 - q_z^2}.
\end{equation}
Stationarity gives
\begin{equation}
q_\perp = \frac{\lambda^2 P}{\mu^2 - \lambda^2}.
\label{eq:d2-stationary}
\end{equation}
When Eq.~(\ref{eq:d2-stationary}) lies below $Q$, substitution collapses the expression,
\begin{equation}
|\partial_\theta \bm{r}_f|^2 = \lambda^2 P^2 + \mu^2 Q^2 + \frac{\lambda^4 P^2}{\mu^2-\lambda^2} = \frac{\lambda^2 \mu^4}{\mu^2 - \lambda^2},
\end{equation}
which is independent of the preparation angle $a$ and gives the first branch of Eq.~(\ref{eq:d2-closed}). Otherwise $q_\perp$ saturates at $Q$, the derivative reduces to $\lambda(P+Q)$, and maximizing over $a$ places the whole Bloch vector in the equatorial plane, $s = 1$, giving $2\lambda^2$. The two branches meet at $\lambda = \mu/\sqrt{2}$, where both equal $\mu^2$, so the closed form is continuous.

We verified Eq.~(\ref{eq:d2-closed}) against a direct numerical maximization of Eq.~(\ref{eq:d2-derivative}) over the preparation angle, the intermediate Euler rotation and the readout direction: the two agree to machine precision at every interrogation time on the grid.

\section{Comparison of optimization objectives}
\label{sec:objective-comparison}

The four objectives of Eq.~(\ref{eq:objectives}) were optimized independently, from the same set of random starts, at $d = 3$, $5$ and $10$. Objectives 1--3 return the same optimum to within $10^{-6}$ relative. This is expected: at fixed $(d,t)$ the cycle time is a constant and cannot affect the maximizer, so objectives 2 and 3 coincide exactly, while objectives 1 and 2 differ only through the factor $1/(1-z^2)$, which is unity at the $z \simeq 0$ working point the optimizer selects.

Objective 4 does reach a distinct optimum. Reducing $\bar n$ in the denominator of Eq.~(\ref{eq:FCph}) lowers the shot noise, which biases the working point towards the dark state, to $z$ between $-0.017$ and $-0.035$. The resulting improvement in $F_C^{(\mathrm{ph})}/t_\mathrm{cyc}$ is at most $0.35\%$, because the bias is limited by the optical contrast of Eq.~(\ref{eq:nbar0-C}). The four objectives are therefore equivalent in practice, and all results below use $F_C^{(z)}$.

\section{Optimizer reliability}
\label{sec:optimizer-reliability}

Because the control landscape is nonconvex, we verified that the reported ceiling is not a local-search artifact. At each of three representative operating points we ran independent L-BFGS-B searches from uniformly random initial angles, with no warm starts, and separately differential evolution (DE) followed by L-BFGS-B polishing:

\begin{center}
\footnotesize
\setlength{\tabcolsep}{3pt}
\begin{tabular}{@{}lcccc@{}}
\toprule
case & restarts & best $F_C^{(z)}$ & spread (best 10) & DE ratio \\
\midrule
$d=3$, $t/T_2^*{=}2.0$  & 100 & 143.6 & $1.3{\times}10^{-10}$ & 0.99999 \\
$d=5$, $t/T_2^*{=}2.5$  & 100 & 156.4 & $1.0{\times}10^{-9}$  & 1.00000 \\
$d=10$, $t/T_2^*{=}3.0$ & 200 & 165.6 & $3.5{\times}10^{-6}$  & 0.99999 \\
\bottomrule
\end{tabular}
\end{center}

Of the 400 independent restarts, all but one converged to the same optimum to within $10^{-4}$ relative; the single exception, at $d=5$, landed $3.9\%$ below. Differential evolution reproduces the same optima to $1.1\times10^{-5}$. The $25$--$27\%$ figure is therefore a genuine ceiling, and CMA-ES or basinhopping are not required. The cumulative distribution of restart outcomes is given in Appendix~\ref{sec:supplemental} (Fig.~\ref{fig:numerics}(a)).

\section{Supplemental figures}
\label{sec:supplemental}

This appendix collects the supporting numerical checks referred to in the main text. These figures document convergence, reliability and the extent of the parameter scans.

\begin{figure*}[!tp]
\centering
\includegraphics[width=0.62\textwidth]{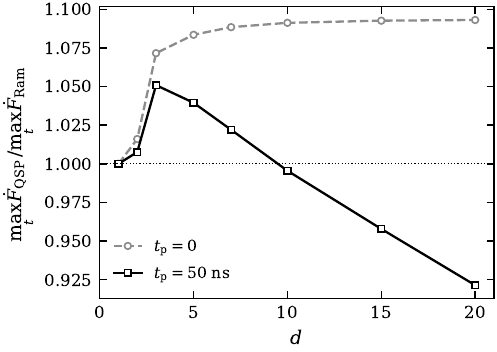}
\caption{Global Fisher-information-rate gain $G_{\dot F}^\mathrm{glob}(d)$ against depth at the nominal overhead $t_\mathrm{ovh} = 2.3~\mu$s, under the two pulse-duration assumptions. With $t_\mathrm{p} = 50$~ns a moderate depth $d \simeq 3$--$5$ is selected and the advantage is lost beyond $d \approx 10$. This is the single-overhead cut of Fig.~\ref{fig:overhead}(b); the restriction does not survive a more expensive readout.}
\label{fig:optimal_depth}
\end{figure*}

\begin{figure*}[!tp]
\centering
\includegraphics[width=\textwidth]{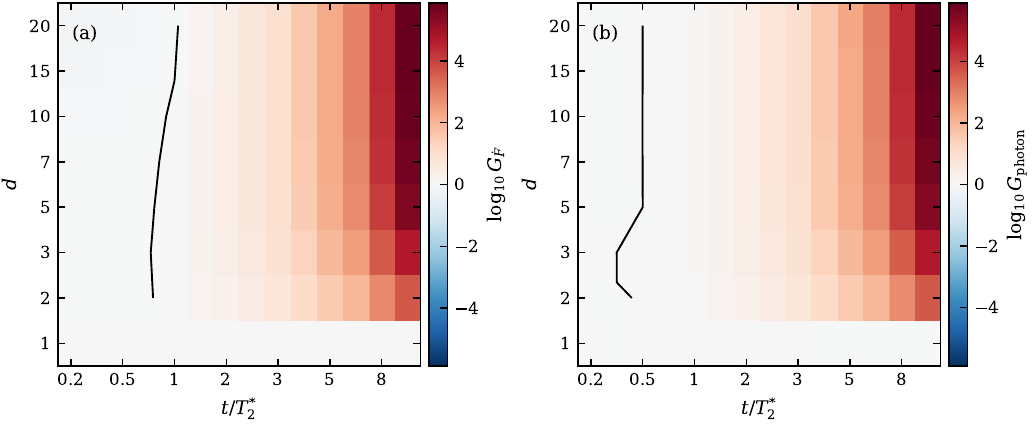}
\caption{Gain maps over the full $(d,t)$ grid. The peak-to-peak numbers of Table~\ref{tab:summary} are cuts through these surfaces along each protocol's own optimal interrogation time.}
\label{fig:gain-maps}
\end{figure*}

The equal-time gain landscape is shown first in Fig.~\ref{fig:gain-maps}, since the peak-to-peak numbers quoted in the main text are particular cuts through it. The ridge running to long interrogation times is the response-reshaping regime analyzed in Sec.~\ref{sec:mechanism}; see Sec.~\ref{sec:mechanism}; the peak-to-peak ratios of Table~\ref{tab:summary} follow each protocol's own optimum instead.

The next figure documents the numerical reliability of the optimization. The restart distribution shows that the reported ceiling is reached from essentially every random initial condition. This rules out a local-search artefact; the derivative step is chosen in the plateau between truncation error at large $\delta T$ and cancellation error at small $\delta T$, which spans more than two decades.

\begin{figure*}[!tp]
\centering
\includegraphics[width=0.48\textwidth]{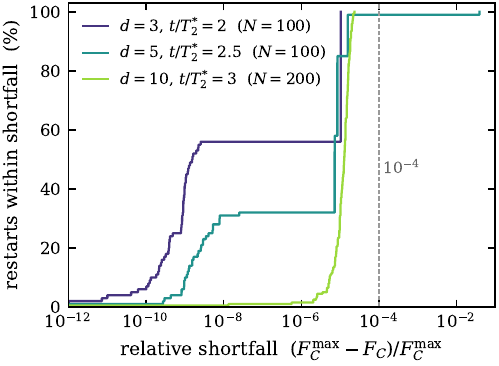}\hfill
\includegraphics[width=0.48\textwidth]{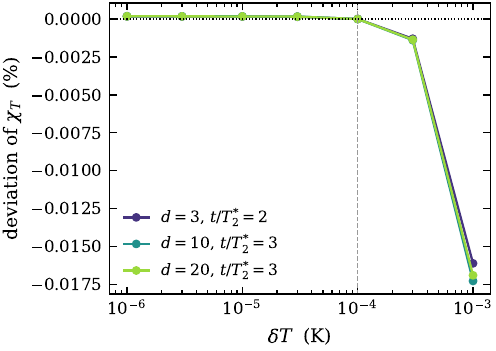}
\caption{Numerical reliability of the optimization. (a) Cumulative distribution of independent optimizer restarts at the three operating points of Appendix~\ref{sec:optimizer-reliability}; all but one of the 400 restarts converge to the same optimum to within $10^{-4}$ relative. (b) Convergence of the symmetric finite-difference derivative with the temperature step $\delta T$, justifying the value $\delta T = 10^{-4}$~K used throughout.}
\label{fig:numerics}
\end{figure*}

\begin{figure*}[!tp]
\centering
\includegraphics[width=\textwidth]{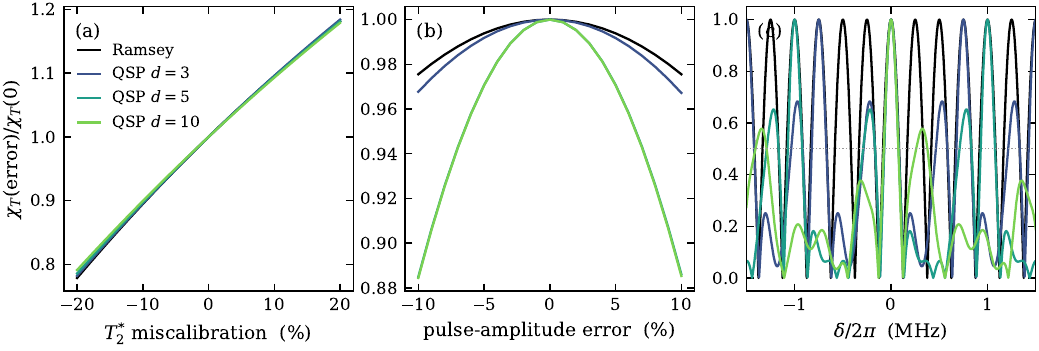}
\caption{Sensitivity retention with the control phases held at their nominal optimum, for the Ramsey baseline and for several depths: (a) $T_2^*$ miscalibration, (b) pulse-amplitude error, (c) detuning. The retention curves of Ramsey and QSP coincide in (a), so the gain ratio is invariant under $T_2^*$ miscalibration. The acceptance bandwidths quoted in Sec.~\ref{sec:robustness} are read off (c) as the contiguous interval around $\delta = 0$ over which the retention exceeds $1/2$.}
\label{fig:robustness-curves}
\end{figure*}

The last figure, Fig.~\ref{fig:robustness-curves}, gives the continuous versions of the robustness numbers tabulated in Sec.~\ref{sec:robustness}. The retention curves for Ramsey and for the QSP depths lie on top of one another under $T_2^*$ miscalibration, which is the statement that the gain ratio is invariant under that error; the amplitude-error panel shows the mild additional cost of the extra pulses; and the detuning scan gives the moderately narrower acceptance window of the deeper sequences, which follows from their longer effective interrogation time.

\clearpage

\bibliographystyle{apsrev4-2}
\bibliography{references}

\begin{thebibliography}{44}%
\makeatletter
\providecommand \@ifxundefined [1]{%
 \@ifx{#1\undefined}
}%
\providecommand \@ifnum [1]{%
 \ifnum #1\expandafter \@firstoftwo
 \else \expandafter \@secondoftwo
 \fi
}%
\providecommand \@ifx [1]{%
 \ifx #1\expandafter \@firstoftwo
 \else \expandafter \@secondoftwo
 \fi
}%
\providecommand \natexlab [1]{#1}%
\providecommand \enquote  [1]{``#1''}%
\providecommand \bibnamefont  [1]{#1}%
\providecommand \bibfnamefont [1]{#1}%
\providecommand \citenamefont [1]{#1}%
\providecommand \href@noop [0]{\@secondoftwo}%
\providecommand \href [0]{\begingroup \@sanitize@url \@href}%
\providecommand \@href[1]{\@@startlink{#1}\@@href}%
\providecommand \@@href[1]{\endgroup#1\@@endlink}%
\providecommand \@sanitize@url [0]{\catcode `\\12\catcode `\$12\catcode
  `\&12\catcode `\#12\catcode `\^12\catcode `\_12\catcode `\%12\relax}%
\providecommand \@@startlink[1]{}%
\providecommand \@@endlink[0]{}%
\providecommand \url  [0]{\begingroup\@sanitize@url \@url }%
\providecommand \@url [1]{\endgroup\@href {#1}{\urlprefix }}%
\providecommand \urlprefix  [0]{URL }%
\providecommand \Eprint [0]{\href }%
\providecommand \doibase [0]{https://doi.org/}%
\providecommand \selectlanguage [0]{\@gobble}%
\providecommand \bibinfo  [0]{\@secondoftwo}%
\providecommand \bibfield  [0]{\@secondoftwo}%
\providecommand \translation [1]{[#1]}%
\providecommand \BibitemOpen [0]{}%
\providecommand \bibitemStop [0]{}%
\providecommand \bibitemNoStop [0]{.\EOS\space}%
\providecommand \EOS [0]{\spacefactor3000\relax}%
\providecommand \BibitemShut  [1]{\csname bibitem#1\endcsname}%
\let\auto@bib@innerbib\@empty
\bibitem [{\citenamefont {Brites}\ \emph {et~al.}(2012)\citenamefont {Brites},
  \citenamefont {Lima}, \citenamefont {Silva}, \citenamefont {Mill{\'a}n},
  \citenamefont {Amaral}, \citenamefont {Palacio},\ and\ \citenamefont
  {Carlos}}]{brites2012}%
  \BibitemOpen
  \bibfield  {author} {\bibinfo {author} {\bibfnamefont {C.~D.~S.}\
  \bibnamefont {Brites}}, \bibinfo {author} {\bibfnamefont {P.~P.}\
  \bibnamefont {Lima}}, \bibinfo {author} {\bibfnamefont {N.~J.~O.}\
  \bibnamefont {Silva}}, \bibinfo {author} {\bibfnamefont {A.}~\bibnamefont
  {Mill{\'a}n}}, \bibinfo {author} {\bibfnamefont {V.~S.}\ \bibnamefont
  {Amaral}}, \bibinfo {author} {\bibfnamefont {F.}~\bibnamefont {Palacio}},\
  and\ \bibinfo {author} {\bibfnamefont {L.~D.}\ \bibnamefont {Carlos}},\
  }\href {https://doi.org/10.1039/C2NR30663H} {\bibfield  {journal} {\bibinfo
  {journal} {Nanoscale}\ }\textbf {\bibinfo {volume} {4}},\ \bibinfo {pages}
  {4799} (\bibinfo {year} {2012})}\BibitemShut {NoStop}%
\bibitem [{\citenamefont {Kucsko}\ \emph {et~al.}(2013)\citenamefont {Kucsko},
  \citenamefont {Maurer}, \citenamefont {Yao}, \citenamefont {Kubo},
  \citenamefont {Noh}, \citenamefont {Lo}, \citenamefont {Park},\ and\
  \citenamefont {Lukin}}]{kucsko2013}%
  \BibitemOpen
  \bibfield  {author} {\bibinfo {author} {\bibfnamefont {G.}~\bibnamefont
  {Kucsko}}, \bibinfo {author} {\bibfnamefont {P.~C.}\ \bibnamefont {Maurer}},
  \bibinfo {author} {\bibfnamefont {N.~Y.}\ \bibnamefont {Yao}}, \bibinfo
  {author} {\bibfnamefont {M.}~\bibnamefont {Kubo}}, \bibinfo {author}
  {\bibfnamefont {H.~J.}\ \bibnamefont {Noh}}, \bibinfo {author} {\bibfnamefont
  {P.~K.}\ \bibnamefont {Lo}}, \bibinfo {author} {\bibfnamefont
  {H.}~\bibnamefont {Park}},\ and\ \bibinfo {author} {\bibfnamefont {M.~D.}\
  \bibnamefont {Lukin}},\ }\href {https://doi.org/10.1038/nature12373}
  {\bibfield  {journal} {\bibinfo  {journal} {Nature}\ }\textbf {\bibinfo
  {volume} {500}},\ \bibinfo {pages} {54} (\bibinfo {year} {2013})}\BibitemShut
  {NoStop}%
\bibitem [{\citenamefont {Toyli}\ \emph {et~al.}(2013)\citenamefont {Toyli},
  \citenamefont {de~las Casas}, \citenamefont {Christle}, \citenamefont
  {Dobrovitski},\ and\ \citenamefont {Awschalom}}]{toyli2013}%
  \BibitemOpen
  \bibfield  {author} {\bibinfo {author} {\bibfnamefont {D.~M.}\ \bibnamefont
  {Toyli}}, \bibinfo {author} {\bibfnamefont {C.~F.}\ \bibnamefont {de~las
  Casas}}, \bibinfo {author} {\bibfnamefont {D.~J.}\ \bibnamefont {Christle}},
  \bibinfo {author} {\bibfnamefont {V.~V.}\ \bibnamefont {Dobrovitski}},\ and\
  \bibinfo {author} {\bibfnamefont {D.~D.}\ \bibnamefont {Awschalom}},\ }\href
  {https://doi.org/10.1073/pnas.1306825110} {\bibfield  {journal} {\bibinfo
  {journal} {Proc. Natl. Acad. Sci. USA}\ }\textbf {\bibinfo {volume} {110}},\
  \bibinfo {pages} {8417} (\bibinfo {year} {2013})}\BibitemShut {NoStop}%
\bibitem [{\citenamefont {Neumann}\ \emph {et~al.}(2013)\citenamefont
  {Neumann}, \citenamefont {Jakobi}, \citenamefont {Dolde}, \citenamefont
  {Burk}, \citenamefont {Reuter}, \citenamefont {Waldherr}, \citenamefont
  {Honert}, \citenamefont {Wolf}, \citenamefont {Brunner}, \citenamefont
  {Shim}, \citenamefont {Suter}, \citenamefont {Sumiya}, \citenamefont
  {Isoya},\ and\ \citenamefont {Wrachtrup}}]{neumann2013}%
  \BibitemOpen
  \bibfield  {author} {\bibinfo {author} {\bibfnamefont {P.}~\bibnamefont
  {Neumann}}, \bibinfo {author} {\bibfnamefont {I.}~\bibnamefont {Jakobi}},
  \bibinfo {author} {\bibfnamefont {F.}~\bibnamefont {Dolde}}, \bibinfo
  {author} {\bibfnamefont {C.}~\bibnamefont {Burk}}, \bibinfo {author}
  {\bibfnamefont {R.}~\bibnamefont {Reuter}}, \bibinfo {author} {\bibfnamefont
  {G.}~\bibnamefont {Waldherr}}, \bibinfo {author} {\bibfnamefont
  {J.}~\bibnamefont {Honert}}, \bibinfo {author} {\bibfnamefont
  {T.}~\bibnamefont {Wolf}}, \bibinfo {author} {\bibfnamefont {A.}~\bibnamefont
  {Brunner}}, \bibinfo {author} {\bibfnamefont {J.~H.}\ \bibnamefont {Shim}},
  \bibinfo {author} {\bibfnamefont {D.}~\bibnamefont {Suter}}, \bibinfo
  {author} {\bibfnamefont {H.}~\bibnamefont {Sumiya}}, \bibinfo {author}
  {\bibfnamefont {J.}~\bibnamefont {Isoya}},\ and\ \bibinfo {author}
  {\bibfnamefont {J.}~\bibnamefont {Wrachtrup}},\ }\href
  {https://doi.org/10.1021/nl401216y} {\bibfield  {journal} {\bibinfo
  {journal} {Nano Lett.}\ }\textbf {\bibinfo {volume} {13}},\ \bibinfo {pages}
  {2738} (\bibinfo {year} {2013})}\BibitemShut {NoStop}%
\bibitem [{\citenamefont {Simpson}\ \emph {et~al.}(2017)\citenamefont
  {Simpson}, \citenamefont {Morrisroe}, \citenamefont {McCoey}, \citenamefont
  {Lombard}, \citenamefont {Mendis}, \citenamefont {Treussart}, \citenamefont
  {Hall}, \citenamefont {Petrou},\ and\ \citenamefont
  {Hollenberg}}]{simpson2017}%
  \BibitemOpen
  \bibfield  {author} {\bibinfo {author} {\bibfnamefont {D.~A.}\ \bibnamefont
  {Simpson}}, \bibinfo {author} {\bibfnamefont {E.}~\bibnamefont {Morrisroe}},
  \bibinfo {author} {\bibfnamefont {J.~M.}\ \bibnamefont {McCoey}}, \bibinfo
  {author} {\bibfnamefont {A.~H.}\ \bibnamefont {Lombard}}, \bibinfo {author}
  {\bibfnamefont {D.~C.}\ \bibnamefont {Mendis}}, \bibinfo {author}
  {\bibfnamefont {F.}~\bibnamefont {Treussart}}, \bibinfo {author}
  {\bibfnamefont {L.~T.}\ \bibnamefont {Hall}}, \bibinfo {author}
  {\bibfnamefont {S.}~\bibnamefont {Petrou}},\ and\ \bibinfo {author}
  {\bibfnamefont {L.~C.~L.}\ \bibnamefont {Hollenberg}},\ }\href
  {https://doi.org/10.1021/acsnano.7b04850} {\bibfield  {journal} {\bibinfo
  {journal} {ACS Nano}\ }\textbf {\bibinfo {volume} {11}},\ \bibinfo {pages}
  {12077} (\bibinfo {year} {2017})}\BibitemShut {NoStop}%
\bibitem [{\citenamefont {Mamin}\ \emph {et~al.}(2013)\citenamefont {Mamin},
  \citenamefont {Kim}, \citenamefont {Sherwood}, \citenamefont {Rettner},
  \citenamefont {Ohno}, \citenamefont {Awschalom},\ and\ \citenamefont
  {Rugar}}]{mamin2013}%
  \BibitemOpen
  \bibfield  {author} {\bibinfo {author} {\bibfnamefont {H.~J.}\ \bibnamefont
  {Mamin}}, \bibinfo {author} {\bibfnamefont {M.}~\bibnamefont {Kim}}, \bibinfo
  {author} {\bibfnamefont {M.~H.}\ \bibnamefont {Sherwood}}, \bibinfo {author}
  {\bibfnamefont {C.~T.}\ \bibnamefont {Rettner}}, \bibinfo {author}
  {\bibfnamefont {K.}~\bibnamefont {Ohno}}, \bibinfo {author} {\bibfnamefont
  {D.~D.}\ \bibnamefont {Awschalom}},\ and\ \bibinfo {author} {\bibfnamefont
  {D.}~\bibnamefont {Rugar}},\ }\href {https://doi.org/10.1126/science.1231540}
  {\bibfield  {journal} {\bibinfo  {journal} {Science}\ }\textbf {\bibinfo
  {volume} {339}},\ \bibinfo {pages} {557} (\bibinfo {year}
  {2013})}\BibitemShut {NoStop}%
\bibitem [{\citenamefont {Staudacher}\ \emph {et~al.}(2013)\citenamefont
  {Staudacher}, \citenamefont {Shi}, \citenamefont {Pezzagna}, \citenamefont
  {Meijer}, \citenamefont {Du}, \citenamefont {Meriles}, \citenamefont
  {Reinhard},\ and\ \citenamefont {Wrachtrup}}]{staudacher2013}%
  \BibitemOpen
  \bibfield  {author} {\bibinfo {author} {\bibfnamefont {T.}~\bibnamefont
  {Staudacher}}, \bibinfo {author} {\bibfnamefont {F.}~\bibnamefont {Shi}},
  \bibinfo {author} {\bibfnamefont {S.}~\bibnamefont {Pezzagna}}, \bibinfo
  {author} {\bibfnamefont {J.}~\bibnamefont {Meijer}}, \bibinfo {author}
  {\bibfnamefont {J.}~\bibnamefont {Du}}, \bibinfo {author} {\bibfnamefont
  {C.~A.}\ \bibnamefont {Meriles}}, \bibinfo {author} {\bibfnamefont
  {F.}~\bibnamefont {Reinhard}},\ and\ \bibinfo {author} {\bibfnamefont
  {J.}~\bibnamefont {Wrachtrup}},\ }\href
  {https://doi.org/10.1126/science.1231675} {\bibfield  {journal} {\bibinfo
  {journal} {Science}\ }\textbf {\bibinfo {volume} {339}},\ \bibinfo {pages}
  {561} (\bibinfo {year} {2013})}\BibitemShut {NoStop}%
\bibitem [{\citenamefont {Zhang}\ \emph {et~al.}(2018)\citenamefont {Zhang},
  \citenamefont {Liu}, \citenamefont {Leong}, \citenamefont {Liu},
  \citenamefont {Kwok}, \citenamefont {Ngai}, \citenamefont {Liu},\ and\
  \citenamefont {Li}}]{zhang2018}%
  \BibitemOpen
  \bibfield  {author} {\bibinfo {author} {\bibfnamefont {T.}~\bibnamefont
  {Zhang}}, \bibinfo {author} {\bibfnamefont {G.-Q.}\ \bibnamefont {Liu}},
  \bibinfo {author} {\bibfnamefont {W.-H.}\ \bibnamefont {Leong}}, \bibinfo
  {author} {\bibfnamefont {C.-F.}\ \bibnamefont {Liu}}, \bibinfo {author}
  {\bibfnamefont {M.-H.}\ \bibnamefont {Kwok}}, \bibinfo {author}
  {\bibfnamefont {T.}~\bibnamefont {Ngai}}, \bibinfo {author} {\bibfnamefont
  {R.-B.}\ \bibnamefont {Liu}},\ and\ \bibinfo {author} {\bibfnamefont
  {Q.}~\bibnamefont {Li}},\ }\href {https://doi.org/10.1038/s41467-018-05673-9}
  {\bibfield  {journal} {\bibinfo  {journal} {Nature Commun.}\ }\textbf
  {\bibinfo {volume} {9}},\ \bibinfo {pages} {3188} (\bibinfo {year}
  {2018})}\BibitemShut {NoStop}%
\bibitem [{\citenamefont {Rendler}\ \emph {et~al.}(2017)\citenamefont
  {Rendler}, \citenamefont {Neburkova}, \citenamefont {Zemek}, \citenamefont
  {Kotek}, \citenamefont {Zappe}, \citenamefont {Chu}, \citenamefont {Cigler},\
  and\ \citenamefont {Wrachtrup}}]{rendler2017}%
  \BibitemOpen
  \bibfield  {author} {\bibinfo {author} {\bibfnamefont {T.}~\bibnamefont
  {Rendler}}, \bibinfo {author} {\bibfnamefont {J.}~\bibnamefont {Neburkova}},
  \bibinfo {author} {\bibfnamefont {O.}~\bibnamefont {Zemek}}, \bibinfo
  {author} {\bibfnamefont {J.}~\bibnamefont {Kotek}}, \bibinfo {author}
  {\bibfnamefont {A.}~\bibnamefont {Zappe}}, \bibinfo {author} {\bibfnamefont
  {Z.}~\bibnamefont {Chu}}, \bibinfo {author} {\bibfnamefont {P.}~\bibnamefont
  {Cigler}},\ and\ \bibinfo {author} {\bibfnamefont {J.}~\bibnamefont
  {Wrachtrup}},\ }\href {https://doi.org/10.1038/ncomms14701} {\bibfield
  {journal} {\bibinfo  {journal} {Nature Commun.}\ }\textbf {\bibinfo {volume}
  {8}},\ \bibinfo {pages} {14701} (\bibinfo {year} {2017})}\BibitemShut
  {NoStop}%
\bibitem [{\citenamefont {Wang}\ \emph {et~al.}(2018)\citenamefont {Wang},
  \citenamefont {Liu}, \citenamefont {Leong}, \citenamefont {Zeng},
  \citenamefont {Feng}, \citenamefont {Li}, \citenamefont {Dolde},
  \citenamefont {Fedder}, \citenamefont {Wrachtrup}, \citenamefont {Cui},
  \citenamefont {Yang}, \citenamefont {Li},\ and\ \citenamefont
  {Liu}}]{wang2018}%
  \BibitemOpen
  \bibfield  {author} {\bibinfo {author} {\bibfnamefont {N.}~\bibnamefont
  {Wang}}, \bibinfo {author} {\bibfnamefont {G.-Q.}\ \bibnamefont {Liu}},
  \bibinfo {author} {\bibfnamefont {W.-H.}\ \bibnamefont {Leong}}, \bibinfo
  {author} {\bibfnamefont {H.}~\bibnamefont {Zeng}}, \bibinfo {author}
  {\bibfnamefont {X.}~\bibnamefont {Feng}}, \bibinfo {author} {\bibfnamefont
  {S.-H.}\ \bibnamefont {Li}}, \bibinfo {author} {\bibfnamefont
  {F.}~\bibnamefont {Dolde}}, \bibinfo {author} {\bibfnamefont
  {H.}~\bibnamefont {Fedder}}, \bibinfo {author} {\bibfnamefont
  {J.}~\bibnamefont {Wrachtrup}}, \bibinfo {author} {\bibfnamefont {X.-D.}\
  \bibnamefont {Cui}}, \bibinfo {author} {\bibfnamefont {S.}~\bibnamefont
  {Yang}}, \bibinfo {author} {\bibfnamefont {Q.}~\bibnamefont {Li}},\ and\
  \bibinfo {author} {\bibfnamefont {R.-B.}\ \bibnamefont {Liu}},\ }\href
  {https://doi.org/10.1103/PhysRevX.8.011042} {\bibfield  {journal} {\bibinfo
  {journal} {Phys. Rev. X}\ }\textbf {\bibinfo {volume} {8}},\ \bibinfo {pages}
  {011042} (\bibinfo {year} {2018})}\BibitemShut {NoStop}%
\bibitem [{\citenamefont {Liu}\ \emph {et~al.}(2021)\citenamefont {Liu},
  \citenamefont {Leong}, \citenamefont {Xia}, \citenamefont {Feng},
  \citenamefont {Finkler}, \citenamefont {Denisenko}, \citenamefont
  {Wrachtrup}, \citenamefont {Li},\ and\ \citenamefont {Liu}}]{liu2021}%
  \BibitemOpen
  \bibfield  {author} {\bibinfo {author} {\bibfnamefont {C.-F.}\ \bibnamefont
  {Liu}}, \bibinfo {author} {\bibfnamefont {W.-H.}\ \bibnamefont {Leong}},
  \bibinfo {author} {\bibfnamefont {K.}~\bibnamefont {Xia}}, \bibinfo {author}
  {\bibfnamefont {X.}~\bibnamefont {Feng}}, \bibinfo {author} {\bibfnamefont
  {A.}~\bibnamefont {Finkler}}, \bibinfo {author} {\bibfnamefont
  {A.}~\bibnamefont {Denisenko}}, \bibinfo {author} {\bibfnamefont
  {J.}~\bibnamefont {Wrachtrup}}, \bibinfo {author} {\bibfnamefont
  {Q.}~\bibnamefont {Li}},\ and\ \bibinfo {author} {\bibfnamefont {R.-B.}\
  \bibnamefont {Liu}},\ }\href {https://doi.org/10.1093/nsr/nwaa194} {\bibfield
   {journal} {\bibinfo  {journal} {Natl. Sci. Rev.}\ }\textbf {\bibinfo
  {volume} {8}},\ \bibinfo {pages} {nwaa194} (\bibinfo {year}
  {2021})}\BibitemShut {NoStop}%
\bibitem [{\citenamefont {Taylor}\ \emph {et~al.}(2008)\citenamefont {Taylor},
  \citenamefont {Cappellaro}, \citenamefont {Childress}, \citenamefont {Jiang},
  \citenamefont {Budker}, \citenamefont {Hemmer}, \citenamefont {Yacoby},
  \citenamefont {Walsworth},\ and\ \citenamefont {Lukin}}]{taylor2008}%
  \BibitemOpen
  \bibfield  {author} {\bibinfo {author} {\bibfnamefont {J.~M.}\ \bibnamefont
  {Taylor}}, \bibinfo {author} {\bibfnamefont {P.}~\bibnamefont {Cappellaro}},
  \bibinfo {author} {\bibfnamefont {L.}~\bibnamefont {Childress}}, \bibinfo
  {author} {\bibfnamefont {L.}~\bibnamefont {Jiang}}, \bibinfo {author}
  {\bibfnamefont {D.}~\bibnamefont {Budker}}, \bibinfo {author} {\bibfnamefont
  {P.~R.}\ \bibnamefont {Hemmer}}, \bibinfo {author} {\bibfnamefont
  {A.}~\bibnamefont {Yacoby}}, \bibinfo {author} {\bibfnamefont
  {R.}~\bibnamefont {Walsworth}},\ and\ \bibinfo {author} {\bibfnamefont
  {M.~D.}\ \bibnamefont {Lukin}},\ }\href {https://doi.org/10.1038/nphys1075}
  {\bibfield  {journal} {\bibinfo  {journal} {Nature Phys.}\ }\textbf {\bibinfo
  {volume} {4}},\ \bibinfo {pages} {810} (\bibinfo {year} {2008})}\BibitemShut
  {NoStop}%
\bibitem [{\citenamefont {Yu}\ \emph {et~al.}(2005)\citenamefont {Yu},
  \citenamefont {Kang}, \citenamefont {Chang}, \citenamefont {Chen},\ and\
  \citenamefont {Yu}}]{yu2005}%
  \BibitemOpen
  \bibfield  {author} {\bibinfo {author} {\bibfnamefont {S.-J.}\ \bibnamefont
  {Yu}}, \bibinfo {author} {\bibfnamefont {M.-W.}\ \bibnamefont {Kang}},
  \bibinfo {author} {\bibfnamefont {H.-C.}\ \bibnamefont {Chang}}, \bibinfo
  {author} {\bibfnamefont {K.-M.}\ \bibnamefont {Chen}},\ and\ \bibinfo
  {author} {\bibfnamefont {Y.-C.}\ \bibnamefont {Yu}},\ }\href
  {https://doi.org/10.1021/ja0567081} {\bibfield  {journal} {\bibinfo
  {journal} {J. Am. Chem. Soc.}\ }\textbf {\bibinfo {volume} {127}},\ \bibinfo
  {pages} {17604} (\bibinfo {year} {2005})}\BibitemShut {NoStop}%
\bibitem [{\citenamefont {Degen}\ \emph {et~al.}(2017)\citenamefont {Degen},
  \citenamefont {Reinhard},\ and\ \citenamefont {Cappellaro}}]{degen2017}%
  \BibitemOpen
  \bibfield  {author} {\bibinfo {author} {\bibfnamefont {C.~L.}\ \bibnamefont
  {Degen}}, \bibinfo {author} {\bibfnamefont {F.}~\bibnamefont {Reinhard}},\
  and\ \bibinfo {author} {\bibfnamefont {P.}~\bibnamefont {Cappellaro}},\
  }\href {https://doi.org/10.1103/RevModPhys.89.035002} {\bibfield  {journal}
  {\bibinfo  {journal} {Rev. Mod. Phys.}\ }\textbf {\bibinfo {volume} {89}},\
  \bibinfo {pages} {035002} (\bibinfo {year} {2017})}\BibitemShut {NoStop}%
\bibitem [{\citenamefont {Awschalom}\ \emph {et~al.}(2018)\citenamefont
  {Awschalom}, \citenamefont {Hanson}, \citenamefont {Wrachtrup},\ and\
  \citenamefont {Zhou}}]{awschalom2018}%
  \BibitemOpen
  \bibfield  {author} {\bibinfo {author} {\bibfnamefont {D.~D.}\ \bibnamefont
  {Awschalom}}, \bibinfo {author} {\bibfnamefont {R.}~\bibnamefont {Hanson}},
  \bibinfo {author} {\bibfnamefont {J.}~\bibnamefont {Wrachtrup}},\ and\
  \bibinfo {author} {\bibfnamefont {B.~B.}\ \bibnamefont {Zhou}},\ }\href
  {https://doi.org/10.1038/s41566-018-0232-2} {\bibfield  {journal} {\bibinfo
  {journal} {Nature Photon.}\ }\textbf {\bibinfo {volume} {12}},\ \bibinfo
  {pages} {516} (\bibinfo {year} {2018})}\BibitemShut {NoStop}%
\bibitem [{\citenamefont {Bar-Gill}\ \emph {et~al.}(2013)\citenamefont
  {Bar-Gill}, \citenamefont {Pham}, \citenamefont {Jarmola}, \citenamefont
  {Budker},\ and\ \citenamefont {Walsworth}}]{bargill2013}%
  \BibitemOpen
  \bibfield  {author} {\bibinfo {author} {\bibfnamefont {N.}~\bibnamefont
  {Bar-Gill}}, \bibinfo {author} {\bibfnamefont {L.~M.}\ \bibnamefont {Pham}},
  \bibinfo {author} {\bibfnamefont {A.}~\bibnamefont {Jarmola}}, \bibinfo
  {author} {\bibfnamefont {D.}~\bibnamefont {Budker}},\ and\ \bibinfo {author}
  {\bibfnamefont {R.~L.}\ \bibnamefont {Walsworth}},\ }\href
  {https://doi.org/10.1038/ncomms2771} {\bibfield  {journal} {\bibinfo
  {journal} {Nature Commun.}\ }\textbf {\bibinfo {volume} {4}},\ \bibinfo
  {pages} {1743} (\bibinfo {year} {2013})}\BibitemShut {NoStop}%
\bibitem [{\citenamefont {Poggiali}\ \emph {et~al.}(2018)\citenamefont
  {Poggiali}, \citenamefont {Cappellaro},\ and\ \citenamefont
  {Fabbri}}]{poggiali2018}%
  \BibitemOpen
  \bibfield  {author} {\bibinfo {author} {\bibfnamefont {F.}~\bibnamefont
  {Poggiali}}, \bibinfo {author} {\bibfnamefont {P.}~\bibnamefont
  {Cappellaro}},\ and\ \bibinfo {author} {\bibfnamefont {N.}~\bibnamefont
  {Fabbri}},\ }\href {https://doi.org/10.1103/PhysRevX.8.021059} {\bibfield
  {journal} {\bibinfo  {journal} {Physical Review X}\ }\textbf {\bibinfo
  {volume} {8}},\ \bibinfo {pages} {021059} (\bibinfo {year}
  {2018})}\BibitemShut {NoStop}%
\bibitem [{\citenamefont {Oshnik}\ \emph {et~al.}(2022)\citenamefont {Oshnik},
  \citenamefont {Rembold}, \citenamefont {Calarco}, \citenamefont {Montangero},
  \citenamefont {Neu},\ and\ \citenamefont {M{\"u}ller}}]{oshnik2022}%
  \BibitemOpen
  \bibfield  {author} {\bibinfo {author} {\bibfnamefont {N.}~\bibnamefont
  {Oshnik}}, \bibinfo {author} {\bibfnamefont {P.}~\bibnamefont {Rembold}},
  \bibinfo {author} {\bibfnamefont {T.}~\bibnamefont {Calarco}}, \bibinfo
  {author} {\bibfnamefont {S.}~\bibnamefont {Montangero}}, \bibinfo {author}
  {\bibfnamefont {E.}~\bibnamefont {Neu}},\ and\ \bibinfo {author}
  {\bibfnamefont {M.~M.}\ \bibnamefont {M{\"u}ller}},\ }\href
  {https://doi.org/10.1103/PhysRevA.106.013107} {\bibfield  {journal} {\bibinfo
   {journal} {Physical Review A}\ }\textbf {\bibinfo {volume} {106}},\ \bibinfo
  {pages} {013107} (\bibinfo {year} {2022})}\BibitemShut {NoStop}%
\bibitem [{\citenamefont {Hecht}\ \emph {et~al.}(2025)\citenamefont {Hecht},
  \citenamefont {Saurav}, \citenamefont {Vlachos}, \citenamefont {Lidar},\ and\
  \citenamefont {Levenson-Falk}}]{hecht2025}%
  \BibitemOpen
  \bibfield  {author} {\bibinfo {author} {\bibfnamefont {M.~O.}\ \bibnamefont
  {Hecht}}, \bibinfo {author} {\bibfnamefont {K.}~\bibnamefont {Saurav}},
  \bibinfo {author} {\bibfnamefont {E.}~\bibnamefont {Vlachos}}, \bibinfo
  {author} {\bibfnamefont {D.~A.}\ \bibnamefont {Lidar}},\ and\ \bibinfo
  {author} {\bibfnamefont {E.~M.}\ \bibnamefont {Levenson-Falk}},\ }\href
  {https://doi.org/10.1038/s41467-025-58947-4} {\bibfield  {journal} {\bibinfo
  {journal} {Nature Communications}\ }\textbf {\bibinfo {volume} {16}},\
  \bibinfo {pages} {3754} (\bibinfo {year} {2025})}\BibitemShut {NoStop}%
\bibitem [{\citenamefont {Low}\ and\ \citenamefont {Chuang}(2017)}]{low2017}%
  \BibitemOpen
  \bibfield  {author} {\bibinfo {author} {\bibfnamefont {G.~H.}\ \bibnamefont
  {Low}}\ and\ \bibinfo {author} {\bibfnamefont {I.~L.}\ \bibnamefont
  {Chuang}},\ }\href {https://doi.org/10.1103/PhysRevLett.118.010501}
  {\bibfield  {journal} {\bibinfo  {journal} {Phys. Rev. Lett.}\ }\textbf
  {\bibinfo {volume} {118}},\ \bibinfo {pages} {010501} (\bibinfo {year}
  {2017})}\BibitemShut {NoStop}%
\bibitem [{\citenamefont {Rembold}\ \emph {et~al.}(2020)\citenamefont
  {Rembold}, \citenamefont {Oshnik}, \citenamefont {M{\"u}ller}, \citenamefont
  {Montangero}, \citenamefont {Calarco},\ and\ \citenamefont
  {Neu}}]{rembold2020}%
  \BibitemOpen
  \bibfield  {author} {\bibinfo {author} {\bibfnamefont {P.}~\bibnamefont
  {Rembold}}, \bibinfo {author} {\bibfnamefont {N.}~\bibnamefont {Oshnik}},
  \bibinfo {author} {\bibfnamefont {M.~M.}\ \bibnamefont {M{\"u}ller}},
  \bibinfo {author} {\bibfnamefont {S.}~\bibnamefont {Montangero}}, \bibinfo
  {author} {\bibfnamefont {T.}~\bibnamefont {Calarco}},\ and\ \bibinfo {author}
  {\bibfnamefont {E.}~\bibnamefont {Neu}},\ }\href
  {https://doi.org/10.1116/5.0006785} {\bibfield  {journal} {\bibinfo
  {journal} {AVS Quantum Science}\ }\textbf {\bibinfo {volume} {2}},\ \bibinfo
  {pages} {024701} (\bibinfo {year} {2020})}\BibitemShut {NoStop}%
\bibitem [{\citenamefont {Cai}\ \emph {et~al.}(2014)\citenamefont {Cai},
  \citenamefont {Jelezko},\ and\ \citenamefont {Plenio}}]{cai2014}%
  \BibitemOpen
  \bibfield  {author} {\bibinfo {author} {\bibfnamefont {J.}~\bibnamefont
  {Cai}}, \bibinfo {author} {\bibfnamefont {F.}~\bibnamefont {Jelezko}},\ and\
  \bibinfo {author} {\bibfnamefont {M.~B.}\ \bibnamefont {Plenio}},\ }\href
  {https://doi.org/10.1038/ncomms5065} {\bibfield  {journal} {\bibinfo
  {journal} {Nature Commun.}\ }\textbf {\bibinfo {volume} {5}},\ \bibinfo
  {pages} {4065} (\bibinfo {year} {2014})}\BibitemShut {NoStop}%
\bibitem [{\citenamefont {Doherty}\ \emph {et~al.}(2013)\citenamefont
  {Doherty}, \citenamefont {Manson}, \citenamefont {Delaney}, \citenamefont
  {Jelezko}, \citenamefont {Wrachtrup},\ and\ \citenamefont
  {Hollenberg}}]{doherty2013}%
  \BibitemOpen
  \bibfield  {author} {\bibinfo {author} {\bibfnamefont {M.~W.}\ \bibnamefont
  {Doherty}}, \bibinfo {author} {\bibfnamefont {N.~B.}\ \bibnamefont {Manson}},
  \bibinfo {author} {\bibfnamefont {P.}~\bibnamefont {Delaney}}, \bibinfo
  {author} {\bibfnamefont {F.}~\bibnamefont {Jelezko}}, \bibinfo {author}
  {\bibfnamefont {J.}~\bibnamefont {Wrachtrup}},\ and\ \bibinfo {author}
  {\bibfnamefont {L.~C.~L.}\ \bibnamefont {Hollenberg}},\ }\href
  {https://doi.org/10.1016/j.physrep.2013.02.001} {\bibfield  {journal}
  {\bibinfo  {journal} {Phys. Rep.}\ }\textbf {\bibinfo {volume} {528}},\
  \bibinfo {pages} {1} (\bibinfo {year} {2013})}\BibitemShut {NoStop}%
\bibitem [{\citenamefont {Rondin}\ \emph {et~al.}(2014)\citenamefont {Rondin},
  \citenamefont {Tetienne}, \citenamefont {Hingant}, \citenamefont {Roch},
  \citenamefont {Maletinsky},\ and\ \citenamefont {Jacques}}]{rondin2014}%
  \BibitemOpen
  \bibfield  {author} {\bibinfo {author} {\bibfnamefont {L.}~\bibnamefont
  {Rondin}}, \bibinfo {author} {\bibfnamefont {J.-P.}\ \bibnamefont
  {Tetienne}}, \bibinfo {author} {\bibfnamefont {T.}~\bibnamefont {Hingant}},
  \bibinfo {author} {\bibfnamefont {J.-F.}\ \bibnamefont {Roch}}, \bibinfo
  {author} {\bibfnamefont {P.}~\bibnamefont {Maletinsky}},\ and\ \bibinfo
  {author} {\bibfnamefont {V.}~\bibnamefont {Jacques}},\ }\href
  {https://doi.org/10.1088/0034-4885/77/5/056503} {\bibfield  {journal}
  {\bibinfo  {journal} {Rep. Prog. Phys.}\ }\textbf {\bibinfo {volume} {77}},\
  \bibinfo {pages} {056503} (\bibinfo {year} {2014})}\BibitemShut {NoStop}%
\bibitem [{\citenamefont {Schild}(1992)}]{schild1992}%
  \BibitemOpen
  \bibfield  {author} {\bibinfo {author} {\bibfnamefont {H.~G.}\ \bibnamefont
  {Schild}},\ }\href {https://doi.org/10.1016/0079-6700(92)90023-R} {\bibfield
  {journal} {\bibinfo  {journal} {Prog. Polym. Sci.}\ }\textbf {\bibinfo
  {volume} {17}},\ \bibinfo {pages} {163} (\bibinfo {year} {1992})}\BibitemShut
  {NoStop}%
\bibitem [{\citenamefont {Heskins}\ and\ \citenamefont
  {Guillet}(1968)}]{heskins1968}%
  \BibitemOpen
  \bibfield  {author} {\bibinfo {author} {\bibfnamefont {M.}~\bibnamefont
  {Heskins}}\ and\ \bibinfo {author} {\bibfnamefont {J.~E.}\ \bibnamefont
  {Guillet}},\ }\href {https://doi.org/10.1080/10601326808051910} {\bibfield
  {journal} {\bibinfo  {journal} {J. Macromol. Sci. Chem.}\ }\textbf {\bibinfo
  {volume} {A2}},\ \bibinfo {pages} {1441} (\bibinfo {year}
  {1968})}\BibitemShut {NoStop}%
\bibitem [{\citenamefont {Levitt}(1986)}]{levitt1986}%
  \BibitemOpen
  \bibfield  {author} {\bibinfo {author} {\bibfnamefont {M.~H.}\ \bibnamefont
  {Levitt}},\ }\href {https://doi.org/10.1016/0079-6565(86)80005-X} {\bibfield
  {journal} {\bibinfo  {journal} {Progress in Nuclear Magnetic Resonance
  Spectroscopy}\ }\textbf {\bibinfo {volume} {18}},\ \bibinfo {pages} {61}
  (\bibinfo {year} {1986})}\BibitemShut {NoStop}%
\bibitem [{\citenamefont {Wimperis}(1994)}]{wimperis1994}%
  \BibitemOpen
  \bibfield  {author} {\bibinfo {author} {\bibfnamefont {S.}~\bibnamefont
  {Wimperis}},\ }\href {https://doi.org/10.1006/jmra.1994.1159} {\bibfield
  {journal} {\bibinfo  {journal} {Journal of Magnetic Resonance, Series A}\
  }\textbf {\bibinfo {volume} {109}},\ \bibinfo {pages} {221} (\bibinfo {year}
  {1994})}\BibitemShut {NoStop}%
\bibitem [{\citenamefont {Yudin}\ \emph {et~al.}(2010)\citenamefont {Yudin},
  \citenamefont {Taichenachev}, \citenamefont {Oates}, \citenamefont {Barber},
  \citenamefont {Lemke}, \citenamefont {Ludlow}, \citenamefont {Sterr},
  \citenamefont {Lisdat},\ and\ \citenamefont {Riehle}}]{yudin2010}%
  \BibitemOpen
  \bibfield  {author} {\bibinfo {author} {\bibfnamefont {V.~I.}\ \bibnamefont
  {Yudin}}, \bibinfo {author} {\bibfnamefont {A.~V.}\ \bibnamefont
  {Taichenachev}}, \bibinfo {author} {\bibfnamefont {C.~W.}\ \bibnamefont
  {Oates}}, \bibinfo {author} {\bibfnamefont {Z.~W.}\ \bibnamefont {Barber}},
  \bibinfo {author} {\bibfnamefont {N.~D.}\ \bibnamefont {Lemke}}, \bibinfo
  {author} {\bibfnamefont {A.~D.}\ \bibnamefont {Ludlow}}, \bibinfo {author}
  {\bibfnamefont {U.}~\bibnamefont {Sterr}}, \bibinfo {author} {\bibfnamefont
  {C.}~\bibnamefont {Lisdat}},\ and\ \bibinfo {author} {\bibfnamefont
  {F.}~\bibnamefont {Riehle}},\ }\href
  {https://doi.org/10.1103/PhysRevA.82.011804} {\bibfield  {journal} {\bibinfo
  {journal} {Physical Review A}\ }\textbf {\bibinfo {volume} {82}},\ \bibinfo
  {pages} {011804(R)} (\bibinfo {year} {2010})}\BibitemShut {NoStop}%
\bibitem [{\citenamefont {Zanon-Willette}\ \emph {et~al.}(2016)\citenamefont
  {Zanon-Willette}, \citenamefont {Minissale}, \citenamefont {Yudin},\ and\
  \citenamefont {Taichenachev}}]{zanonwillette2016}%
  \BibitemOpen
  \bibfield  {author} {\bibinfo {author} {\bibfnamefont {T.}~\bibnamefont
  {Zanon-Willette}}, \bibinfo {author} {\bibfnamefont {M.}~\bibnamefont
  {Minissale}}, \bibinfo {author} {\bibfnamefont {V.~I.}\ \bibnamefont
  {Yudin}},\ and\ \bibinfo {author} {\bibfnamefont {A.~V.}\ \bibnamefont
  {Taichenachev}},\ }\href {https://doi.org/10.1088/1742-6596/723/1/012057}
  {\bibfield  {journal} {\bibinfo  {journal} {Journal of Physics: Conference
  Series}\ }\textbf {\bibinfo {volume} {723}},\ \bibinfo {pages} {012057}
  (\bibinfo {year} {2016})}\BibitemShut {NoStop}%
\bibitem [{\citenamefont {Zanon-Willette}\ \emph {et~al.}(2018)\citenamefont
  {Zanon-Willette}, \citenamefont {Lefevre}, \citenamefont {Metzdorff},
  \citenamefont {Sillitoe}, \citenamefont {Almonacil}, \citenamefont
  {Minissale}, \citenamefont {de~Clercq}, \citenamefont {Taichenachev},
  \citenamefont {Yudin},\ and\ \citenamefont {Arimondo}}]{zanonwillette2018}%
  \BibitemOpen
  \bibfield  {author} {\bibinfo {author} {\bibfnamefont {T.}~\bibnamefont
  {Zanon-Willette}}, \bibinfo {author} {\bibfnamefont {R.}~\bibnamefont
  {Lefevre}}, \bibinfo {author} {\bibfnamefont {R.}~\bibnamefont {Metzdorff}},
  \bibinfo {author} {\bibfnamefont {N.}~\bibnamefont {Sillitoe}}, \bibinfo
  {author} {\bibfnamefont {S.}~\bibnamefont {Almonacil}}, \bibinfo {author}
  {\bibfnamefont {M.}~\bibnamefont {Minissale}}, \bibinfo {author}
  {\bibfnamefont {E.}~\bibnamefont {de~Clercq}}, \bibinfo {author}
  {\bibfnamefont {A.~V.}\ \bibnamefont {Taichenachev}}, \bibinfo {author}
  {\bibfnamefont {V.~I.}\ \bibnamefont {Yudin}},\ and\ \bibinfo {author}
  {\bibfnamefont {E.}~\bibnamefont {Arimondo}},\ }\href
  {https://doi.org/10.1088/1361-6633/aac9e9} {\bibfield  {journal} {\bibinfo
  {journal} {Reports on Progress in Physics}\ }\textbf {\bibinfo {volume}
  {81}},\ \bibinfo {pages} {094401} (\bibinfo {year} {2018})}\BibitemShut
  {NoStop}%
\bibitem [{\citenamefont {Low}\ \emph {et~al.}(2016)\citenamefont {Low},
  \citenamefont {Yoder},\ and\ \citenamefont {Chuang}}]{lowyoderchuang2016}%
  \BibitemOpen
  \bibfield  {author} {\bibinfo {author} {\bibfnamefont {G.~H.}\ \bibnamefont
  {Low}}, \bibinfo {author} {\bibfnamefont {T.~J.}\ \bibnamefont {Yoder}},\
  and\ \bibinfo {author} {\bibfnamefont {I.~L.}\ \bibnamefont {Chuang}},\
  }\href {https://doi.org/10.1103/PhysRevX.6.041067} {\bibfield  {journal}
  {\bibinfo  {journal} {Physical Review X}\ }\textbf {\bibinfo {volume} {6}},\
  \bibinfo {pages} {041067} (\bibinfo {year} {2016})}\BibitemShut {NoStop}%
\bibitem [{\citenamefont {Aiello}\ \emph {et~al.}(2013)\citenamefont {Aiello},
  \citenamefont {Hirose},\ and\ \citenamefont {Cappellaro}}]{aiello2013}%
  \BibitemOpen
  \bibfield  {author} {\bibinfo {author} {\bibfnamefont {C.~D.}\ \bibnamefont
  {Aiello}}, \bibinfo {author} {\bibfnamefont {M.}~\bibnamefont {Hirose}},\
  and\ \bibinfo {author} {\bibfnamefont {P.}~\bibnamefont {Cappellaro}},\
  }\href {https://doi.org/10.1038/ncomms2375} {\bibfield  {journal} {\bibinfo
  {journal} {Nature Communications}\ }\textbf {\bibinfo {volume} {4}},\
  \bibinfo {pages} {1419} (\bibinfo {year} {2013})}\BibitemShut {NoStop}%
\bibitem [{\citenamefont {Giovannetti}\ \emph {et~al.}(2011)\citenamefont
  {Giovannetti}, \citenamefont {Lloyd},\ and\ \citenamefont
  {Maccone}}]{giovannetti2011}%
  \BibitemOpen
  \bibfield  {author} {\bibinfo {author} {\bibfnamefont {V.}~\bibnamefont
  {Giovannetti}}, \bibinfo {author} {\bibfnamefont {S.}~\bibnamefont {Lloyd}},\
  and\ \bibinfo {author} {\bibfnamefont {L.}~\bibnamefont {Maccone}},\ }\href
  {https://doi.org/10.1038/nphoton.2011.35} {\bibfield  {journal} {\bibinfo
  {journal} {Nature Photon.}\ }\textbf {\bibinfo {volume} {5}},\ \bibinfo
  {pages} {222} (\bibinfo {year} {2011})}\BibitemShut {NoStop}%
\bibitem [{\citenamefont {Zhou}\ \emph {et~al.}(2018)\citenamefont {Zhou},
  \citenamefont {Zhang}, \citenamefont {Preskill},\ and\ \citenamefont
  {Jiang}}]{zhou2018}%
  \BibitemOpen
  \bibfield  {author} {\bibinfo {author} {\bibfnamefont {S.}~\bibnamefont
  {Zhou}}, \bibinfo {author} {\bibfnamefont {M.}~\bibnamefont {Zhang}},
  \bibinfo {author} {\bibfnamefont {J.}~\bibnamefont {Preskill}},\ and\
  \bibinfo {author} {\bibfnamefont {L.}~\bibnamefont {Jiang}},\ }\href
  {https://doi.org/10.1038/s41467-017-02510-3} {\bibfield  {journal} {\bibinfo
  {journal} {Nature Commun.}\ }\textbf {\bibinfo {volume} {9}},\ \bibinfo
  {pages} {78} (\bibinfo {year} {2018})}\BibitemShut {NoStop}%
\bibitem [{\citenamefont {Sekatski}\ \emph {et~al.}(2017)\citenamefont
  {Sekatski}, \citenamefont {Skotiniotis}, \citenamefont {Ko{\l}ody{\'n}ski},\
  and\ \citenamefont {D{\"u}r}}]{sekatski2017}%
  \BibitemOpen
  \bibfield  {author} {\bibinfo {author} {\bibfnamefont {P.}~\bibnamefont
  {Sekatski}}, \bibinfo {author} {\bibfnamefont {M.}~\bibnamefont
  {Skotiniotis}}, \bibinfo {author} {\bibfnamefont {J.}~\bibnamefont
  {Ko{\l}ody{\'n}ski}},\ and\ \bibinfo {author} {\bibfnamefont
  {W.}~\bibnamefont {D{\"u}r}},\ }\href
  {https://doi.org/10.22331/q-2017-09-06-27} {\bibfield  {journal} {\bibinfo
  {journal} {Quantum}\ }\textbf {\bibinfo {volume} {1}},\ \bibinfo {pages} {27}
  (\bibinfo {year} {2017})}\BibitemShut {NoStop}%
\bibitem [{\citenamefont {Jiang}\ \emph {et~al.}(2009)\citenamefont {Jiang},
  \citenamefont {Hodges}, \citenamefont {Maze}, \citenamefont {Maurer},
  \citenamefont {Taylor}, \citenamefont {Cory}, \citenamefont {Hemmer},
  \citenamefont {Walsworth}, \citenamefont {Yacoby}, \citenamefont {Zibrov},\
  and\ \citenamefont {Lukin}}]{jiang2009}%
  \BibitemOpen
  \bibfield  {author} {\bibinfo {author} {\bibfnamefont {L.}~\bibnamefont
  {Jiang}}, \bibinfo {author} {\bibfnamefont {J.~S.}\ \bibnamefont {Hodges}},
  \bibinfo {author} {\bibfnamefont {J.~R.}\ \bibnamefont {Maze}}, \bibinfo
  {author} {\bibfnamefont {P.}~\bibnamefont {Maurer}}, \bibinfo {author}
  {\bibfnamefont {J.~M.}\ \bibnamefont {Taylor}}, \bibinfo {author}
  {\bibfnamefont {D.~G.}\ \bibnamefont {Cory}}, \bibinfo {author}
  {\bibfnamefont {P.~R.}\ \bibnamefont {Hemmer}}, \bibinfo {author}
  {\bibfnamefont {R.~L.}\ \bibnamefont {Walsworth}}, \bibinfo {author}
  {\bibfnamefont {A.}~\bibnamefont {Yacoby}}, \bibinfo {author} {\bibfnamefont
  {A.~S.}\ \bibnamefont {Zibrov}},\ and\ \bibinfo {author} {\bibfnamefont
  {M.~D.}\ \bibnamefont {Lukin}},\ }\href
  {https://doi.org/10.1126/science.1176496} {\bibfield  {journal} {\bibinfo
  {journal} {Science}\ }\textbf {\bibinfo {volume} {326}},\ \bibinfo {pages}
  {267} (\bibinfo {year} {2009})}\BibitemShut {NoStop}%
\bibitem [{\citenamefont {Zhao}\ \emph {et~al.}(2024)\citenamefont {Zhao},
  \citenamefont {Xu}, \citenamefont {Shi}, \citenamefont {Chen}, \citenamefont
  {Kong}, \citenamefont {Yang}, \citenamefont {Wang}, \citenamefont {Ye},
  \citenamefont {Yu}, \citenamefont {Wang}, \citenamefont {Xie}, \citenamefont
  {Shi},\ and\ \citenamefont {Du}}]{zhao2024}%
  \BibitemOpen
  \bibfield  {author} {\bibinfo {author} {\bibfnamefont {Z.}~\bibnamefont
  {Zhao}}, \bibinfo {author} {\bibfnamefont {S.}~\bibnamefont {Xu}}, \bibinfo
  {author} {\bibfnamefont {Q.}~\bibnamefont {Shi}}, \bibinfo {author}
  {\bibfnamefont {Y.}~\bibnamefont {Chen}}, \bibinfo {author} {\bibfnamefont
  {X.}~\bibnamefont {Kong}}, \bibinfo {author} {\bibfnamefont {Z.}~\bibnamefont
  {Yang}}, \bibinfo {author} {\bibfnamefont {M.}~\bibnamefont {Wang}}, \bibinfo
  {author} {\bibfnamefont {X.}~\bibnamefont {Ye}}, \bibinfo {author}
  {\bibfnamefont {P.}~\bibnamefont {Yu}}, \bibinfo {author} {\bibfnamefont
  {Y.}~\bibnamefont {Wang}}, \bibinfo {author} {\bibfnamefont {T.}~\bibnamefont
  {Xie}}, \bibinfo {author} {\bibfnamefont {F.}~\bibnamefont {Shi}},\ and\
  \bibinfo {author} {\bibfnamefont {J.}~\bibnamefont {Du}},\ }\href
  {https://doi.org/10.1126/sciadv.adp9228} {\bibfield  {journal} {\bibinfo
  {journal} {Science Advances}\ }\textbf {\bibinfo {volume} {10}},\ \bibinfo
  {pages} {eadp9228} (\bibinfo {year} {2024})}\BibitemShut {NoStop}%
\bibitem [{\citenamefont {Jarmola}\ \emph {et~al.}(2012)\citenamefont
  {Jarmola}, \citenamefont {Acosta}, \citenamefont {Jensen}, \citenamefont
  {Chemerisov},\ and\ \citenamefont {Budker}}]{jarmola2012}%
  \BibitemOpen
  \bibfield  {author} {\bibinfo {author} {\bibfnamefont {A.}~\bibnamefont
  {Jarmola}}, \bibinfo {author} {\bibfnamefont {V.~M.}\ \bibnamefont {Acosta}},
  \bibinfo {author} {\bibfnamefont {K.}~\bibnamefont {Jensen}}, \bibinfo
  {author} {\bibfnamefont {S.}~\bibnamefont {Chemerisov}},\ and\ \bibinfo
  {author} {\bibfnamefont {D.}~\bibnamefont {Budker}},\ }\href
  {https://doi.org/10.1103/PhysRevLett.108.197601} {\bibfield  {journal}
  {\bibinfo  {journal} {Phys. Rev. Lett.}\ }\textbf {\bibinfo {volume} {108}},\
  \bibinfo {pages} {197601} (\bibinfo {year} {2012})}\BibitemShut {NoStop}%
\bibitem [{\citenamefont {Acosta}\ \emph {et~al.}(2010)\citenamefont {Acosta},
  \citenamefont {Bauch}, \citenamefont {Ledbetter}, \citenamefont {Waxman},
  \citenamefont {Bouchard},\ and\ \citenamefont {Budker}}]{acosta2010}%
  \BibitemOpen
  \bibfield  {author} {\bibinfo {author} {\bibfnamefont {V.~M.}\ \bibnamefont
  {Acosta}}, \bibinfo {author} {\bibfnamefont {E.}~\bibnamefont {Bauch}},
  \bibinfo {author} {\bibfnamefont {M.~P.}\ \bibnamefont {Ledbetter}}, \bibinfo
  {author} {\bibfnamefont {A.}~\bibnamefont {Waxman}}, \bibinfo {author}
  {\bibfnamefont {L.-S.}\ \bibnamefont {Bouchard}},\ and\ \bibinfo {author}
  {\bibfnamefont {D.}~\bibnamefont {Budker}},\ }\href
  {https://doi.org/10.1103/PhysRevLett.104.070801} {\bibfield  {journal}
  {\bibinfo  {journal} {Phys. Rev. Lett.}\ }\textbf {\bibinfo {volume} {104}},\
  \bibinfo {pages} {070801} (\bibinfo {year} {2010})}\BibitemShut {NoStop}%
\bibitem [{\citenamefont {Rosskopf}\ \emph {et~al.}(2017)\citenamefont
  {Rosskopf}, \citenamefont {Zopes}, \citenamefont {Boss},\ and\ \citenamefont
  {Degen}}]{rosskopf2017}%
  \BibitemOpen
  \bibfield  {author} {\bibinfo {author} {\bibfnamefont {T.}~\bibnamefont
  {Rosskopf}}, \bibinfo {author} {\bibfnamefont {J.}~\bibnamefont {Zopes}},
  \bibinfo {author} {\bibfnamefont {J.~M.}\ \bibnamefont {Boss}},\ and\
  \bibinfo {author} {\bibfnamefont {C.~L.}\ \bibnamefont {Degen}},\ }\href
  {https://doi.org/10.1038/s41534-017-0030-6} {\bibfield  {journal} {\bibinfo
  {journal} {npj Quantum Inf.}\ }\textbf {\bibinfo {volume} {3}},\ \bibinfo
  {pages} {33} (\bibinfo {year} {2017})}\BibitemShut {NoStop}%
\bibitem [{\citenamefont {Escher}\ \emph {et~al.}(2011)\citenamefont {Escher},
  \citenamefont {de~Matos~Filho},\ and\ \citenamefont
  {Davidovich}}]{escher2011}%
  \BibitemOpen
  \bibfield  {author} {\bibinfo {author} {\bibfnamefont {B.~M.}\ \bibnamefont
  {Escher}}, \bibinfo {author} {\bibfnamefont {R.~L.}\ \bibnamefont
  {de~Matos~Filho}},\ and\ \bibinfo {author} {\bibfnamefont {L.}~\bibnamefont
  {Davidovich}},\ }\href {https://doi.org/10.1038/nphys1958} {\bibfield
  {journal} {\bibinfo  {journal} {Nature Physics}\ }\textbf {\bibinfo {volume}
  {7}},\ \bibinfo {pages} {406} (\bibinfo {year} {2011})}\BibitemShut {NoStop}%
\bibitem [{\citenamefont {Bonato}\ \emph {et~al.}(2016)\citenamefont {Bonato},
  \citenamefont {Blok}, \citenamefont {Dinani}, \citenamefont {Berry},
  \citenamefont {Markham}, \citenamefont {Twitchen},\ and\ \citenamefont
  {Hanson}}]{bonato2016}%
  \BibitemOpen
  \bibfield  {author} {\bibinfo {author} {\bibfnamefont {C.}~\bibnamefont
  {Bonato}}, \bibinfo {author} {\bibfnamefont {M.~S.}\ \bibnamefont {Blok}},
  \bibinfo {author} {\bibfnamefont {H.~T.}\ \bibnamefont {Dinani}}, \bibinfo
  {author} {\bibfnamefont {D.~W.}\ \bibnamefont {Berry}}, \bibinfo {author}
  {\bibfnamefont {M.~L.}\ \bibnamefont {Markham}}, \bibinfo {author}
  {\bibfnamefont {D.~J.}\ \bibnamefont {Twitchen}},\ and\ \bibinfo {author}
  {\bibfnamefont {R.}~\bibnamefont {Hanson}},\ }\href
  {https://doi.org/10.1038/nnano.2015.261} {\bibfield  {journal} {\bibinfo
  {journal} {Nature Nanotechnol.}\ }\textbf {\bibinfo {volume} {11}},\ \bibinfo
  {pages} {247} (\bibinfo {year} {2016})}\BibitemShut {NoStop}%
\bibitem [{\citenamefont {Barry}\ \emph {et~al.}(2020)\citenamefont {Barry},
  \citenamefont {Schloss}, \citenamefont {Bauch}, \citenamefont {Turner},
  \citenamefont {Hart}, \citenamefont {Pham},\ and\ \citenamefont
  {Walsworth}}]{barry2020}%
  \BibitemOpen
  \bibfield  {author} {\bibinfo {author} {\bibfnamefont {J.~F.}\ \bibnamefont
  {Barry}}, \bibinfo {author} {\bibfnamefont {J.~M.}\ \bibnamefont {Schloss}},
  \bibinfo {author} {\bibfnamefont {E.}~\bibnamefont {Bauch}}, \bibinfo
  {author} {\bibfnamefont {M.~J.}\ \bibnamefont {Turner}}, \bibinfo {author}
  {\bibfnamefont {C.~A.}\ \bibnamefont {Hart}}, \bibinfo {author}
  {\bibfnamefont {L.~M.}\ \bibnamefont {Pham}},\ and\ \bibinfo {author}
  {\bibfnamefont {R.~L.}\ \bibnamefont {Walsworth}},\ }\href
  {https://doi.org/10.1103/RevModPhys.92.015004} {\bibfield  {journal}
  {\bibinfo  {journal} {Rev. Mod. Phys.}\ }\textbf {\bibinfo {volume} {92}},\
  \bibinfo {pages} {015004} (\bibinfo {year} {2020})}\BibitemShut {NoStop}%
\end{thebibliography}%

\end{document}